\documentclass{aa}  

\usepackage{amsmath}
\usepackage{graphicx}
\usepackage{graphics}
\usepackage{subfigure}
\usepackage{txfonts}
\usepackage{natbib}
\usepackage{xcolor}
\usepackage{soul}
\usepackage{tablefootnote}
\usepackage{pdflscape}
\usepackage{tabularx}
\usepackage{threeparttable}
\usepackage{rotating}
\usepackage{longtable}
\usepackage{wrapfig}
\usepackage{soul}

\usepackage{hyperref}

\bibpunct{(}{)}{;}{a}{}{,}

\newcommand{\hi}{H~{\sc i}}
\newcommand{\hii}{H~{\sc ii}}
\newcommand{\ha}{\ifmmode {\rm H}\alpha \else H$\alpha$\fi}
\newcommand{\hb}{\ifmmode {\rm H}\beta \else H$\beta$\fi}
\newcommand{\lya}{\ifmmode {\rm Ly}\alpha \else Ly$\alpha$\fi}

\newcommand{\ebv}{\ifmmode E_{\rm B-V} \else $E_{\rm B-V}$\fi}
\newcommand{\av}{\ifmmode A_{\rm V} \else $A_{\rm V}$\fi}
\newcommand{\alphaCO}{\ifmmode \alpha_{\rm CO} \else $\alpha_{\rm CO}$\fi}

\def\msun{\ifmmode M_{\odot} \else M$_{\odot}$\fi}
\def\msunyr{\ifmmode M_{\odot} {\rm yr}^{-1} \else M$_{\odot}$ yr$^{-1}$\fi}
\def\zsun{\ifmmode Z_{\odot} \else Z$_{\odot}$\fi}
\def\lsun{\ifmmode L_{\odot} \else L$_{\odot}$\fi}
\def\mup{\ifmmode M_{\rm up} \else M$_{\rm up}$\fi}
\def\mlow{\ifmmode M_{\rm low} \else M$_{\rm low}$\fi}

\def\aap{A\&A}

\def\apjl{ApJL}
\def\apjs{ApJS}

\newcommand{\oh}{\ifmmode 12 + \log({\rm O/H}) \else$12 + \log({\rm O/H})$\fi}
\newcommand{\nii}{[N~{\sc ii}]}
\newcommand{\niii}{[N~{\sc iii}]}
\newcommand{\oi}{[O~{\sc i}]}

\newcommand{\oiii}{[O~{\sc iii}]}
\newcommand{\oiv}{[O~{\sc iv}]}

\newcommand{\siii}{[S~{\sc iii}]}
\newcommand{\siv}{[S~{\sc iv}]}
\newcommand{\ci}{[C~{\sc i}]}
\newcommand{\cii}{[C~{\sc ii}]}

\newcommand{\neiii}{[Ne~{\sc iii}]}
\newcommand{\neii}{[Ne~{\sc ii}]}
\newcommand{\nev}{[Ne~{\sc v}]}

\newcommand{\feii}{[Fe~{\sc ii}]}
\newcommand{\feiii}{[Fe~{\sc iii}]}

\newcommand{\silii}{[Si~{\sc ii}]}
\newcommand{\arii}{[Ar~{\sc ii}]}
\newcommand{\ariii}{[Ar~{\sc iii}]}

\def\fesc{\ifmmode f_{\rm esc} \else $f_{\rm esc}$\fi}
\def\feschii{\ifmmode f_{\rm esc,HII} \else $f_{\rm esc,HII}$\fi}

\defcitealias{Ramambason2022}{R22}
\defcitealias{madden_tracing_2020}{M20}
\defcitealias{Bolatto_2013}{B13}

\begin{document}
  \title{Modeling the molecular gas content and CO-to-H$_2$ conversion factors in low-metallicity star-forming dwarf galaxies}
  \subtitle{}
  \author{L. Ramambason\inst{1,2}, 
  V. Lebouteiller\inst{2}, S. C. Madden\inst{2}, F. Galliano\inst{2}, C. T. Richardson\inst{3}, A. Saintonge\inst{4}, I. De Looze\inst{4,5}, M. Chevance\inst{1,6}, N. P. Abel\inst{7}, S. Hernandez\inst{8}, J. Braine\inst{9}}
  \institute{Institut fur Theoretische Astrophysik, Zentrum für Astronomie, Universität Heidelberg, Albert-Ueberle-Str. 2, D-69120 Heidelberg, Germany\\
  email: \texttt{lise.ramambason@uni-heidelberg.de}
  \and Université Paris Cité, Université Paris-Saclay, CEA, CNRS, AIM, F-91191, Gif-sur-Yvette, France
  \and Physics Department, Elon University, 100 Campus Drive CB 2625, Elon, NC, 27244, USA
  \and Department of Physics \& Astronomy, University College London, Gower Street, London WC1E 6BT, UK
  \and Sterrenkundig Observatorium, Ghent University, Krijgslaan 281 – S9, B-9000 Ghent, Belgium
  \and Cosmic Origins Of Life (COOL) Research DAO, coolresearch.io
  \and University of Cincinnati, Clermont College, 4200 Clermont College Drive, Batavia, OH 45103, USA
  \and AURA for ESA, Space Telescope Science Institute, 3700 San Martin Drive, Baltimore, MD 21218, USA
  \and Laboratoire d’Astrophysique de Bordeaux, Univ. Bordeaux, CNRS, B18N, Allée Geoffroy Saint-Hilaire, 33615 Pessac, France}

\date{Received 26 June 2023 / Accepted 6 October 2023}

\abstract{
Low-metallicity dwarf galaxies often show no or little CO emission, despite the intense star formation observed in local samples. Both simulations and resolved observations indicate that molecular gas in low-metallicity galaxies may reside in small dense clumps, surrounded by a substantial amount of more diffuse gas, not traced by CO. Constraining the relative importance of CO-bright versus CO-dark H$_2$ star-forming reservoirs is crucial to understand how star formation proceeds at low metallicity.
}
{
We put to the test classically used single component radiative transfer models and compare their results to those obtained assuming an increasingly complex structure of the interstellar gas, mimicking an inhomogeneous distribution of clouds with various physical properties.
}
{
Using the Bayesian code MULTIGRIS, we compute representative models of the interstellar medium as combinations of several gas components, each with a specific set of physical parameters. We introduce physically-motivated models assuming power-law distributions for the density, ionization parameter, and the depth of molecular clouds.
}
{
This new modeling framework allows reproducing simultaneously the spectral constraints from the ionized gas, neutral atomic gas, and molecular gas in 18 galaxies from the Dwarf Galaxy Survey.
We confirm the presence of a predominantly CO-dark molecular reservoir in low-metallicity galaxies. The predicted total H$_2$ mass is best traced by \cii 158$\mu$m and, to a lesser extent, by \ci\  609$\mu$m, rather than by CO(1-0). 
We examine the CO-to-H$_2$ conversion factor ($\alphaCO$) vs. metallicity relation and find that its dispersion increases significantly when different  geometries of the gas are considered. We define a “clumpiness” parameter that anti-correlates with \cii/CO and explains the dispersion of the $\alphaCO$ vs. metallicity relation. We find that low-metallicity galaxies with high clumpiness parameters may have $\alphaCO$ values as low as the Galactic value, even at low-metallicity.
}
{
We identify the clumpiness of molecular gas as a key parameter to understand variations of geometry-sensitive quantities, such as $\alphaCO$. This new modeling framework enables the derivation of constraints on the internal cloud distribution of unresolved galaxies, based only on their integrated spectra.
}


\authorrunning{L. Ramambason et al.}
\titlerunning{CO-dark gas in the DGS}
\maketitle

\section{Introduction}
\label{section_intro}

Quantifying the total amount of molecular gas hosted within a galaxy is an important step to understand how stars form in different environments. The presence of cold molecular gas is thought to be a necessary element to fuel the formation of new stars through the fragmentation of dense molecular clouds \citep[e.g.,][]{Chevance_life_MC_2022}. Its direct detection is complicated by the fact that H$_2$ has no electric dipole moment. While rovibrational H$_2$ emission can be detected in the infrared, the latter only traces a warm (E/k $\gtrsim$ 510\,K, \citealt{Roueff_2019}) H$_2$ component rather than the total H$_2$ mass. As an alternative, CO emission has been extensively used to probe the colder H$_2$ component in large samples of local galaxies with instruments such as IRAM, APEX (e.g.,  HERACLES \citealt{Leroy_heracles_2009}, xCOLD gas survey; \citealt{Saintonge_xcoldgass_2017},  \citealt{Montoya_Arroyave_2023}), ALMA \citep[e.g. ALMA-PHANGS,][]{Leroy_CO_phangs_2021}, in systems at intermediate redshift \citep[z < 1, e.g.,;][]{Freundlich_2019}, and up to redshifts of $\sim 6$ \citep[e.g.,][]{Ginolfi_2017, Boogaard_2023}.

Nevertheless, converting the CO emission into estimates of the total molecular gas masses is not straightforward. A first complication comes from the fact that the CO-to-H$_2$ conversion factor (hereafter $\alphaCO$ in pc$^{-2}$(K km s$^{-1}$)$^{-1}$) is sensitive to internal variations of the physical properties of the interstellar medium (ISM), rendering the interpretation of an averaged galactic value difficult. At galactic scales, the $\alphaCO$ conversion factor can be derived as a statistical average for the populations of CO-emitting clouds, assuming, in particular, that they are virialized and nonoverlapping \citep[e.g.,][hereafter \citetalias{Bolatto_2013}]{Dickman_1986, Bolatto_2013}. Nevertheless, important cloud-to-cloud variations are expected within the ISM of galaxies \citep[e.g.,][]{Sun_2020, Sun_prop_clouds_2022}, making the $\alphaCO$ values particularly sensitive to the spatial variations within galaxies. 

In particular, radial gradients of $\alphaCO$ have been observed in local spiral galaxies \citep[e.g.,][]{Teng_2022, den_Brok_2023} with lower conversion factors derived in galactic centers \citep[e.g.,][]{Sandstrom_2013}. 
Environmental effects may also impact star formation mechanisms, leading to dependencies of the $\alphaCO$ values on global galactic parameters \citep[e.g.,][]{Accurso_2017}. In particular, both the
metallicity and dust content of galaxies strongly impact the $\alphaCO$ values \citep[e.g.][]{Glover_MacLow_2011, Schruba_2012, Bolatto_2013,Genzel_2015, Amorin_2016,Accurso_2017,Tacconi_2018, madden_tracing_2020}. Recently, \cite{Hirashita_2023} has suggested that the $\alphaCO$ conversion factor may be sensitive not only to the dust-to-gas mass
ratio but also to the size distribution of dust grains, which impacts the formation and destruction pathways of both H$_2$ and
CO molecules. In low-metallicity and dust-poor galaxies, the radiation field may penetrate deep into the molecular cloud envelopes and photodissociate CO molecules, while H$_2$ may remain self-shielded. As a result, large amounts of “CO-dark” \citep[e.g.,][]{Wolfire_2010} H$_2$ gas, invisible in CO, have been reported in low-metallicity galaxies \citep[e.g.,][]{Poglitsch_1995, Madden_1997, Schruba_2012, Amorin_2016, Schruba_2017, Cormier_2014, madden_tracing_2020}.

Several methods have been explored to recover the fraction of molecular gas hidden in this CO-dark gas component. The direct detections of high-rotational level H$_2$ in the mid-IR can be used to infer the warm and total H$_2$ masses, based on assumptions on the distribution of H$_2$ rotational temperatures \citep{Togi_2016}. This method is especially promising in the context of JWST, which enables new detections of mid-IR H$_2$ lines with an unprecedented sensitivity \citep[e.g.,][]{Hernandez_2023}. At low metallicities, however, direct H$_2$ detection remains challenging and alternative methods relying on indirect CO-dark gas tracers are needed.  While Galactic studies can rely on numerous indirect tracers, including for example $\gamma$-ray emission \cite[e.g.,][]{Grenier_2005,Ackermann_2012, Remy_2018,Hayashi_2019} or molecular absorption lines \citep[e.g.,][]{Liszt_Lucas_oh_1996, Lucas_Liszt_hco_1996, Allen_2015, Nguyen_2018}, studies of external galaxies must rely on the direct emission of luminous tracers associated with H$_2$ reservoirs. In particular, the
dust continuum has been classically used to estimate the total H$_2$ mass in massive galaxies \citep[e.g.,][]{Magnelli_2012, Sandstrom_2013, Genzel_2015, Tacconi_2018, Zavala_2022}, as well as in the small Local Group spiral galaxy M33 \citep[$Z \sim 1/2 Z_{\odot}$; e.g.,][]{Braine_2010, Gratier_2017} and in the SMC \citep[][]{Tokuda_2021}. 

The \cii 158$\mu$m emission line also provides a widely used proxy of the molecular gas content \citep[e.g.,][]{Zanella_2018,Bethermin_2020, Dessauges-Zavadsky_2020}, including in dwarf galaxies \citep[e.g.,][]{Jameson_2018, madden_tracing_2020} and a potentially promising proxy for high-redshift studies \citep[e.g.,][]{Vizgan_2022}. More recently, the \ci\ emission line has also been identified as a potential tracer of the molecular gas \citep[e.g.,][]{Jiao_2019,Crocker_2019, madden_tracing_2020,Dunne_2021,Dunne_2022}, although its detection remains relatively more challenging.

A proper calibration of the $\alphaCO$ conversion factor, accounting for the CO-dark component, is crucial to solve long-standing debates regarding star formation mechanisms in low-metallicity environments. Among other, the lack of H$_2$ and CO emission in star-forming low-metallicity galaxies \citep[e.g.,][]{Tacconi_1987, Taylor_1998, Cormier_2014, Cormier_2017, Leroy_2007} could indicate unusually high star formation efficiencies at low-metallicity \citep[e.g.,][]{Turner_2015}. Alternative scenarios include the existence of mechanisms preventing the formation of H$_2$ molecules, in particular due to the smaller amount of dust grains on which H$_2$ form, or their disruption in the aftermath of star formation. It also could possibly serve as evidence for the existence of star-formation pathways directly in the neutral atomic gas \citep[e.g.,][]{Glover_Clark_necessay_2012, Glover_clark_sf_2012}, with important implications for star formation at high-redshift.

While primordial galaxies remain beyond the reach of current CO-observing facilities, local low-metallicity dwarf galaxies with CO measurements are ideal to investigate those questions. In \cite{madden_tracing_2020} (hereafter \citetalias{madden_tracing_2020}), we analyzed a sample of nearby star forming low-metallicity galaxies from the Dwarf Galaxy Survey \citep[DGS;][]{Madden_2013}. Using the wealth of available infrared spectral tracers to constrain radiative transfer models, we inferred the total H$_2$ mass in each galaxy of this sample. Our results suggest that most of their molecular mass may reside in CO-dark layers. Accounting for this CO-dark component, \citetalias{madden_tracing_2020} report that these dwarf galaxies fall back on the Kennicutt-Schmidt (KS) relation linking star formation rates (SFR) and stellar masses, from which they were significantly offset when accounting only for the CO-visible H$_2$ mass. 

While radiative transfer models, such as those used in \citetalias{madden_tracing_2020} enable a deeper understanding of the underlying physics of the ISM than empirical studies, they rely on strong modeling assumptions. In particular, the Cloudy models used in \citetalias{madden_tracing_2020} assumed a simple geometry, with the emission arising from the neutral ISM (atomic and molecular phase) matched by a single component (single metallicity and single ionizing source). This simplification of the actual complexity of the multiphase ISM was necessary to keep a reasonably low number of free parameters and to use the CO emission as a direct constraint on the visual extinction of molecular clouds (A$_V$). Nevertheless, it prevented performing a fully consistent optimization of the free parameters, since A$_V$ was manually adjusted a posteriori to match the observed CO emission. Under those assumptions, \citetalias{madden_tracing_2020} found a strong anti-correlation of A$_V$ with the \cii/CO emission line ratio and an anti-correlation between $\alphaCO$ and A$_V$. Because visual extinction correlates by design with the gas-phase metallicity in single component models, the latter results also imply an anti-correlation of the $\alphaCO$ with metallicity. \citetalias{madden_tracing_2020} report a negative slope of the $\alphaCO$ vs. metallicity relation, with a narrow dispersion ($<0.32$\,dex) of the DGS galaxies around it.

In the current study, we explore a more realistic set-up by relaxing the geometrical constraints imposed in \citetalias{madden_tracing_2020}. We use a new modeling framework, enabling the combination of multiple ISM components. This “topological” representation follows that introduced in \cite{pequignot_heating_nodate} and extensively applied to galaxies, including those drawn from the DGS \citep{2012_Cormier, polles_modeling_2019, cormier_herschel_2019, lebouteiller_neutral_2017}. Those models require the simultaneous determination of numerous free parameters, which is difficult to robustly achieve with frequentist methods such as a $\chi^2$ minimization. This difficulty was leveraged by the development of MULTIGRIS\footnote{ \href{https://gitlab.com/multigris/mgris}{https://gitlab.com/multigris/mgris}} \citep{MULTIGRIS2022}, which provides a new framework for model combination based on Bayesian statistics. 
While this new tool has successfully allowed the modeling of the ionized and neutral ISM of the DGS galaxies with unprecedented detailed models (up to four components; \citealt{LebouteillerRamambason2022, Ramambason2022}), this representation can still be improved. In particular, the use of statistical distributions to parameterize the combination of numerous components has been introduced in several recent studies to describe the hydrogen density and ionization parameter \citep{richardson_interpreting_2014,richardson_interpreting_2016, richardson_addressing_2019}, as well as the visual extinction \citep{Bisbas_2021} of clouds distributed within the ISM. These distribution functions provide a different representation, closer to what is predicted by simulations of a gravoturbulent star-forming ISM \citep[e.g.,][]{Offner_2014, Burkhart_2018, Burkhart_2019, Appel_2023} and observed in the nearby universe \citep[e.g.,][]{Brunt_2015, Lombardi_2015}

The work presented here uses the new ISM modeling framework from MULTIGRIS to revisit the \citetalias{madden_tracing_2020} results, assuming a more complex geometry. The data used in this analysis, including updated CO measurements, is presented in Section \ref{section_observations}. In Section \ref{section_models}, we describe the different architectures of models (single and multicomponent). We compare their results to previous models from \citetalias{madden_tracing_2020} and motivate the choice of an optimal architecture in Section \ref{section_results}. We then focus on the results obtained using the power-law models in Section \ref{section_results_plaw}. The physical interpretations, caveats, and possible improvements of our study are discussed in Section \ref{section_discussion}. Our main findings are summarized in Section \ref{section_conclusion}.

\section{Data}
\label{section_observations}
\subsection{Sample}
\label{section_sample}

\begin{table}[t!]
\begin{threeparttable}
\centering
\caption[Integrated measurements in the DGS.]{Integrated CO(1-0) measurements in the DGS galaxies.}
\label{table_CO_dgs}

\begin{tabular}{p{0.01\textwidth}p{0.07\textwidth}p{0.03\textwidth}p{0.08\textwidth}p{0.12\textwidth}p{0.01\textwidth}}
\# & Galaxy & $\theta ['']$ & K\,km\,s$^{-1}$ & 10$^{-20} \times$ W\,m$^{-2}$ & Ref.\\ \hline 

\hline \hline
 1 & Haro\,2 & 55 & 1.13$\pm$0.26 & 14.25$\pm$3.28 & (1)\\
\hline
 2 & Haro\,3 & 22 & 3.07$\pm$0.29 & 6.19$\pm$0.58 & (2)\\
\hline
 3 & Haro\,11 & - & - & 1.32$\pm$0.04 & (3)\\
\hline
 4 & He2\,10 & 55 & 4.85$\pm$0.24 & 61.18$\pm$3.03 & (5)\\
\hline
 5 & II\,Zw\,40 & 45 & 0.46$\pm$0.10 & 3.88$\pm$0.84 & (6)\\
\hline
 6 & I\,Zw\,18 & - & - & 0.0018$\pm$0.0005 & (7)\\
\hline
 7 & Mrk\,209 & 45 & 0.45$\pm$0.10 & 3.80$\pm$0.84 & (6)\\
\hline
 8 & Mrk\,930 & 22 & 0.14$\pm$0.07 & 0.28$\pm$0.14 & (1)\\
\hline
 9 & Mrk\,1089 & 22 & 1.53$\pm$0.30 & 3.10$\pm$0.61 & (4)\\
\hline
 10 & NGC\,1569 & 55 & 0.69$\pm$0.10 & 8.64$\pm$1.31 & (8)\\
\hline
 11 & NGC\,1140 & 22 &  1.00$\pm$0.13 & 2.01$\pm$0.26 & (2)\\
\hline
 12 & NGC\,1705 & - & - & 0.43$\pm$0.21$^{*}$ & (9)\\
\hline
 13 & NGC\,5253 & - & - & 3.72$\pm$0.37$^{*}$ & (9)\\
\hline
 14 & NGC\,625 & 30 & 2.05$\pm$0.63 & 7.69$\pm$2.36 & (4)\\
\hline
 15 & SBS\,0335 & 1 & $\leq 2.63$ & $\leq$ 0.01 & (10)\\
\hline
 16 & UM\,448 & 22 & 1.04$\pm$0.22 & 2.09$\pm$0.44 & (2)\\
\hline
 17 & UM\,461 & 55 & $\leq 0.78$ & $\leq$ 9.84 & (11)\\
\hline
 18 & VII\,Zw\,403 & 55 & $\leq 0.99$ & $\leq$ 12.49 & (12)\\
\hline
\end{tabular}
\begin{tablenotes}
   \footnotesize
   \item When available, we report the corresponding beam size $\theta$ used in the conversion from K km s$^{-1}$ to W\,m$^{-2}$. Most measurements correspond to CO(1-0) direct detections, except for I\,Zw\,18 and SBS\,0335-52. The uncertainties and upper limits are reported at 3$\sigma$, unless stated otherwise.
   \item (1): \cite{Thronson_Bally_1987}
   \item (2): \cite{Hunt_2015}
   \item (3): \cite{Gao_Haro11_2022}
   \item (4): \cite{Cormier_2014}
   \item (5): \cite{Kobulnicky_1995}
   \item (6): \cite{Young_1995}
   \item (7): Conversion based on CO(2-1); \cite{Zhou_co_2021}. 
   \item (8): \cite{Greve_1996}
   \item (9): \cite{Hunt_2023}. 
   \item (10): Conversion based on CO(3-2); \cite{Hunt_2014}. 
   \item (11): \cite{Sage_1992}
   \item (12): \cite{Leroy_2005}
   \item *: We consider a 50\% uncertainty for NGC\,1705 to account for the discrepancy between the two values reported for 7-m and 12-m integrated fluxes. For NGC\,5253, we adopt a 10\% uncertainty.
\end{tablenotes}

\end{threeparttable}
\end{table}

Our sample is drawn from the DGS survey \citep{Madden_2013} which gathers observation of 50 nearby (0.5--191 Mpc) star-forming dwarf galaxies. This sample spans a large range of physical conditions and in particular a wide range of sub-solar\footnote{We use the values from \cite{Asplund_2009} as solar references with the total mass fraction of metals $Z_{\odot}$=0.0134 and the oxygen abundance (O/H)$_{\odot} = 4.9 \times 10^{-4}$; i.e., \oh$_{\odot}$ = 8.69.} metallicities from \oh=7.14 ($\sim$ 1/35 Z$_\odot$) up to 8.43 ($\sim$ 1/2 Z$_\odot$).
This sample has been observed both in the far-infrared and submillimeter domains with the \textit{Herschel} Space Telescope, as well as in the mid-infrared (MIR) domain with the \textit{Spitzer} observatory. The wealth of emission lines arising from the different phases of the ISM makes it an ideal sample to constrain in detail the ISM structure. We restrict our sample to 18 compact galaxies for which either a detection or an upper limit in CO is available for the whole galaxy. We excluded galaxies for which only partial regions were observed. 

In Table \ref{table_CO_dgs}, we list the CO(1-0) measurements and upper limits taken from the literature, with their associated uncertainties. This table is based on that used in \citetalias{madden_tracing_2020} but includes updated measurements and their reported uncertainties. Most measurements are expressed in K\,km\,s$^{-1}$ and are reported with the instrumental beam $\theta$, expressed in arcseconds. We also report the CO conversion to integrated flux in W m$^{-2}$ based on \cite{Solomon_1997}, assuming that the angular size of the source was negligible compared to the telescope beam: $\Omega_{\rm S} << \Omega_{\ \rm b}$. Hence, we approximate the solid angle of the source convolved with the telescope beam $\Omega_{\rm S * b} \approx \Omega_{\rm b} \approx \theta^2$, with $\theta$ the full width at half maximum of the telescope beam in arcsecs. 

This assumption is equivalent to fixing arbitrarily the size of the source, which is unknown. 
It introduces uncertainties linked to the potential presence of CO emission outside the beam, which could lead to underestimation of the molecular gas masses. Most importantly, \cite{Cormier_2014} stress that this uncertainty may hamper the comparison between different CO line transitions, corresponding to different beam sizes, since their ratio is quite sensitive to the choice of beam used for the reduction. In the current study, we hence focus only on the CO(1-0) line, for which the largest number of detections is available. We note that transitions of higher energy level (e.g., CO(2-1) and CO(3-2)), as well as emission from $^{13}$CO have also been detected in a few galaxies, which we do not include in the current study. Nevertheless, for the two lowest metallicity galaxies in our sample for which no CO(1-0) detection is reported, we use conversions based on other transitions. For SBS\,0335-052, we use the CO(3-2) detection reported in \cite{Hunt_2014} to estimate CO(1-0) luminosity, assuming an area of $1\arcsec$ subtended by the source. For I\,Zw\,18, we use the CO(2-1) detection reported in \cite{Zhou_co_2021} at 3.5$\sigma$ significance, which they used to estimate a CO(1-0) luminosity, assuming an optically thick and thermalized emission. 

Finally, we include the two recent CO(1-0) detections with the ALMA 12-m array for Haro\,11 and NGC\,1705, reported in \cite{Gao_Haro11_2022} and \cite{Hunt_2023}, respectively. Considering the new detections rather than the previous upper limits on CO lowers the predicted $\alphaCO$ values by a factor $\sim$72 for Haro\,11 and a factor $\sim$6 for NGC 1705. Nevertheless, the predictions obtained for all the main quantities discussed throughout the paper (emission lines, masses, and conversion factors) remain compatible with the previous values within errorbars.\footnote{Errorbars are defined based on the upper and lower-bounds of the High Density Probability Interval at 94\%.}. While the exact values of physical parameters are slightly modified (e.g., up to a factor $\sim$3 for the predicted H$_2$ masses), it does not change any of the trends discussed throughout the paper.

\cite{Hunt_2023} also report another detection obtained for NGC\,1705  with the ACA 7-m array, which is a factor of two lower. To account for this uncertainty on the integrated flux, we consider a large $\sigma$ value of half the ALMA 12-m detection we consider a detection uncertainty of 50\% of the total flux. Two new measurements for NGC\,625 and NGC\,5253 are also provided in \cite{Hunt_2023}. We include the updated measurement for NGC\,5253, assuming an uncertainty of 10\%, as the new value differ significantly from the value from \cite{Taylor_1998} previously used in \citetalias{madden_tracing_2020} (lower by a factor 2.5). Using the new detection yields an H$_2$ mass lower by a factor 2.2, and a $\alphaCO$ value higher by a factor 1.14. As for Haro\,11 and NGC\,1705, those variations are smaller than the errorbars associated with our predictions. The ALMA 12-m integrated luminosity reported in \cite{Hunt_2023} for NGC\,625 corresponds to a luminosity of $6.59 \times 10^{-20}$\,W/m$^2$, slightly lower but compatible within errorbars with the measurement from \cite{Cormier_2014} used in the current study. 

\subsection{Tracers of multiphase ISM}
\label{sect_tracer_ism}
\begin{table}[t]
    \centering
    \caption{IR tracers used as constraints and corresponding ionization potentials\protect\footnotemark[1] for ionic lines.}
    \begin{tabular}{p{0.15\textwidth}p{0.25\textwidth}}
         & Tracers \\
        \hline
        \hline
        Molecules & CO(1-0)$\lambda$2.6\,mm \\
        Neutral and ionized & \oi$\lambda\lambda$63,145$\mu$m\\
        gas tracers & \feii$\lambda\lambda$17,26$\mu$m (7.9eV), \\
         & \silii$\lambda$35$\mu$m (8.2eV),\\
         & \cii$\lambda$158$\mu$m (11.3eV),\\  
         & Hu$\alpha \lambda$12$\mu$m (13.6eV),\\
         & \nii$\lambda\lambda$122,205$\mu$m (14.5eV),\\
         & \arii$\lambda$7$\mu$m (15.7eV),\\
         & \feiii$\lambda$23$\mu$m (16.2eV),\\
         & \neii$\lambda$13$\mu$m (21.6eV),\\
         & \siii$\lambda\lambda$19,33$\mu$m (23.3eV),\\
         & \ariii$\lambda\lambda$9,22$\mu$m (27.6eV),\\
         & \niii$\lambda$57$\mu$m (29.6eV), \\
         & \siv$\lambda$11$\mu$m (34.7eV),\\
         & \oiii$\lambda$88$\mu$m (35.1eV), \\
         & \neiii$\lambda$16$\mu$m (40.9eV)\\
         &\oiv$\lambda$26$\mu$m (54.9eV),\\ 
         & \nev$\lambda\lambda$14,24$\mu$m (97.1eV), \\
         Total IR luminosity & $L_{\rm TIR}$ (1$\mu$m-1000$\mu$m) \\
         \hline
    \end{tabular}
    \begin{tablenotes}
    \footnotesize
    \item $^1$We report the ionization potentials corresponding to the energy thresholds required to create the ion producing a given emission line, either by deexcitation or by recombination.
    \end{tablenotes}
    \label{tracers}
\end{table}

In Table \ref{tracers}, we list the spectral tracers used as constraints. The tracers used to constrain the ionized gas and the photodissociated regions (PDR) are similar to those used in \cite{Ramambason2022}. In addition to those tracers, the current analysis requires the use of tracers emitted in the molecular gas. As we will discuss in Section \ref{tracer_of_MH2}, the choice of tracers associated with molecular gas is not straightforward, especially in low-metallicity environments. In particular, part of the H$_2$ may form in the PDR due to the self-shielding of H$_2$ molecules. Hence, PDR tracers are particularly useful to estimate the total amount of molecular gas mass. When available, we include the following tracers that partly or totally emit in the PDRs: \oi, \feii, \silii, and \cii.

As opposed to \cite{Ramambason2022}, we do not include the H$_2$ rovibrational lines as constraints, which consist mostly of upper limits, and use CO(1-0) instead. This exclusion is motivated by the coarse radial depth sampling in our models (see Section \ref{sampling_depth}), which does not capture the sharp transition in the H$_2$ cumulative emission. As shown in the Figure \ref{H2_problem} from the Appendix, the emission of H$_2$ may sharply increase before the H/H$_2$ transition (defined in terms of fractional abundances). Because this transition is not well captured, this creates an abrupt jump in our predictions, with models stopping at the ionization front predicting no H$_2$ emission, while all models stopping after the ionization front predict substantial H$_2$ emission. As a result, our model predictions can only match the H$_2$ upper limits with models that are completely deprived of molecular gas and PDR. Nevertheless, we check a posteriori how the predicted H$_2$ luminosities compare with the H$_2$ detections and upper limits, as shown in Figure \ref{H2_obs_pred} in the Appendix. We find that H$_2$\,S(0) and H$_2$\,S(1) upper limits are systematically overpredicted by our models, while this issue is somehow mitigated for the H$_2$ detections. We find that our predictions agree  within 0.5\,dex with all the H$_2$ detections, despite slight overpredictions of H$_2$\,S(0) and H$_2$\,S(1).
The coarse radial sampling does not affect CO predicted emission, which has a smoother radial profile and probes a colder molecular gas reservoir, at larger A$_V$ (see Figure \ref{H2_problem}). This motivates the choice of our sample, described in Section \ref{section_sample}, for which CO(1-0) measurements or upper limits are available. 

\section{Modeling framework}
\label{section_models}

\subsection{SFGX grid}

\subsubsection{Overview}
\label{subsect_sfgx}

The individual models or “components” used in this study are drawn from the “Star-Forming Galaxies with X-ray sources” (SFGX) grid of models presented in \cite{Ramambason2022}, tailored to study star-forming low-metallicity dwarf galaxies. This grid consists of spherical models computed with the photoionization and photodissociation code Cloudy v17.02 \citep{2017_Cloudy_v17} and spans a large metallicity range going from $\sim 1/100\,$Z$_{\odot}$ to 2\,Z$_{\odot}$. Eight parameters were varied, associated with the physical properties of the radiation sources (the stellar luminosity, age of the stellar burst, X-ray luminosity, and temperature of the X-ray source assumed to be a multicolor blackbody) and the gas (the metallicity, the density at the illuminated edge $n$, and the ionization parameter $U$ at the illuminated edge). The ionization parameter at the illuminated edge is defined as follows:
\begin{equation}
    U(R_{\rm in}) = \frac{Q({\rm H^0})}{4 \pi n_0 c R_{\rm in}^2},
\end{equation}
where Q(H$^0$) is the number of ionizing photons emitted per second, $n_0$ the density at the illuminated edge, located at the inner radius $R_{\rm in}$.

The SFGX grid was updated compared to the previous version used in \cite{Ramambason2022} with the addition of the dust-to-gas mass ratio (Z$_{\rm dust}$) as a free parameter (further described in Section \ref{subsec_dtg}).  This additional free parameter increases the size of the grid by a factor of three, leading to a total number of 96\,000 Cloudy models covering 9 free parameters.
Each model is then cut into 17 sub-models stopping at different depths in the cloud, controlled by the “cut” parameter, following a procedure described in Section \ref{sampling_depth}. These submodels allow us to account for both “naked” \hii\ regions, density-bounded or radiation-bounded, which are not associated with any neutral gas, and embedded regions, associated with a layer of neutral gas, with varying thickness controlled by its cut. In the following section, we provide an overview of the characteristics of the SFGX grid which matters the most in the present study, including this cut parameter. We refer to \cite{Ramambason2022} for a complete description of the models. We now briefly recall the characteristics of three key parameters in the current study: the radial density profile of Cloudy models, the depth sampling, and the dust-to-gas mass ratio.

\subsubsection{Radial sampling}
\label{sampling_depth}

\begin{figure}[h!]
\centering
\includegraphics[width=0.45\textwidth]{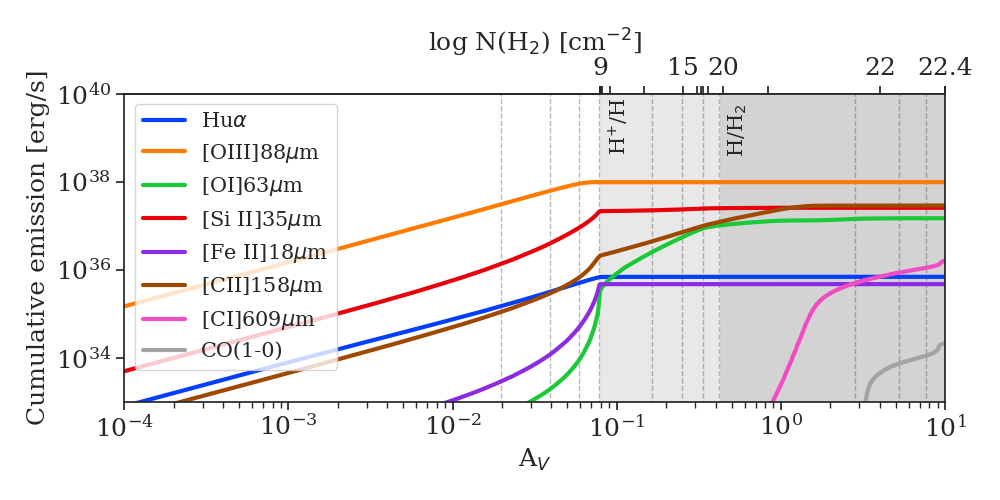}
\includegraphics[width=0.45\textwidth]{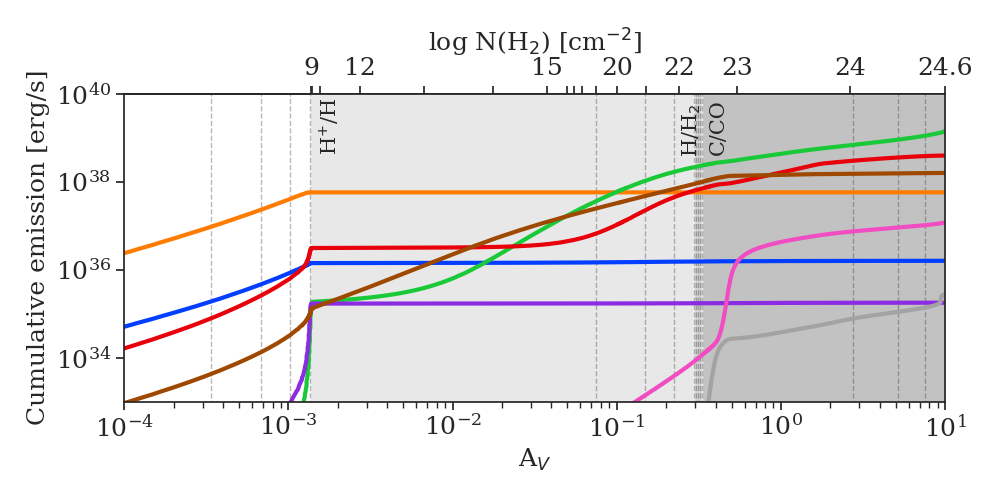}
\caption[Cumulative emission lines profiles of some chosen tracers.]{Cumulative emission lines profiles of some chosen tracers as a function of the visual extinction A$_V$, for two models drawn from the SFGX grid. Both models are computed with a density at the illuminated edge of 100\,cm$^{-2}$, an ionization parameter at the illuminated edge U=$10^{-2}$, an instantaneous stellar burst of 3\,Myrs computed with BPASS stellar atmospheres \citep{eldridge_binary_2017} and no X-ray source. The top row shows a solar metallicity model while the bottom row shows a model with Z=1/10Z$_{\odot}$ and $Z_{\rm dust}=1/155\,Z_{\rm dust, \odot}$, following the prescription from Section \ref{subsec_dtg}. The shaded areas mark the location of the PDR (light gray) and molecular zone (dark gray). The vertical dashed lines show the cuts considered for each model. We note that the C/CO transition is not visible in the first panel since it occurs after A$_V$=10.}
\label{sampling_cut}
\end{figure}

All models are computed until they reach either an A$_V$ of 10 or have their electron temperature drop to 10\,K.\footnote{This second stopping condition ensures that low-metallicity models with little dust converge, even if their maximum A$_V$ remains below 10.} Each initial Cloudy model is used to create 17 sub-models stopping at different A$_V$ controlled by the “cut” parameter, shown in Figures \ref{sampling_cut} and \ref{density_profile}. The transitions are defined in using the relative abundances of hydrogen ($x(\rm H^+)$, $x(\rm H)$, and $x(\rm H_2)$) and carbon ($x(\rm C)$ and $x(\rm CO)$), where $x$ is the fractional abundance, normalized by the total abundance of a given element. The original model is successively cut at the inner radius (cut=0), ionization front (cut=1; $x(\rm H^+)$=$x(\rm H)$), H$_2$ dissociation front (cut=2; $x(\rm H)$=$x(\rm H_2)$), CO dissociation front (cut=3; $x(\rm C)$=$x(\rm CO)$), and outer radius (cut=4).

To sample the different phases (\hii\ region, PDR, CO-dark H$_2$ gas, and CO-bright H$_2$ gas) defined by those cuts, three additional cuts are added between each integer $i$ (cut=$i$+0.25, $i$+0.5, $i$+0.75), equally spaced in A$_V$ between cut=$i$ and cut=$i$+1. We note that stopping the model at a given cut and truncating it a posteriori is not strictly equivalent but results in minor variations of the emission, in the case of spherical models in which gas is only illuminated from one single side. Figure \ref{sampling_cut} illustrates the sampling in depth for two models drawn for the SFGX grid and the cumulative emission of some key tracers used to constrain the depth of a given component (see Section \ref{sect_tracer_ism}). 

\subsubsection{Radial density profile}
\label{subsection_radial_profile}
\begin{figure}[h!]
\centering
\includegraphics[width=0.45\textwidth]{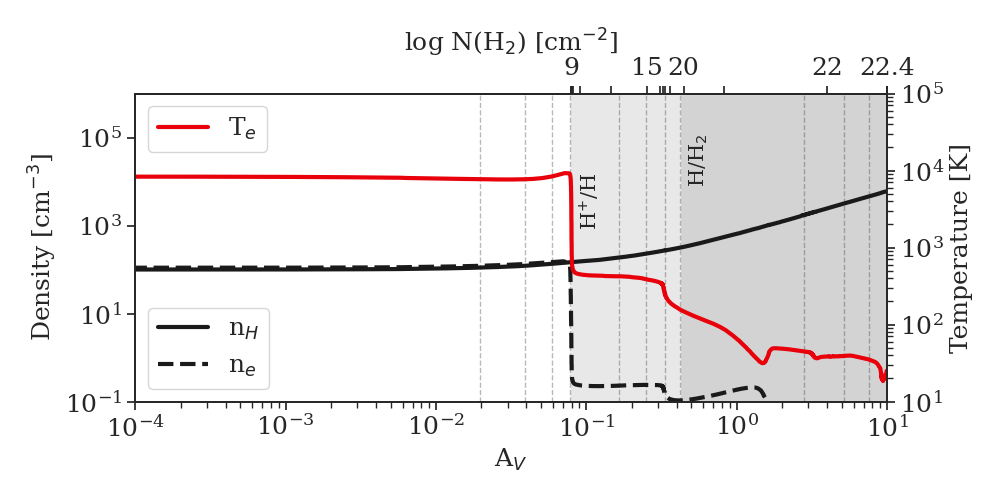}
\includegraphics[width=0.45\textwidth]{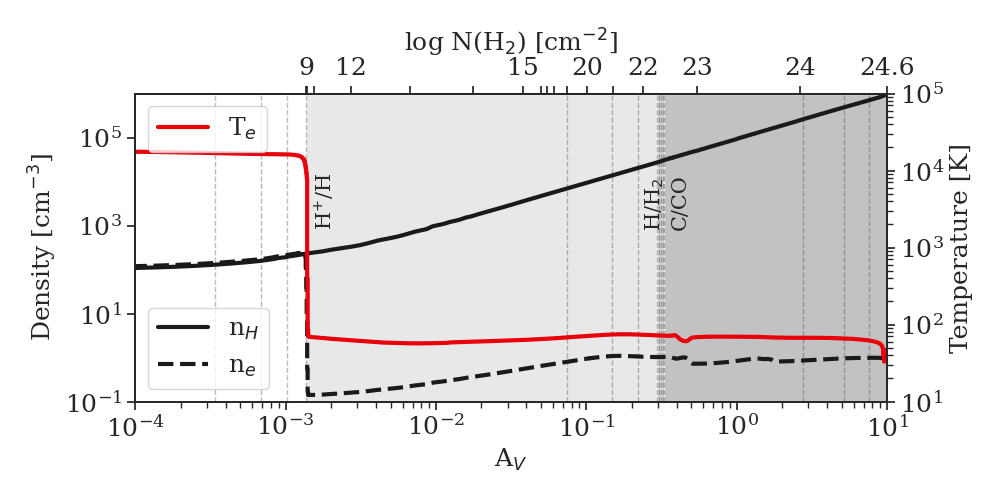}
\caption[Density and temperature radial profiles in two models from the SFGX grid as a function of the visual extinction A$_V$.]{Density and temperature radial profiles for the two models described in Figure \ref{sampling_cut}.}
\label{density_profile}
\end{figure}

The initial hydrogen density, corresponding to the illuminated edge of the models, is varied from 1\,cm$^{-3}$ to 10$^4$\,cm$^{-3}$. To consistently describe the density profile throughout the \hii\ region, the PDR, and the molecular gas, we adopt the same density law as in \cite{cormier_herschel_2019}, in which the hydrogen density is nearly constant in the \hii\ region and scales with the total hydrogen column density above 10$^{21}$\,cm$^{-2}$, as follows:
\begin{equation}
    \displaystyle n(r) = n_0 \times \left(1+\frac{N(\rm H)}{10^{21}}\right);
\end{equation}

where $n_0$ is the hydrogen density at the illuminated edge and $N$(H) is the hydrogen column density.
This law describes a smoothly varying density, matching both the density profile expected in the PDR of confined, dynamically-expanding \hii\ regions \citep{Hosokawa_Inutsuka_2005, Hosokawa_Inutsuka_2006}, and in the interior of turbulent molecular clouds \citep{Wolfire_2010}. 
    
As shown in Figure \ref{density_profile}, this prescription results in a smooth increase of density throughout the PDR and molecular zone, as opposed to the sharp variations expected for, e.g., constant pressure models (see discussion in \citealt{cormier_herschel_2019}). Although this prescription is fixed in our grid, the resulting geometry of the models depends on other parameters such as the metallicity, ionization parameter, or the hardness of the radiation field. In particular, the depth of the PDR is strongly affected by the metallicity. In models with low metal and dust content, photons can propagate deeper and produce a thicker PDR than in solar-like metallicity models, as shown in Figure \ref{density_profile}.

\subsubsection{Dust-to-gas mass ratio}
\label{subsec_dtg}

Dust is included in our Cloudy models following the prescription described in \cite{Ramambason2022}. The radial variation of the dust abundance follows that of the hydrogen density, while the total dust mass is set by assuming a metallicity-dependent dust-to-gas mass (Z$_{\rm dust}$).
We consider three dust-to-gas mass ratio values per metallicity bin, by sampling the Z$_{\rm dust}$ vs. metallicity relation derived in \cite{galliano_nearby_2021}. This calibration relies on the spectral fitting of the infrared continuum for a large sample of DustPedia \citep{Davies_dustpedia_2017} and DGS galaxies, which includes our sample. They provide analytical fits corresponding to median, upper and lower values of the envelope encompassing 95\% of the galaxies used in their analysis. These prescriptions allow us to explore the potential effects of  Z$_{\rm dust}$ variations on our predicted quantities, while restricting the exploration to plausible values of Z$_{\rm dust}$ (within the 95\% envelope) for each metallicity bin. We note that this prescription is more refined than the free power-law fit previously derived in \cite{Remy-Ruyer_2013}, although compatible. Specifically, \cite{Remy-Ruyer_2013} predict that Z$_{\rm dust}$ scales with $\sim Z^{-2}$, when assuming a metallicity-dependent $\alpha_{\rm CO}$. For a metallicity of 1/10\,$Z_{\odot}$, the dust-to-gas mass ratio is scaled by a factor $\sim$1/100, while the prescription from \cite{galliano_nearby_2021} results in a scaling of $\sim$1/155. 
Although our grid is not designed to investigate in detail the dust composition and its radial distribution, which are both fixed, it remains flexible enough to self-consistently reproduce the observed integrated dust masses of galaxies (see Section \ref{subsect_masses}).

\subsection{MULTIGRIS runs}
\label{subsect_mgris_config}

We apply the Bayesian code MULTIGRIS \citep{MULTIGRIS2022} to the sample of galaxies described in Section \ref{section_sample}. This code enables flexible combinations of “components” drawn from a large grid of pre-computed models, here the SFGX grid described in Section \ref{subsect_sfgx}. The code architecture, based on the python package PyMC3 \citep{pymc3_paper, PyMC3}, samples the parameter space assuming a given likelihood and sampling method, chosen by the user to infer posterior probability distribution functions (PDF) of any parameter. In the current study, we adopted a multigaussian likelihood and a Monte Carlo Markov Chain (MCMC) sampler which is ideally suited to deal with multimodal distributions (Sequential Monte Carlo sampler, \citealt{Minson_2013,Ching_2007}). Those choices are motivated and detailed in \cite{LebouteillerRamambason2022} and \cite{Ramambason2022}.

Specifically, we aim at inferring the posterior PDFs of the gas masses associated with different phases of the ISM, namely:
\begin{itemize}
    \item $M$(H$^+$), the total gas mass associated with the ionized reservoir as traced by H$^{+}$.
    \item $M$(H$^0$),  the total gas mass associated with the neutral hydrogen reservoir as traced by H$^0$.
    \item $M$(H$_2$),  the total gas mass associated with the molecular gas reservoir as traced by H$_2$.
    \item $M_{\rm dust}$, the total mass of dust.
\end{itemize}

Those masses are extracted from each individual Cloudy models considered as “components”, following the procedure described in Section \ref{subsect_single_comp}. We also derive those quantities for more complex architectures that consist of either linear combinations of a discrete number of components (see Section \ref{subsec_multi_comp}) or are based on statistical distributions for some key parameters (see Section \ref{broken_plaw_models}).

\subsubsection{Single component models}
\label{subsect_single_comp}
For each Cloudy model, the total gas mass, $M_{\rm H,i}$, associated with a given hydrogen state (H$^+$, H$^0$, or H$_2$) is computed as follows:
\begin{equation}
    \label{eq_masses}
    \begin{cases}
    \displaystyle M({\rm H_i}) = m_{\rm H_i} \int_{R_{\rm in}}^{R_{\rm out}} n_{\rm H_i}(r) 4\pi r^2 dr, \\
    \displaystyle n_{\rm H_i}(r) = x_{\rm H_i}(r) n_{\rm H}(r),
    \end{cases}    
\end{equation}
where $m_{\rm H,i}$ is the mass of the hydrogen ion or molecule (H$^+$, H$^0$, or H$_2$) used as tracer, $n_{\rm H,i}$ the density of this tracer, $x_{\rm H,i}$ the relative abundance of the tracer with respect to total hydrogen, $n_{\rm H}$ the density of total hydrogen, $r$ the radius varying between the inner ($R_{\rm in}$) and outer ($R_{\rm out}$) radius, and $4\pi r^2 dr$ the volume of shell of gas between the radius $r$ and $r+dr$.

We derive an estimate of the total dust mass based on the dust-to-gas mass ratio, $Z_{\rm dust}$, defined in Section \ref{subsec_dtg}. The total dust mass is defined by:
\begin{equation}
    \displaystyle M_{\rm dust} \approx Z_{\rm dust} \times \frac{1}{1-Y_{\odot}} \times (M({\rm H^+}) + M({\rm H^0}) + M({\rm H_2})), \\
\end{equation}
where $Z_{\rm dust}$ is the dust-to-gas mass ratio, $Y_{\odot}$ is the Galactic mass fraction of Helium, and $M$(H$^+$),  $M$(H$^0$), and $M$(H$_2$) are defined by equation \ref{eq_masses}. In the rest of the paper, we adopt $Y_{\odot}$=0.270 \citep{Asplund_2009}, leading to a corrective factor for the helium mass of $\sim 1.36$.

The gas mass definition introduced in Equation \ref{eq_masses} relies on the use of radial abundance profile of hydrogen atoms, under different states. This definition is flexible and can easily be adapted to trace the gas reservoir of ionized, neutral, or molecular gas associated with other elements (e.g., carbon). We define the gas reservoir in different phases as the integrated hydrogen mass, weighted by the fractional abundance of a given tracer as follows:
\begin{equation}
    \label{eq_masses_zone}
    \begin{cases}
    \displaystyle M({\rm H_i})_{X_j} = m_{\rm H_i} \int_{R_{\rm in}}^{R_{\rm out}}n_{X_j} n_{\rm H_i}(r) 4\pi r^2 dr , \\
    \displaystyle n_{\rm H_i}(r) = x_{\rm H_i}(r) n_{\rm H}(r), \\
    \displaystyle n_{X_j}(r) = x_{X_j}(r) n_{X}(r), 
    \end{cases}    
\end{equation}
where $n_{X_j}$ is the density of the element $X$ in a given ionization state (or molecular form) j and $x_{X_j}$ is the fractional abundance of this tracer with respect to the total density of the element $X$. By design, we note that $\sum_{j} x_{X_j}=1$. We apply this formula to derive the mass of H$_2$ in the different carbon phases: $M$(H$_2)_{\rm C^+}$, $M$(H$_2)_{\rm C}$, and $M$(H$_2)_{\rm CO}$. 

\subsubsection{Few independent components}
\label{subsec_multi_comp}

We define a multicomponent architecture as a linear combination of components, using a mixing-weight. Those models represent a galaxy as a mix of $N$ different gas components, associated with independent sets of gas parameters ($U$, $n$, cut, and the mixing-weight; 4$N$ free parameters), sharing similar chemical properties (described by 2 parameters; the gas-phase metallicity and dust-to-gas mass ratio) and illuminated by the same radiation field (i.e., a single representative cluster of stars, described by 4 free  parameters: the stellar age, the total cluster luminosity, the luminosity, and temperature of a potential X-ray source). The number of free parameters for such models, including the mixing-weights, is thus $4N+6$, where $N$ is the number of gas components. We refer the reader to \cite{Ramambason2022} for further details, in which a larger sample of DGS galaxies was modeled with architectures combining up to three components. 

We define multicomponent models by computing linear combinations for the extensive quantities of linear components, in particular the integrated luminosities and gas masses. To first order, we assume that the gas masses scale linearly with the observed luminosity. In other words, our multicomponent models can be considered as luminosity-weighted combination of gas components. We compute the total masses (i.e., M(H$_2$), M(\hi), M(\hii), and M$_{\rm dust}$), and the H$_2$ masses in different carbon phases (i.e., M(H$_2$)$_{\rm C+}$, M(H$_2$)$_{\rm C}$, and M(H$_2$)$_{\rm CO}$) as linear combinations of the masses of each component, such that:
\begin{equation}
    M_{\rm multi}= \sum_{\rm i=1}^{N_{\rm comp}} w_{\rm i} M_{\rm i},
\end{equation}
where $w_{\rm i}$ is the mixing-weight and $M_{\rm i}$ the predicted mass associated with a specific ISM phase, for the $i^{\rm th}$ component. The same formula is applied to derive the combined total luminosity of a given line.

As shown in \cite{Ramambason2022}, multicomponent models with one to three components enable reproducing simultaneously most of the emission lines arising from the \hii\ regions and PDRs in the DGS. Nevertheless, the tailored model developed for I\,Zw\,18 in \cite{LebouteillerRamambason2022} which includes additional tracers of the molecular gas (e.g., CO) has required the addition of a fourth component, corresponding to a component that is deep enough to reach the denser molecular phase in which CO emits, associated with a relatively small mixing-weight. In the current study, we hence follow a similar approach and consider models with up to four independent components.

We use the same selection criterion as in \cite{Ramambason2022} to select the optimal number of components and keep only the best configuration, according to the marginal likelihood metric. The values for the relative marginal likelihoods and the number of components chosen in the best configuration are reported in Table \ref{table_lm}. We note that the models having the largest number of components do not necessarily perform better in terms of marginal likelihood. This comes from the fact that numerous free parameters (e.g., $4\times4+6=22$ free parameters for a 4-component model) penalize models when not enough constraints are available. To solve this issue, we present in the following section another architecture allowing the combination of large numbers of components while keeping a relatively low number of free parameters. 

\subsubsection{Statistical distributions of components}
\label{broken_plaw_models}

While in theory it is possible to apply the linear combination described in Section \ref{subsec_multi_comp} to a large (>4) number of components, the resulting models are associated with lower marginal likelihoods, reflecting the fact that the solution is diluted in an increasingly large parameter space. Instead, we generalize the combination of models by combining components drawn from statistical distributions defined analytically. 

A first simple approach is to describe the key parameters of our models using power-law distributions. This approach was inspired by the Local Optimally emitting Clouds (LOC; \citealt{Baldwin_1995}), previously used to reproduce emission lines from the narrow line region in quasars and star-forming galaxies \citep{Ferguson_1997,richardson_interpreting_2014,richardson_interpreting_2016}. The latter models consider power-laws to describe the distributions of density and radiation field of \hii\ regions, allowing the representation of the integrated emission of a population of clouds with various densities, distributed at various distances from an ionizing source. In the current study, we use the ionization parameter $U$ as a proxy for the strength of the radiation field and consider an additional power-law describing the depth of each component (via the cut parameter, described in Section \ref{sampling_depth}). This is in essence similar to the approach based on A$_V$-PDF used, e.g., in \cite{Bisbas_2019, Bisbas_2023}, although with different choices of prior distributions, which we will further discuss in Section \ref{discussion_powerlaw}. 

\begin{table}[t!]
\begin{threeparttable}
\centering
\caption[Distribution of parameters adopted in the broken power-law configuration.]{Distribution of parameters adopted in the broken power-law configuration.}
\label{table_broken_plaw}

\begin{tabular}{p{0.1\textwidth}p{0.3\textwidth}}

\hline
Parameters & Statistical distribution$^{(1)}$ \\ \hline 
$n$ & $\phi_{n} \propto n^{\alpha_n}$; [$n_{\rm min}$, $n_{\rm max}$]\\
\hline
$U$ & $\phi_U \propto U^{\alpha_U}$; [$U_{\rm min}$, $U_{\rm max}$]\\
\hline
Cut & 
    $\phi_{\rm cut} \propto$ ${\rm cut}^{\alpha_{\rm cut,1}}; [{\rm cut}_{\rm min}, {\rm cut}_{\rm IF}]$\\
    & $\phi_{\rm cut} \propto$ ${\rm cut}^{\alpha_{\rm cut,2}}; [{\rm cut}_{\rm IF}, {\rm cut}_{\rm max}]$\\
\hline
\hline
Random variables & Priors$^{(2)}$ \\ \hline 
$\alpha_n$ & $\mathcal{N}$($\mu$=-1.5, $\sigma$=2)\\
$\log n_{\rm min}$ & $\mathcal{N}$($\mu$=0, $\sigma$=1)\\
$\log n_{\rm max}$ & $\mathcal{N}$($\mu$=4, $\sigma$=1)\\
\hline
$\alpha_U$ & $\mathcal{N}$($\mu$=-1.5, $\sigma$=2)\\
$\log U_{\rm min}$ & $\mathcal{N}$($\mu$=-4, $\sigma$=1)\\
$\log U_{\rm max}$ & $\mathcal{N}$($\mu$=0, $\sigma$=1)\\
\hline
$\alpha_{\rm cut,1}$ & $\mathcal{N}$($\mu$=0, $\sigma$=2)\\
$\alpha_{\rm cut,2}$ & $\mathcal{N}$($\mu$=0, $\sigma$=2)\\
cut$_{\rm min}$ & $\mathcal{N}$($\mu$=0, $\sigma$=1)\\
cut$_{\rm max}$ & $\mathcal{N}$($\mu$=2, $\sigma$=1)\\
\hline
\end{tabular}

\begin{tablenotes}
    \footnotesize
   \item (1) $\phi_X$ is the statistical distribution assumed for parameter $X$.
   \item (2) $\mathcal{N}$($\mu$, $\sigma$) is a Normal distribution with mean $\mu$ and full-width half maximum $\sigma$.
\end{tablenotes}

\end{threeparttable}
\end{table}

As detailed in Table \ref{table_broken_plaw}, we use power-law distributions to describe the three main gas parameters of our models:
\begin{itemize}
    \item the hydrogen density $n$ (defined at the illuminated edge of \hii\ regions),
    \item the radiation field (through the ionization parameter $U$ of \hii\ regions),
    \item the depth of each component (through the cut parameter).
\end{itemize}
This method allows us to integrate over a power-law distribution defined by three free parameters: a slope, a lower-bound, and an upper-bound. The prior distributions set for those parameters are relatively weak and defined as normal distributions, centered on a value $\mu$ with a relatively large $\sigma$. For the “cut” parameter, we consider a broken power-law, allowing different slopes in the ionized gas (below cut$_{\rm IF}=1$, corresponding to the ionizing front) and in the neutral gas (above cut$_{\rm IF}=1$), hence parametrized by four free parameters instead of three. This results in a total of $3\times2+4= 10$ free parameters controlling the power-law distributions for $U$, $n$, and cut.

In practice, power-law distributions could also be considered for other parameters in our models. In the current study, we focus primarily on the gas parameters, and consider a single set of stellar and X-ray source parameters. This setting corresponds to several gas components illuminated by a single population of stars, and can be considered as a generalization of the models with few components organized around a central cluster, described in Section \ref{subsec_multi_comp}. We also assume that no strong metallicity gradients are present in our sample, which is compatible with observations of dwarf galaxies \citep[e.g.,][]{Lagos_Papaderos_2013}. We hence use a single free value for the other 6 parameters of the SFGX grid: the metallicity, stellar age, cluster luminosity, dust-to-gas mass ratio, luminosity, and temperature of a potential X-ray source. 

In total, those new models account for $10+6=16$ free parameters, less than the $18$ parameters necessary to combine 3 discrete components. Nevertheless, the integration of the power-laws effectively performs the linear combination of numerous components, the exact number of which depends on the sampling of the initial grid and varies with the boundaries of the power-laws, from a few tens to a few hundreds of components. Even with the coarse sampling of the SFGX grid, with 5 bins for density and ionization parameter and 17 cuts, the maximum number of components combined using statistical distributions can reach up to 425, when the lower and upper boundaries correspond to the minima/maxima of the grid.

In practice, the integrals over the statistical distributions are calculated on-the-fly during the MCMC sampling process (avoiding the storage of a large precomputed grid, with predefined mixing-weights) using the following formula for each mass (and line luminosity):

\begin{multline}
\label{e1}
    M_{\rm power-law} = \sum_{\theta \notin [U,n,\rm cut]} \sum_{\rm U_{min}}^{\rm U_{max}} \sum_{\rm n_{min}}^{\rm n_{max}} \sum_{\rm cut_{min}}^{\rm cut_{max}}  M(\theta, x,y,z)\\ \times \Phi(x,y,z) d\theta dx dy dz,
\end{multline}

where the integration weight function $\Phi$ is defined as follows:
\begin{equation}
        \Phi = \phi_{U}(x)\phi_{n}(y)\phi_{\rm cut}(z)=
    \begin{cases}
        x^{\alpha_U} y^{\alpha_n}z^{\alpha_{cut,1}}\, \rm \,if\, cut < cut_{\rm IF},\\
        x^{\alpha_U} y^{\alpha_n} z^{\alpha_{cut,2}}\, \rm \,if\, cut > cut_{\rm IF},    
    \end{cases}
\end{equation}
where $x \in [U_{\rm min}, U_{\rm max}]$, y $x \in [n_{\rm min}, n_{\rm max}]$, and $z \in [{\rm cut_{min}}, {\rm cut_{max}}]$. 

Because the masses are stored in $\log_{10}$, the weighting function can be re-expressed as follows:
\begin{multline}
\label{e2}
    10^{\log M_{\rm power-law}} = \sum_{\theta \notin [U,n,\rm cut]} \sum_{\rm U_{min}}^{\rm U_{max}} \sum_{\rm n_{min}}^{\rm n_{max}} \sum_{\rm cut_{min}}^{\rm cut_{max}}  10^{\log M(\log\, \theta, X,Y,Z)+\Psi(X,Y,Z)}\\ d\theta dX dY dZ
\end{multline}
with $X=\log\, x$, $Y=\log\, y$, $Z=\log\, z$, $dx = xdX$, $dy=ydY$, $dz=zdZ$, and $\Psi$ defined as follows:

\begin{multline}
        \Psi(X,Y,Z) = \log \Phi(x,y,z)+\log\, x+\log\, y+\log\, z\\
        = (\alpha_U + 1)X + (\alpha_n + 1)Y + (\alpha_{\rm cut} + 1)Z,
\end{multline}
 
\section{Comparing model architectures}
\label{section_results}

\subsection{Overview}
\label{overview_masses}

\begin{figure*}[h!]
\centering
\includegraphics[width=0.99\textwidth]{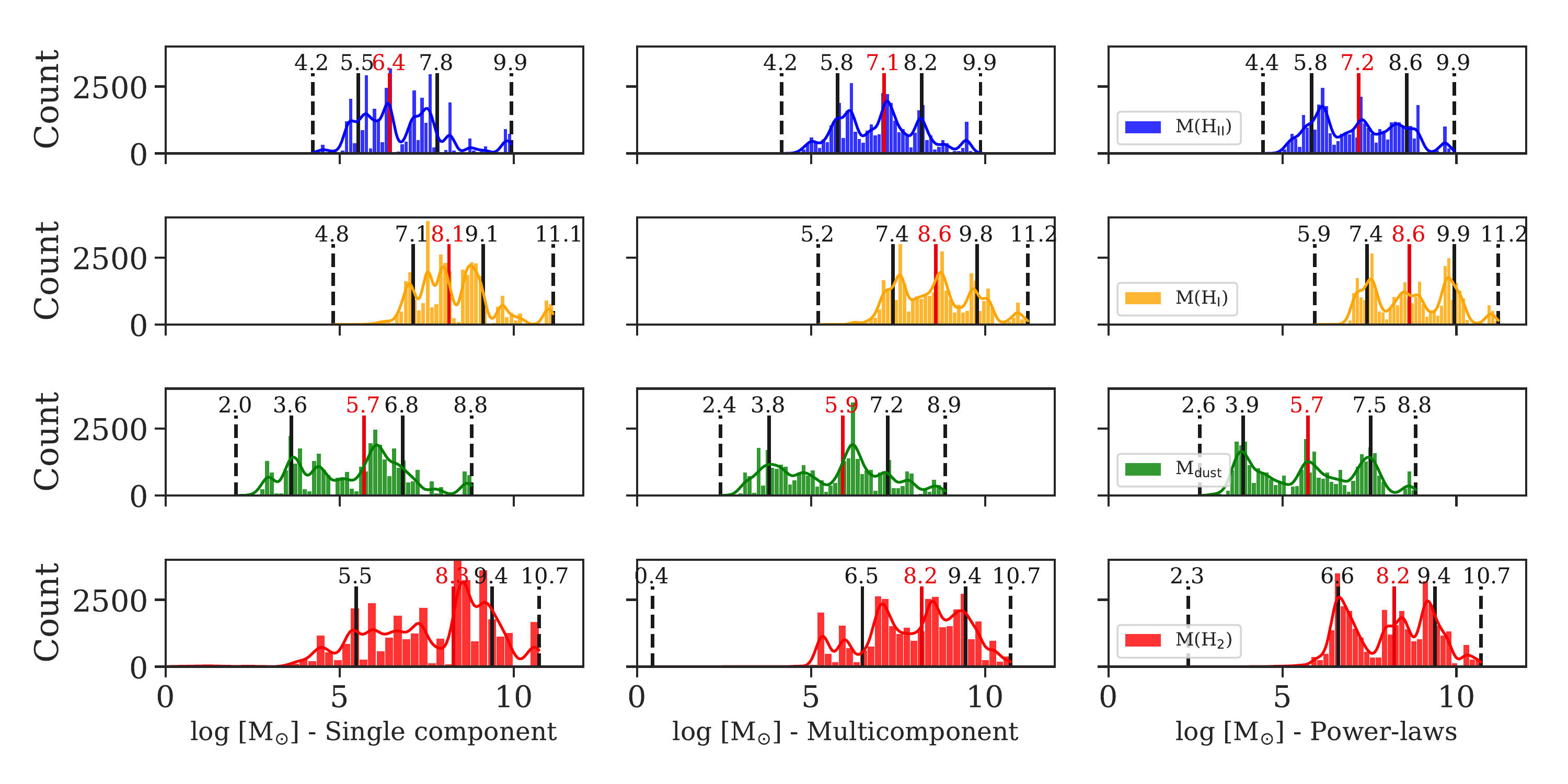}
\caption[Comparison of the \hii, \hi, and H$_2$ masses for different architectures]{Comparison of the \hii, \hi, and H$_2$ masses for different architectures. The histograms are built using the MCMC draws for the whole sample. The vertical lines show the minima and maxima of the distributions (dashed lines), the quantiles at 15\% and 85\% (plain black lines), and the median value (plain red line).}
\label{mass_comparison_all}
\end{figure*}

We now compare the results obtained assuming three different architectures using: single component models (see Section \ref{subsect_single_comp}), multicomponent models with a few (1 to 4) independent components (see Section \ref{subsec_multi_comp}), or multicomponent models with statistical distributions (power-laws; see Section \ref{broken_plaw_models}). In Table \ref{3sigma_values}, we quantify how well each configuration reproduces the set of emission lines used as constraints by looking at the probability P(3$\sigma$) that the prediction from the models falls within 3$\sigma$ of the observed value, with $\sigma$ the error on the detection. We also report the associated posterior predictive p-values \citep[e.g.,][]{Meng_PPP_1994}, which quantify the goodness of the fit for each line. Those p-values are calculated by generating replicated data based on the posterior distributions of parameters and estimating how much they deviate from the observed data \citep[see, e.g.,][]{galliano_nearby_2021}. This metric is classically used in Bayesian statistics, and allows us to flag models associated with overfitting (p-values close to 0) or under fitting (p-values close to 1), for specific sets of lines. The global P(3$\sigma$) and p-values, averaged for all lines, are reasonably good for all architectures. Specifically, the predicted line fluxes agree with the observations at 3$\sigma$ for at least 66\% of the draws for single component models and 71\% of the draws for multicomponent models (few components and statistical distributions), meaning that, on average, they all well reproduce most of the constraints. The subset of PDR lines is more difficult to reproduce in some galaxies, and the P(3$\sigma$) values are globally lower, although most values remain above $\sim$50\% (except for a few galaxies, flagged in Table \ref{3sigma_values}). We note models that well reproduce the observed emission lines do not necessarily correspond to realistic gas structures. However, using increasingly more complex architectures allows us to derive predictions using models which are, a priori, closer to the actual geometry of the ISM.

We now compare the different gas mass distributions obtained with each configuration and motivate the choice of the best configuration. As a first sanity check, we extract the integrated gas masses associated with the H$^+$, H$^0$, H$_2$, and dust reservoirs in our sample (see Section \ref{section_sample}). The histograms obtained for the whole sample are provided in Figure \ref{mass_comparison_all}. We note that all architectures (single component, multicomponent with 1 to 4 components, and multicomponent with power-law distributions) predict integrated masses globally in good agreement with each other and that all models are dominated by H$^0$ and H$_2$ gas reservoirs. We find that the M(\hii)/M(\hi) mass ratios are below 7\% regardless of the architecture we consider. Whether H$_2$ or \hi\ gas reservoir dominates the total gas mass varies from galaxy to galaxy and depend on the architecture. Single-component models predict that H$_2$ mass may dominate in 4 out of 18 galaxies (Haro\,2, Haro\,11, He\,2-10, and Mrk\,1089) with the M(H$_2$)/M(\hi) mass ratios between 0.001 and 13.6. Increasing the number of components leads to more \hi-dominated galaxies; with multicomponent models, we find only one H$_2$-dominated galaxy (He\,2-10) and M(H$_2$)/M(\hi) mass ratios between 0.006 and 7. With the power-law distributions, all the galaxies in our sample are predicted to be \hi-dominated, with M(H$_2$)/M(\hi) between 5\% and 66\%.

\subsection{Atomic hydrogen and dust reservoirs}
\label{subsect_masses}

We then compare the predicted H$^0$ and dust masses (respectively M(\hi) and M$_{\rm dust}$) with previous measurements. The dust measurements were derived in \cite{remy-ruyer_linking_2015} using a phenomenological dust spectral energy distribution (SED) model, accounting for starlight intensity mixing in the ISM, to interpret the whole IR-to-submillimeter observed SED. Their models were applied to a large set of photometric bands available using \textit{Herschel}, \textit{Spitzer}, WISE, and 2MASS, when available. 

For the measured \hi\ masses, we use the data reported in \cite{remy-ruyer_gas--dust_2014} based on estimates using the \hi\ 21\,cm line (see references therein). The latter \hi\ masses were corrected to match the dust photometric aperture. We note that we used spectroscopic data from \textit{Herschel} and \textit{Spitzer}, integrated on the full instrumental apertures \citep{Cormier_2015}, while the photometric data was extracted using specific apertures \citep[see][]{Remy-Ruyer_2013}. Nevertheless, assuming that most of the line emission arises from the galactic body emitting dust tracers, we expect the dust masses we derive to be comparable with those previous measurements.

\begin{figure*}[h!]
\centering
\includegraphics[height=0.32\textwidth]{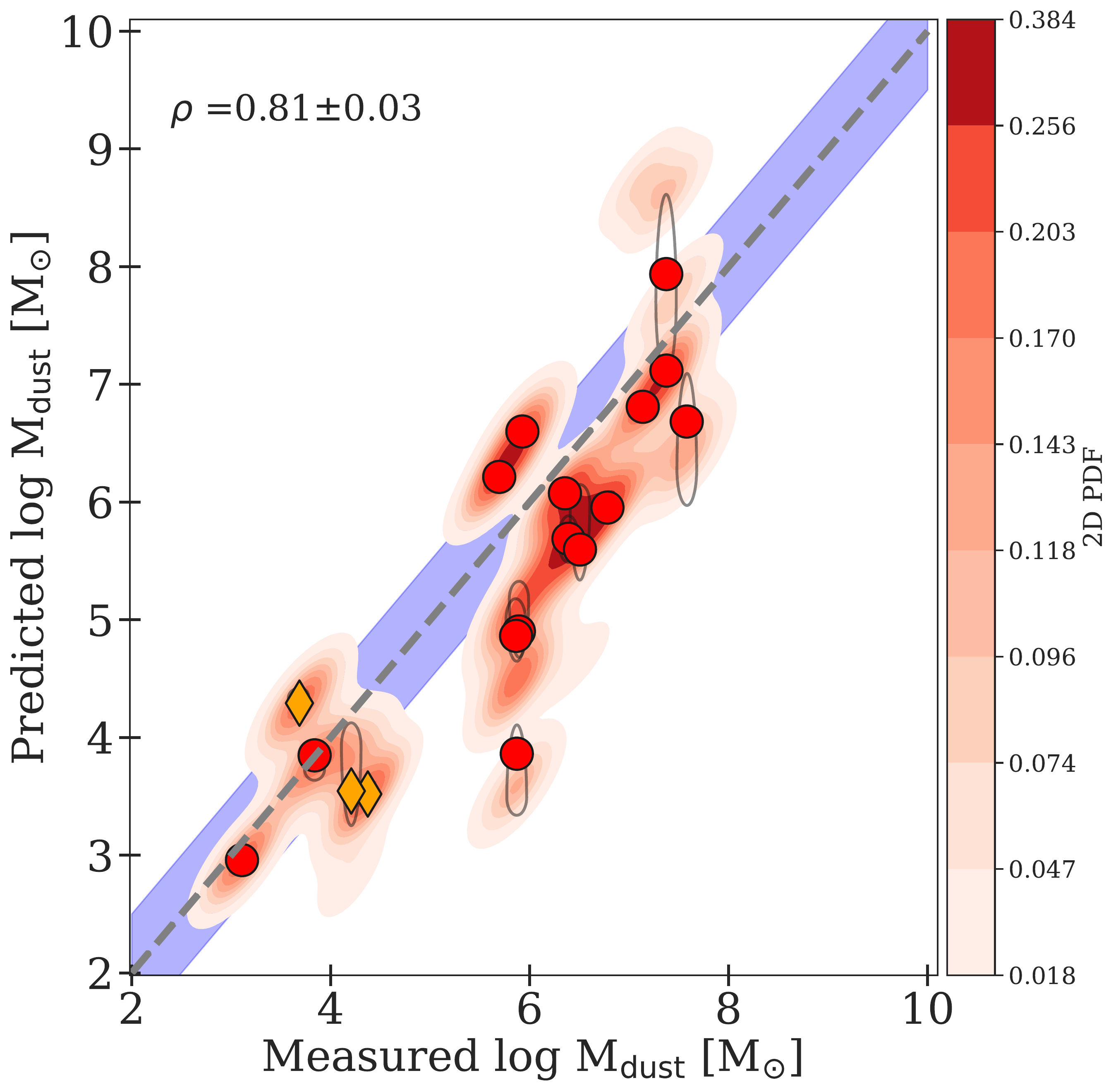}
\includegraphics[height=0.32\textwidth]{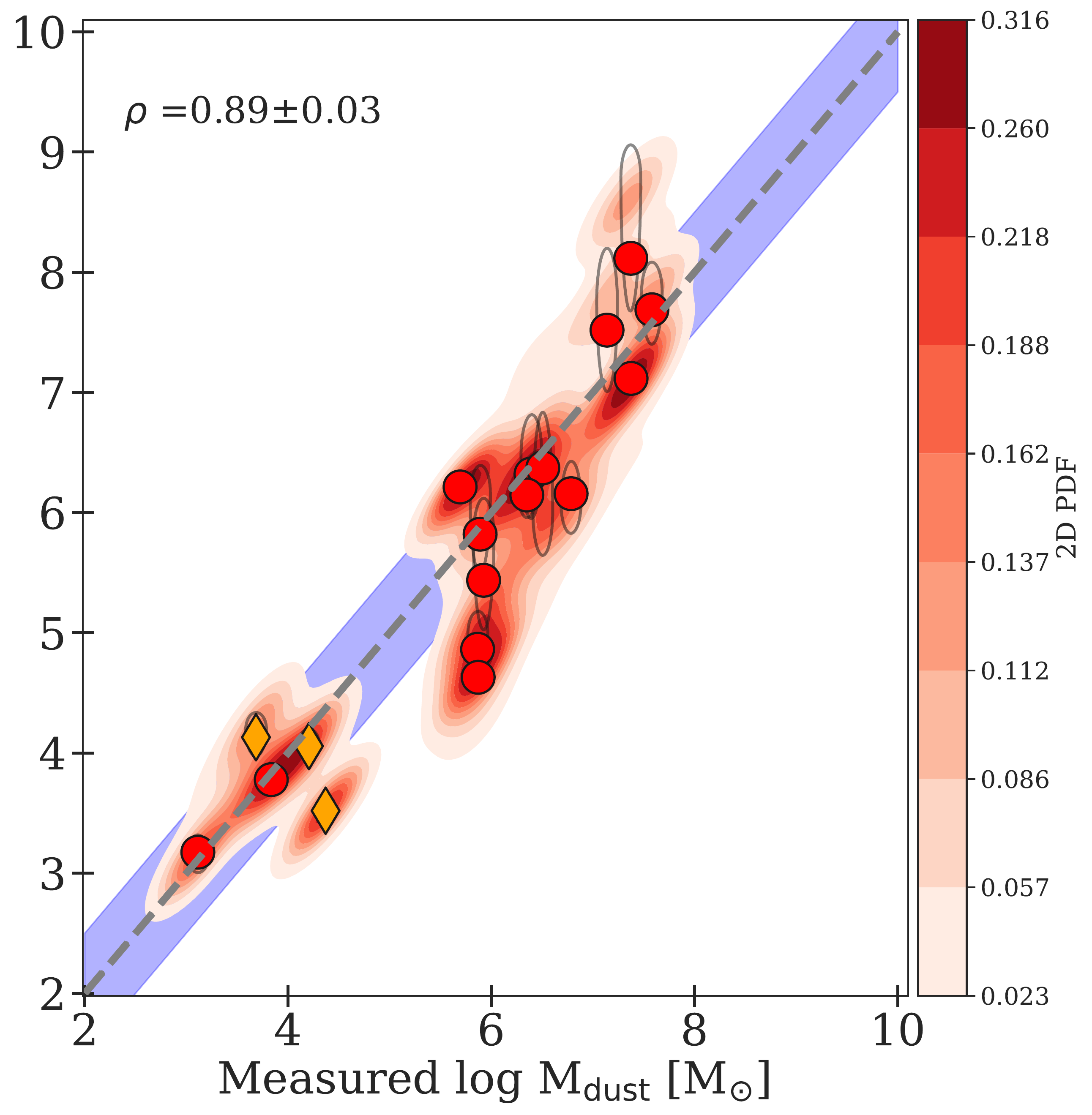}
\includegraphics[height=0.32\textwidth]{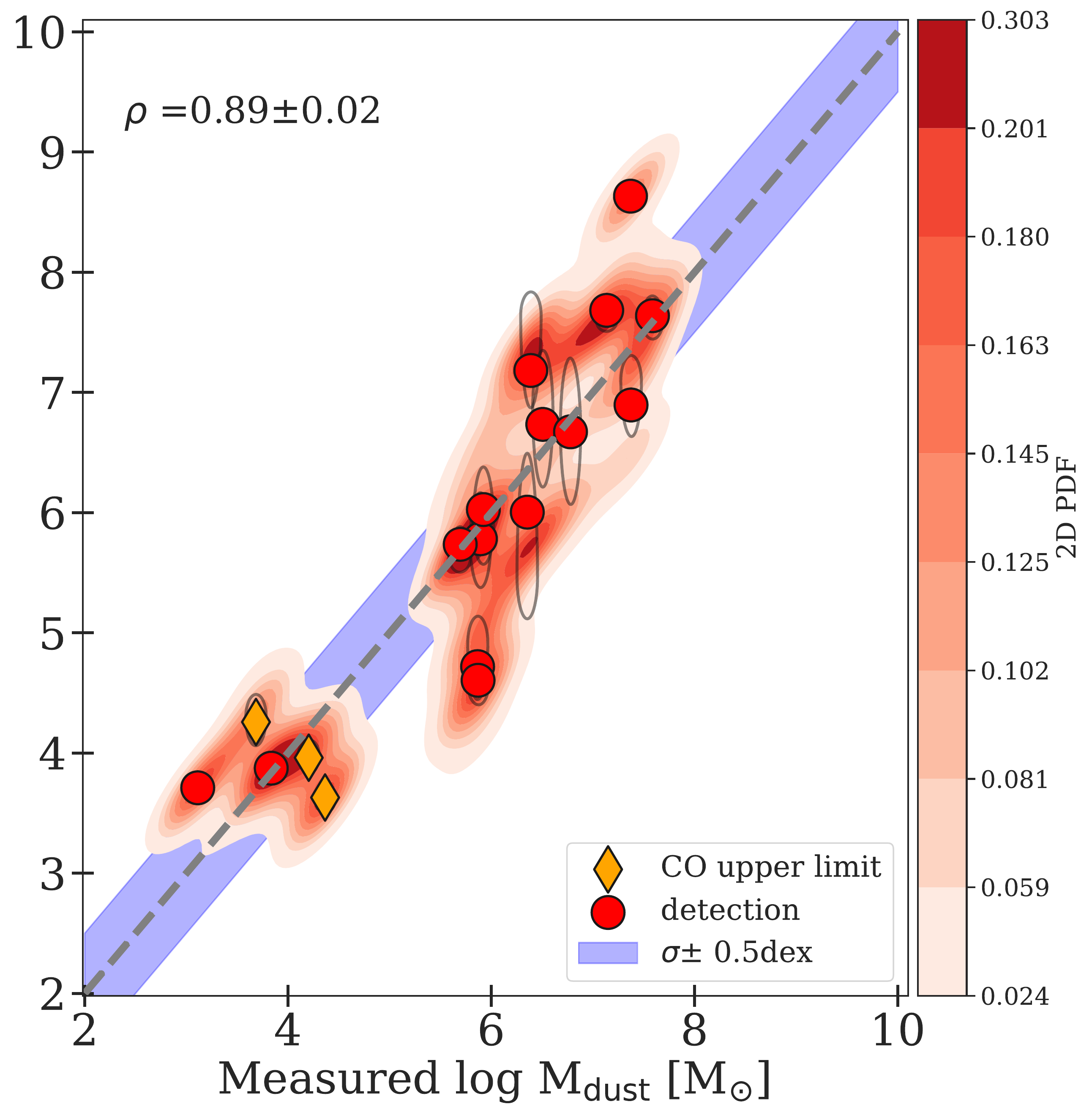}
\includegraphics[height=0.32\textwidth]{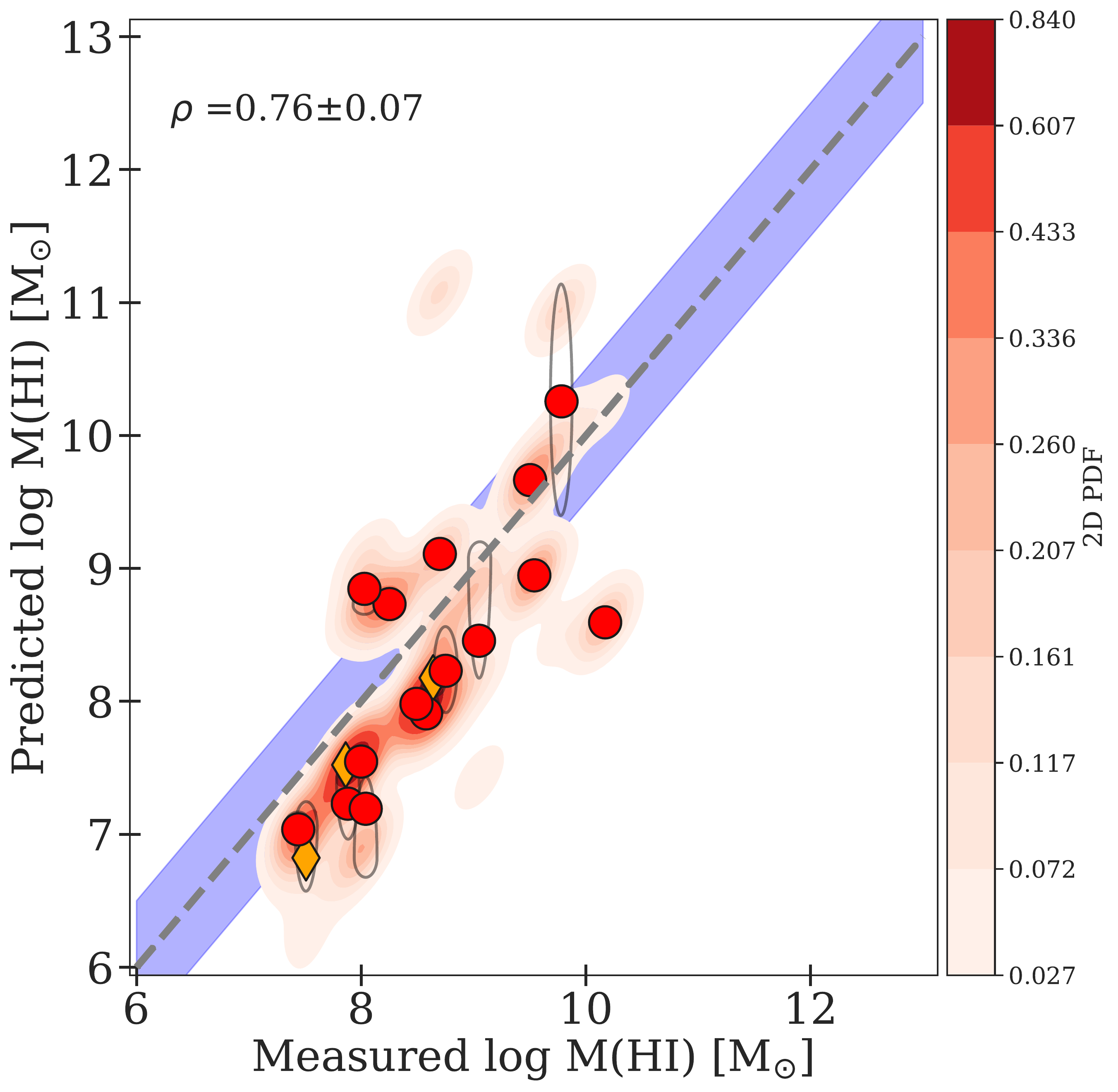}
\includegraphics[height=0.32\textwidth]{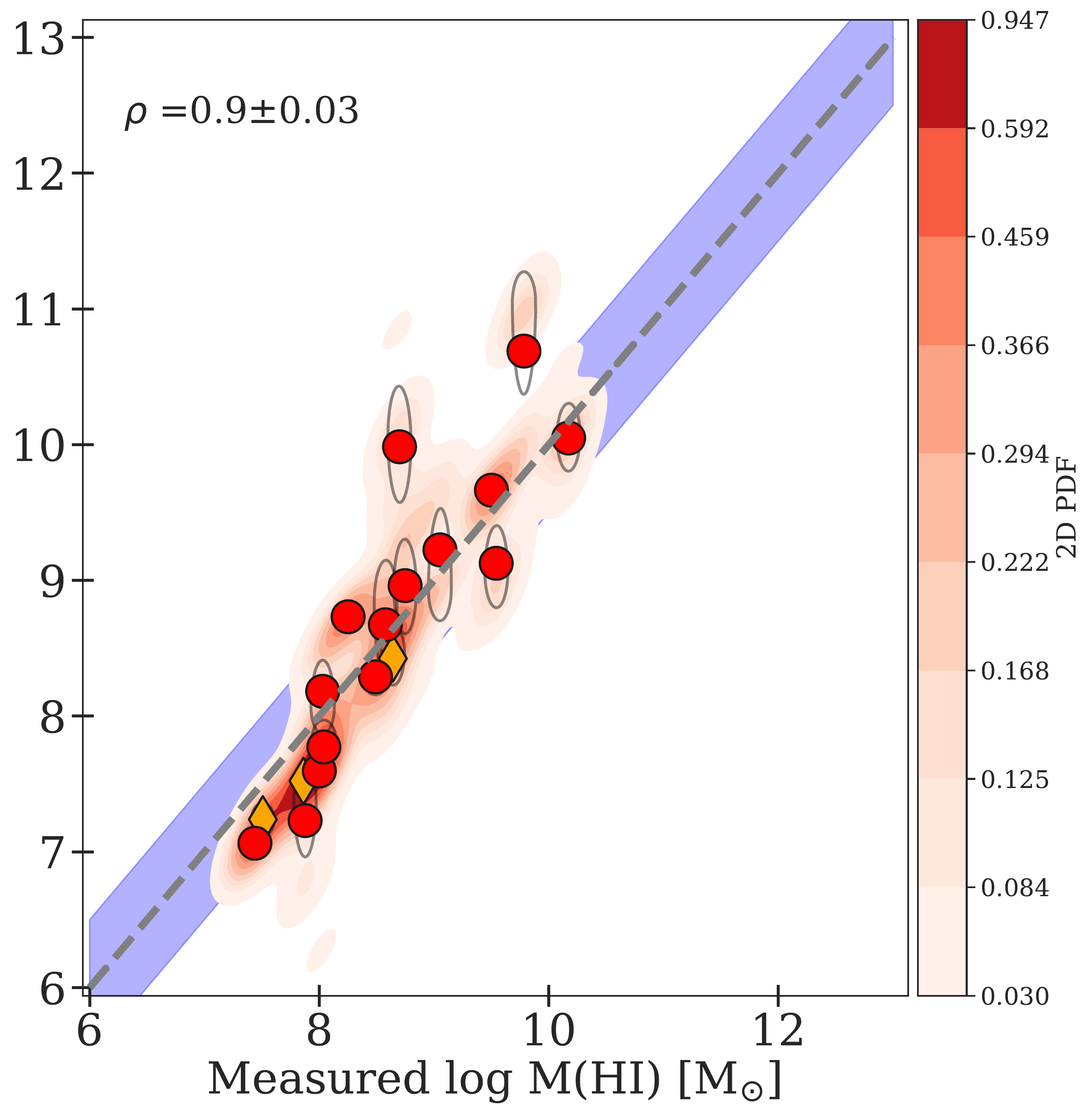}
\includegraphics[height=0.32\textwidth]{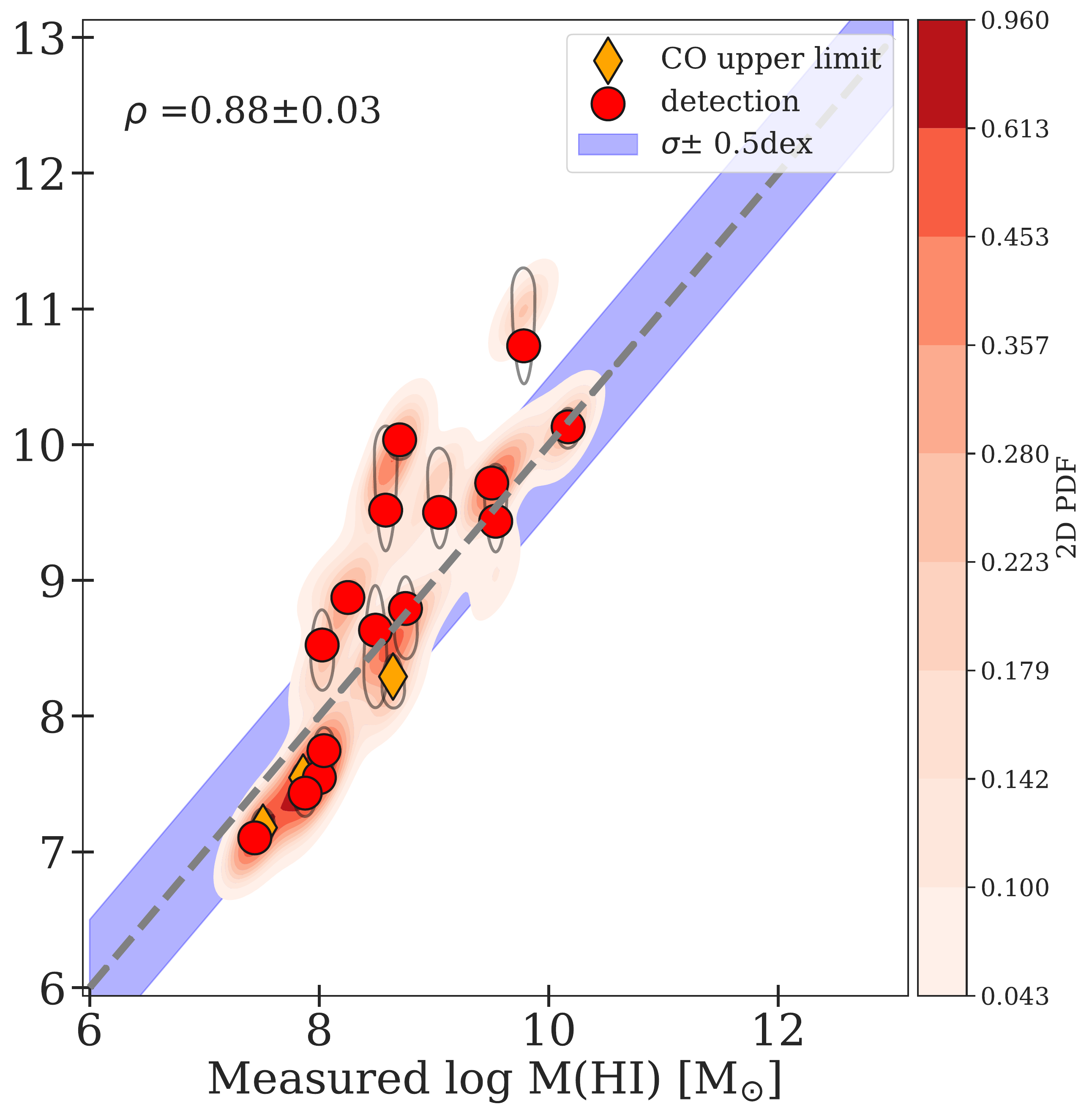}
\caption[M$_{\rm dust}$ and M(H$^0$) predicted by single component, multicomponent, and power-law models]{Dust and \hi\ masses predicted by single component (left), multicomponent (middle), and power-law models (right). The red points mark the location of robust means and the ellipses the 1$\sigma$ uncertainty around the main peak, following the Skewed Uncertainty Estimate formalism described in \cite{galliano_nearby_2021}. We report the mean and standard deviation of the Spearman correlation coefficients ($\rho$) calculated for of the whole sample, at each of the MCMC steps, represented by the shaded Kernel Density Estimate underlying data points. The dashed lines indicate the 1:1 relation and the blue shaded area a deviation of 0.5\,dex around the latter relation.}
\label{masses_relation}
\end{figure*}

In Figure \ref{masses_relation}, we show that for all architectures, the predicted dust and \hi\ masses are globally in good agreement (within 0.5\,dex) with the measurements. We note that this was not the case in \cite{Ramambason2022}, in which the \hi\ masses were systematically underpredicted. This underprediction was driven by the fact that \cite{Ramambason2022} accounted for H$_2$ upper limits, while the latter upper limits are now excluded. We refer the reader to Section \ref{sect_tracer_ism} for a detailed description of the problem regarding H$_2$ upper limits, due to the sharp variations of H$_2$ radial profile. We stress, however, that our H$_2$ predictions remain in good agreement with the detections of H$_2$ (see Figure \ref{H2_obs_pred}) and that this issue only concerns H$_2$ upper limits. 

\subsection{Molecular gas reservoirs}

\subsubsection{Matching the CO emission}
\label{subsect_matching_co}

\begin{figure*}[h!]
\centering
\includegraphics[height=0.32\textwidth]{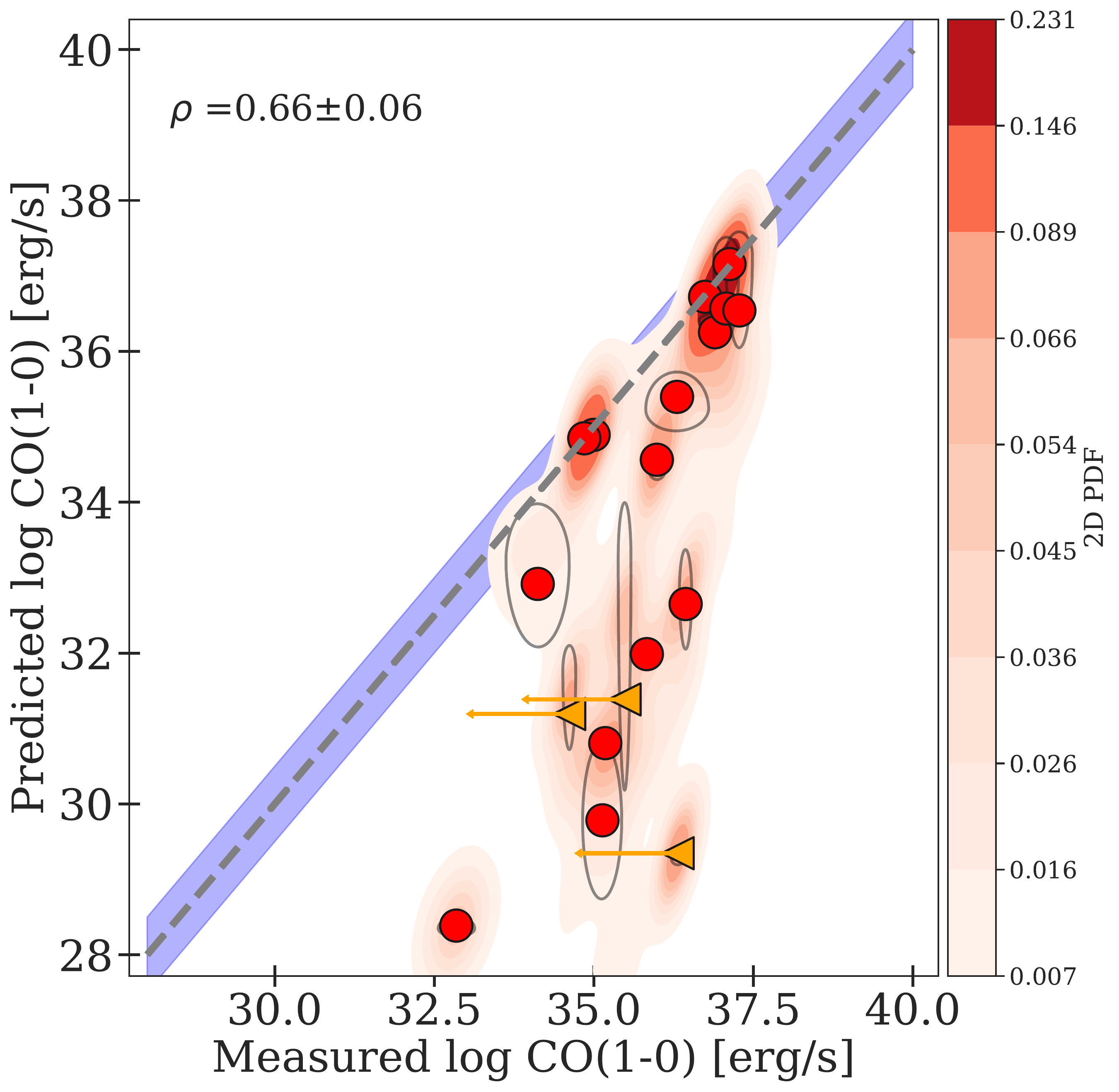}
\includegraphics[height=0.32\textwidth]{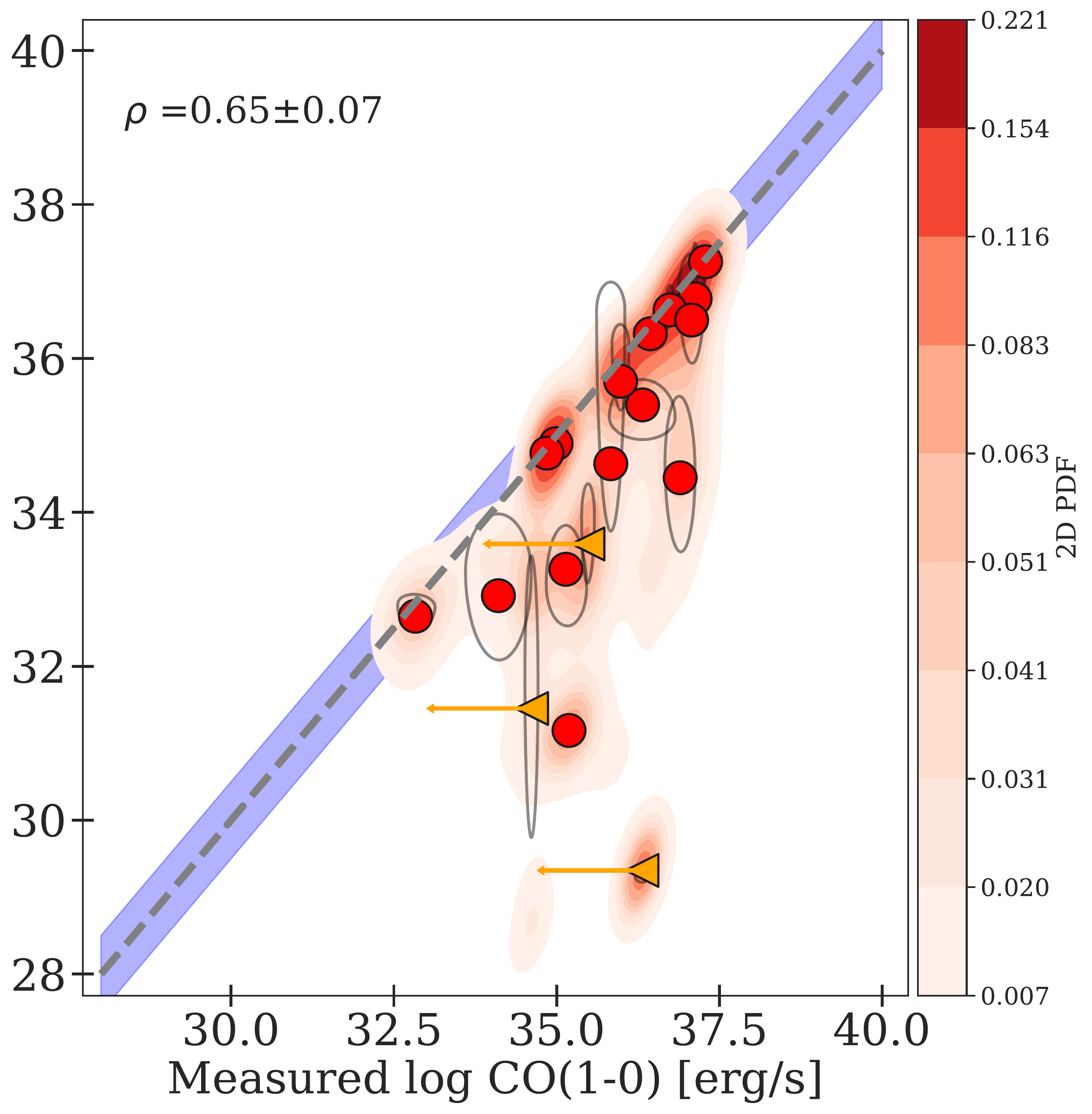}
\includegraphics[height=0.32\textwidth]{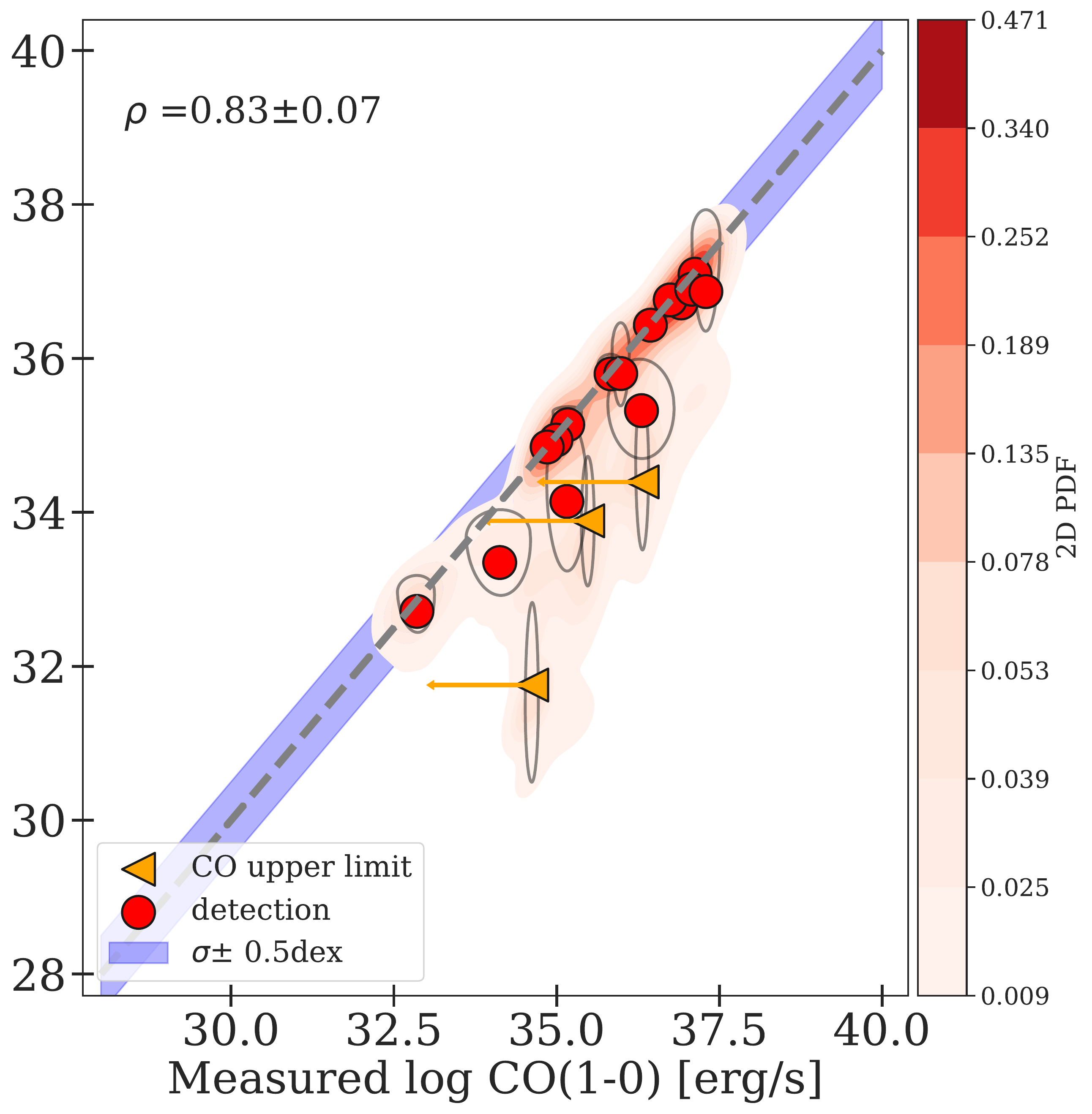}

\caption[Predicted vs. observed CO emission]{Predicted vs. observed CO emission for single-sector model (left column), multisector models (middle column), and power-law models (right column). The shades, symbols, and legends are described in Figure \ref{masses_relation}.}
\label{CO_obs_pred}
\end{figure*}

In Figure \ref{CO_obs_pred}, we plot the predicted CO luminosities versus the measured values, for all architectures. In single component models, the emission of CO is underpredicted by a factor larger than three (0.5\,dex) for 6 out of 15 detected galaxies. Multicomponent models, with one to four components, perform better at reproducing CO (12 out of 15 detections are well reproduced within 0.5\,dex), but they still underestimate CO by a factor of more than 3 in 3 detected galaxies. The multicomponent models based on power-law distributions are the only models to predict CO emission in agreement (within 0.5\,dex) with the observations, for all galaxies of our sample.

This inability of the models with a few (1 to 4) components to match CO emission is also illustrated by the P(3$\sigma$) values for CO only, reported in Table \ref{3sigma_values}. For single component models, the P(3$\sigma$) obtained for CO are low ($\leq 25\%$) for 7/15 galaxies detected in CO, and null in 4/15. The P(3$\sigma$) obtained using multicomponent models are globally larger, but remain below 25\% for 3/15 detections, and null for one galaxy. In other words, the tracers of the ionized and neutral gas are reproduced at the expense of CO. 

We note that this underestimation of CO lines was not present in \citetalias{madden_tracing_2020}, despite their use of a similar single-component approach, because the CO emission was matched a posteriori by adjusting the depth (A$_V$) of each model. While such adjustment successfully allows matching CO, it modifies a posteriori the solution, which may not match anymore the other ionized gas and PDR tracers. Here, on the other hand, we consider A$_V$ as a free parameter and attempt to fit all lines simultaneously during the MCMC sampling. Within our Bayesian framework, we find that matching all constraints, including CO, is impossible for models with only a few components. This result indicates that in some galaxies, the CO emission can only be matched by models accounting for numerous clouds or gas components. This will be further discussed in Section \ref{discussion_clumpiness}.

\subsubsection{CO-bright vs. CO-dark molecular gas}

\begin{figure*}[h!]
\centering
\includegraphics[width=0.99\textwidth]{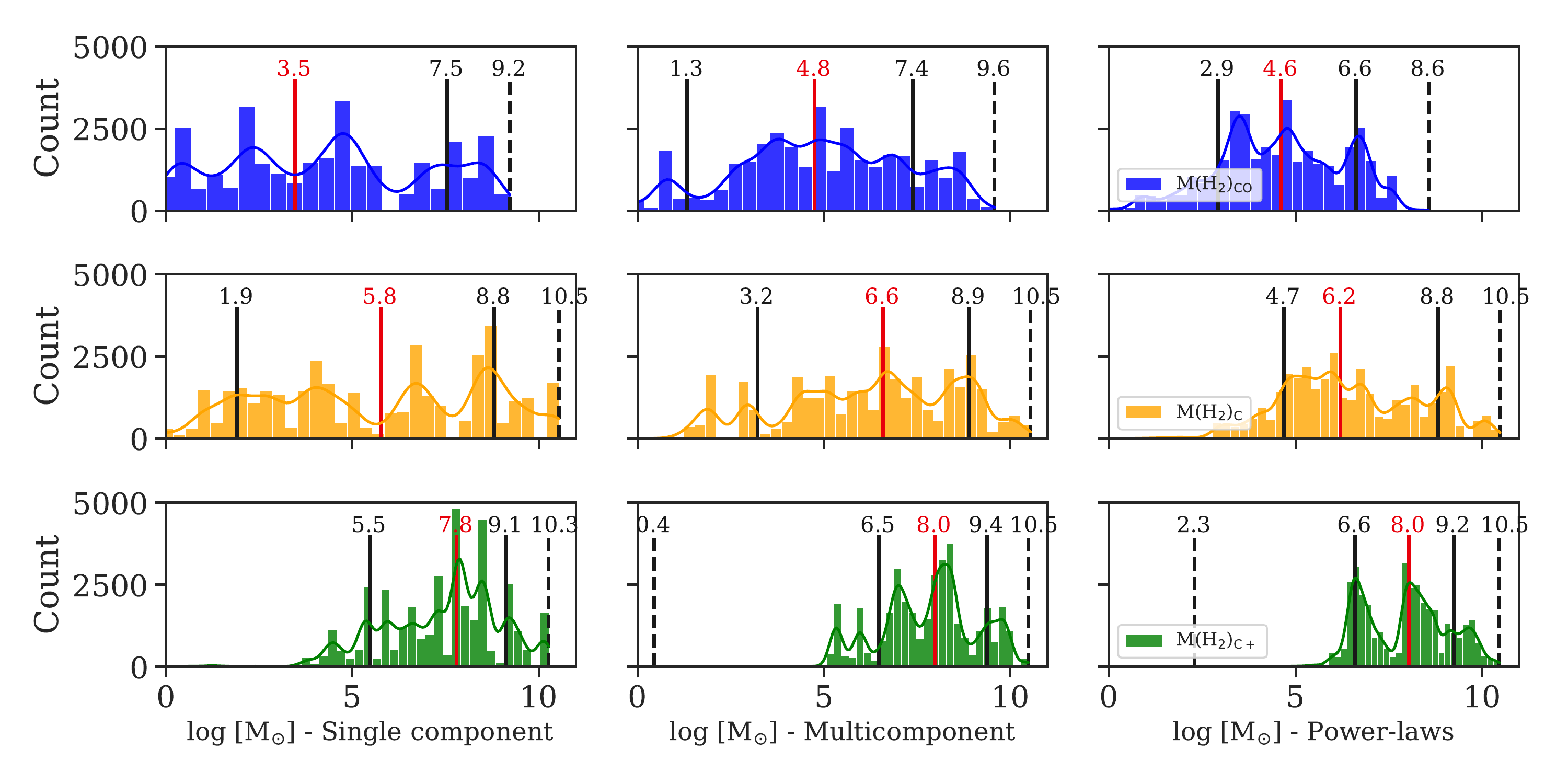}
\caption[Comparison of the M(H$_2$)$_{\rm CO}$, M(H$_2$)$_{\rm C}$, and M(H$_2$)$_{\rm C+}$ for different architectures.]{Comparison of the M(H$_2$)$_{\rm CO}$, M(H$_2$)$_{\rm C}$, and M(H$_2$)$_{\rm C+}$ for different architectures. The histograms are built using the MCMC draws for the whole sample. The vertical lines show the minima and maxima of the distributions (dashed lines), the quantiles at 15\% and 85\%  (plain black lines), and the median value (plain red line).}
\label{mass_H2_all}
\end{figure*}

\begin{figure}[h!]
\centering
\includegraphics[width=0.49\textwidth]{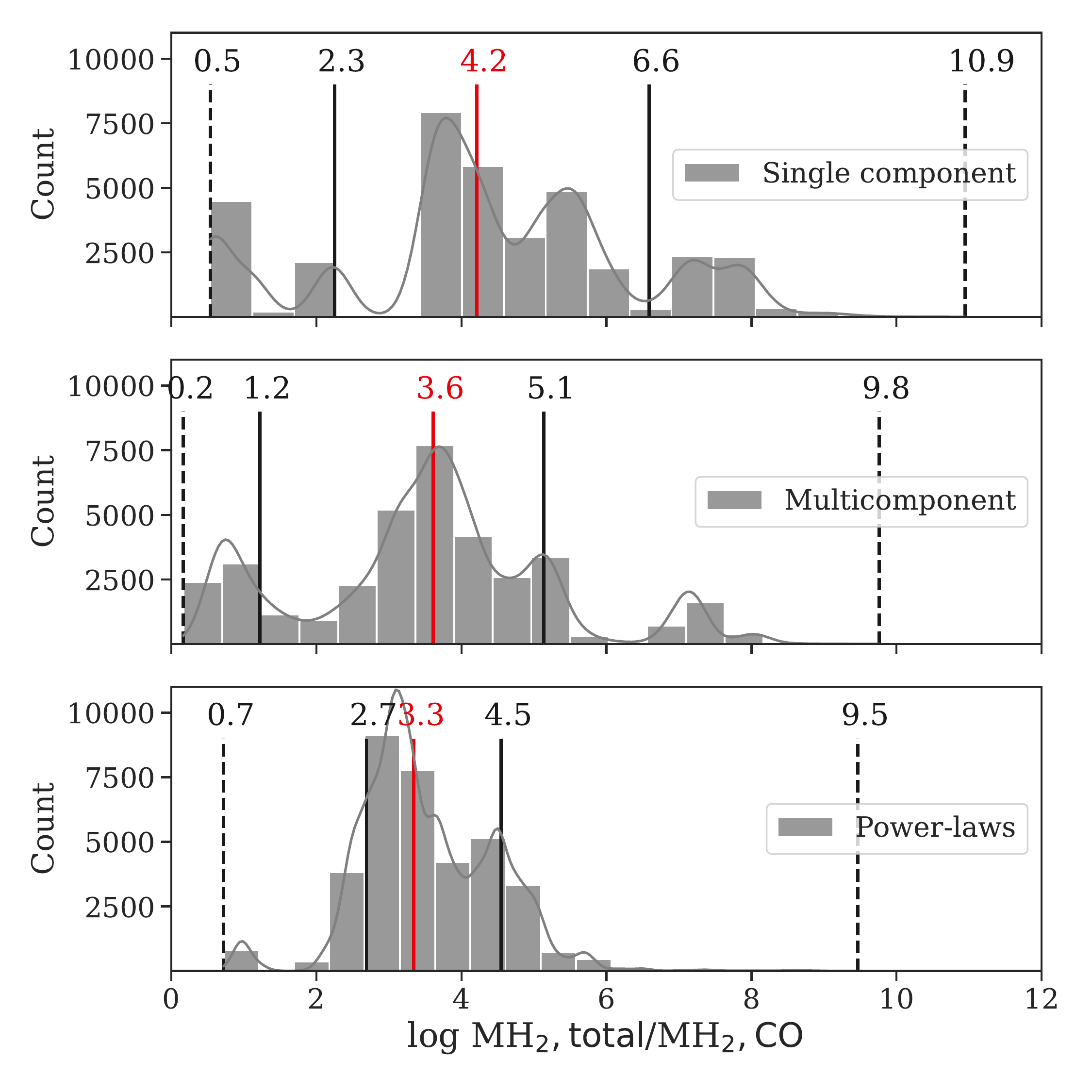}
\caption[Stacked histogram for the whole DGS sample of the total-to-CO-dark H$_2$ masses.]{Stacked histograms of the posterior distribution of the total-to-CO-bright H$_2$ masses for the whole DGS sample. The histograms are built using the MCMC draws for the whole sample. The vertical lines show the minima and maxima of the distributions (dashed lines), the quantiles at 15\% and 85\%  (plain black lines), and the median value (plain red line).}
\label{fdark}
\end{figure}

\begin{figure*}[h!]
\centering
\includegraphics[height=0.32\textwidth]{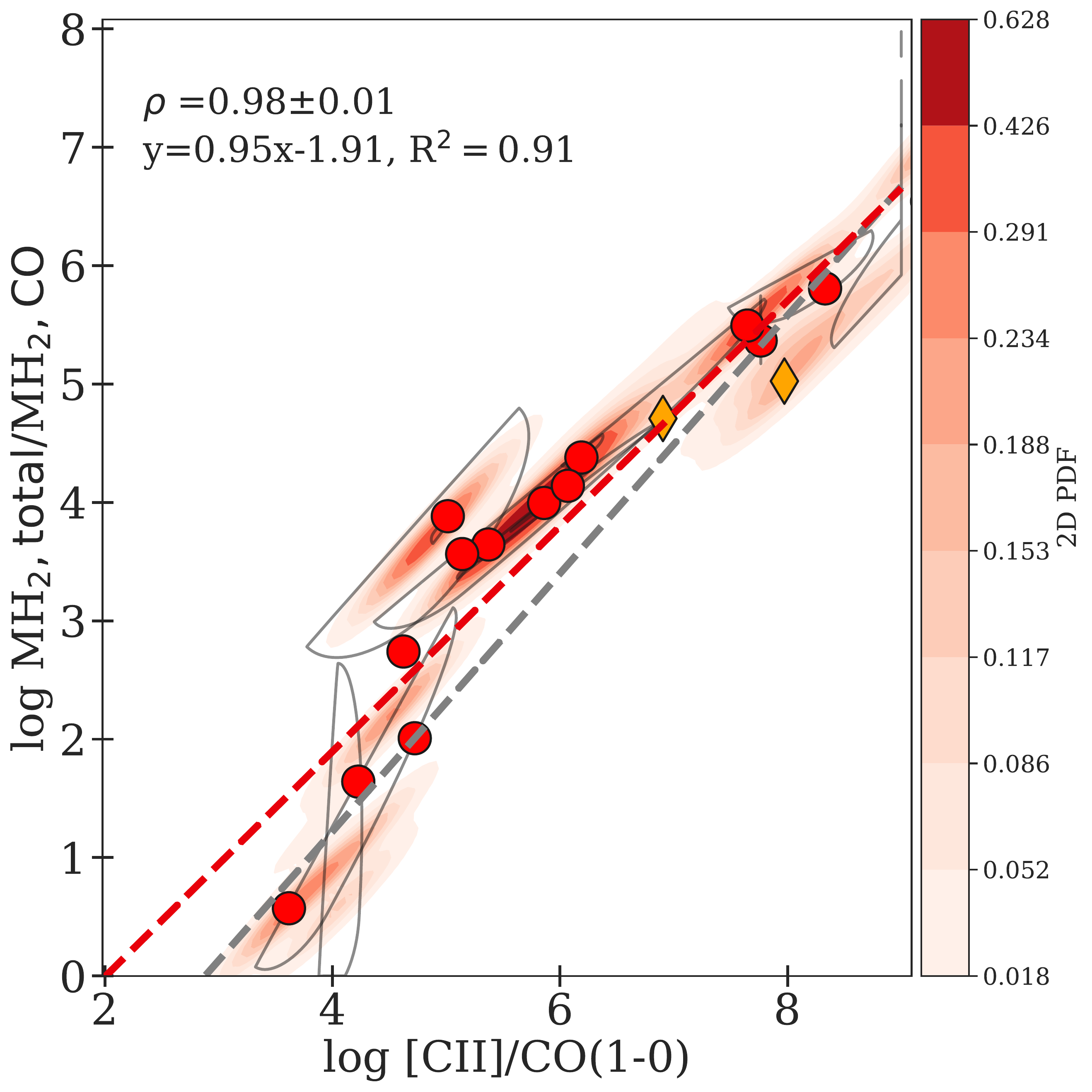}
\includegraphics[height=0.32\textwidth]{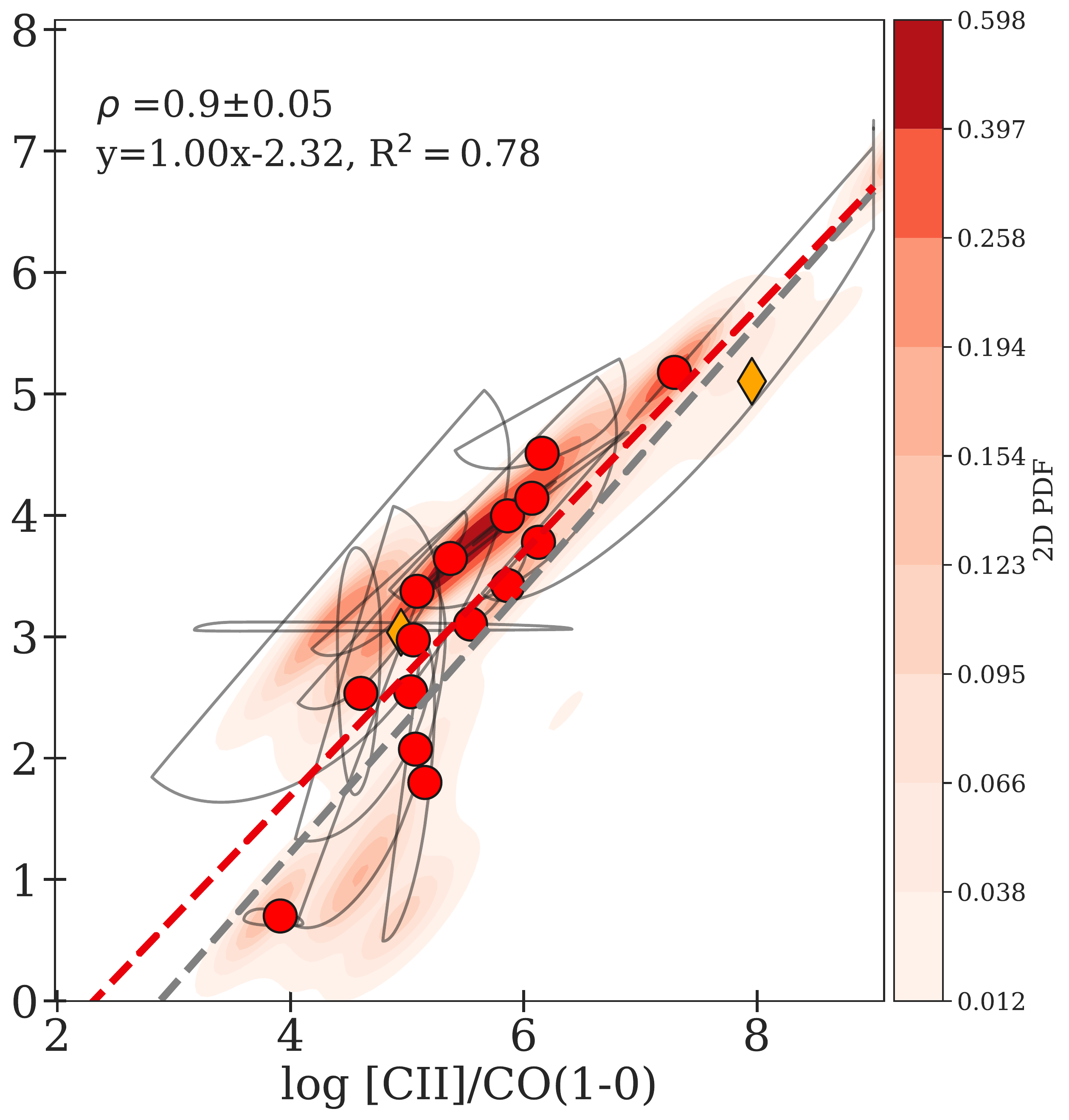}
\includegraphics[height=0.32\textwidth]{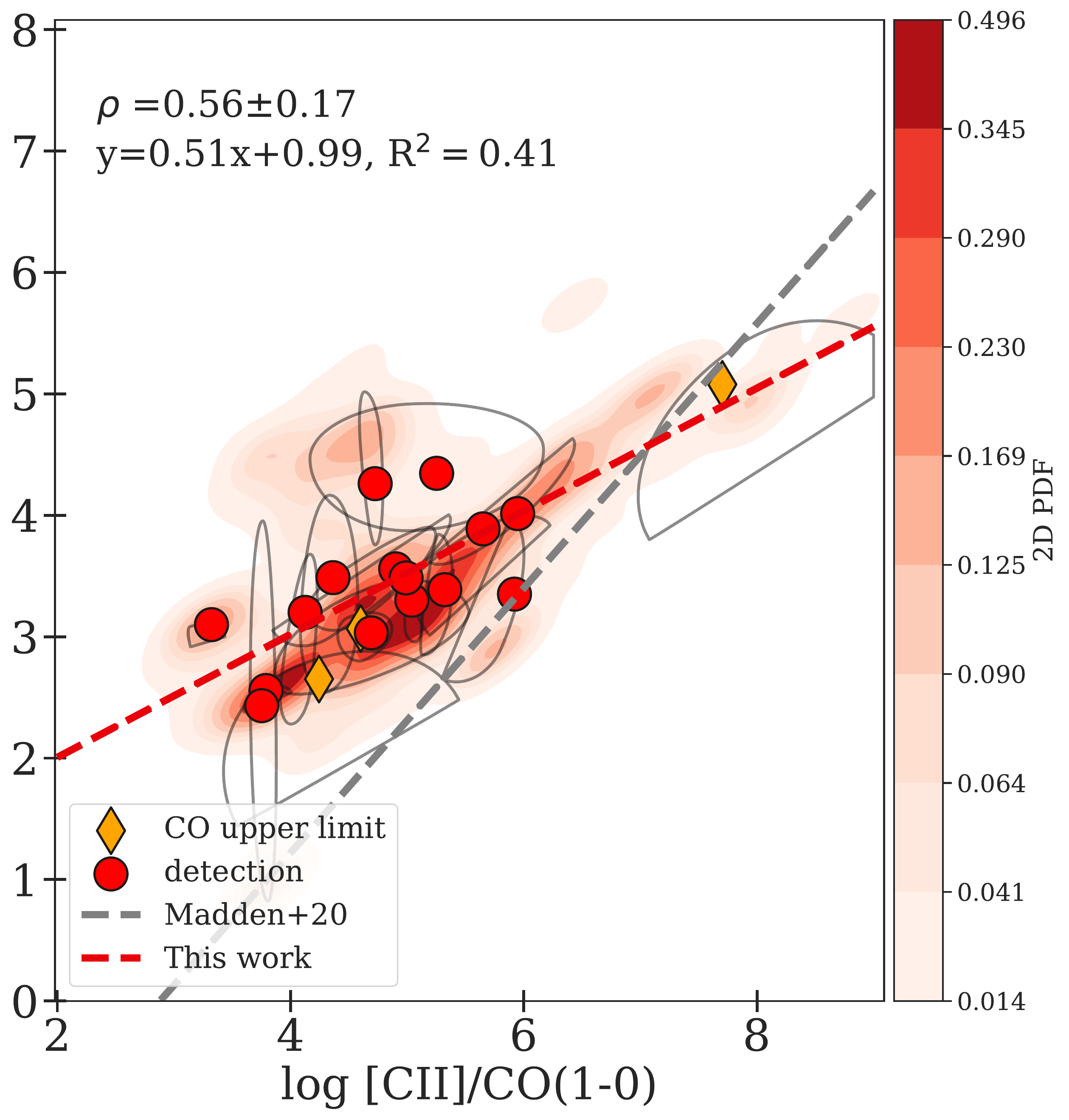}

\caption[MH$_{2, \rm total}$/MH$_{2, \rm CO}$ vs. \cii/CO(1-0)]{MH$_{2, \rm total}$/MH$_{2, \rm CO}$ vs. \cii/CO(1-0) for single component models (left), multicomponent models (middle), and power-law models (right). The shades and symbols are described in Figure \ref{masses_relation}. The dashed gray lines corresponds to the relation from \citetalias{madden_tracing_2020} while the dashed red lines show the linear regressions fitted to the combined posterior distribution for the entire sample. We indicate the equations of linear regressions fitted to the 2D-PDF of the whole sample, and the associated coefficients of determination R$^2$.}
\label{CIICO_all}
\end{figure*}

While the global mass distribution associated with \hii, \hi, and H$_2$ are not significantly sensitive to the choice of a given configuration (see Figure \ref{mass_H2_all}), the amount of CO-dark vs. CO-bright H$_2$ predicted using different geometries may vary significantly. Using Equation \ref{eq_masses}, we extract the masses of H$_2$ associated with C$^+$, C$^0$, and CO. The histograms obtained for the whole sample in each configuration are provided in Figure \ref{mass_H2_all}. Qualitatively, the total H$_2$ masses in our sample are dominated by H$_2$ reservoirs associated with C$^+$, and to a lesser extent with C$^0$, while the reservoirs associated with CO are 2 to 4 orders of magnitude below. 

In a more quantitative way, we examine in Figure \ref{fdark} the ratios of the total H$_2$ mass (M$_{\rm H2, total}$) to CO-bright H$_2$ mass (M$_{\rm H2, CO}$), shown as a histogram for the whole sample. 
On average, we find that the galaxies in our sample are completely dominated by the CO-dark H$_2$ masses. The predicted total H$_2$ masses are on average 100 to 10$^6$ times larger than the H$_2$ masses associated with CO, in all architectures, although larger for single component models. Specifically, we find median total-to-CO-bright ratios of 10$^{4.5}$, 10$^{3.7}$, and 10$^{3.4}$ for single component models, multicomponent models, and power-law models respectively. Those numbers indicate that, on average, nearly 100\% of the H$_2$ gas is CO-dark in our sample. This is linked to the selection of our sample, which consists of extremely CO-faint galaxies and galaxies in which CO is undetected. We find that the predicted fractions of CO-dark gas are at least of 70\%, 37\%, and 80\% in single component models, multicomponent models, and power-law models respectively.

In Figure \ref{CIICO_all}, we plot the total-to-CO-bright H$_2$ mass ratio as a function of \cii/CO for the three architectures. In \citetalias{madden_tracing_2020}, this mass ratio was found to tightly correlate with the \cii/CO. Similarly, we find a tight correlation of M$_{\rm H2, total}$/M$_{\rm H2, CO}$ with \cii/CO for single component models (left panel). Our predictions are slightly shifted away from the relation of \citetalias{madden_tracing_2020} because CO is underpredicted by single-component models, as already pointed out in Section \ref{subsect_matching_co}. The correlation between M$_{\rm H2, total}$/M$_{\rm H2, CO}$ and \cii/CO still holds when a larger number of components are combined. Nevertheless, we find that increasing the number of components flattens the relation and increases its dispersion.
Indeed, purely CO-dark components, corresponding either to diffuse CO-dark reservoirs or CO-dark clumps, may be included in the models when combining components, which is not the case with a single component model where the CO-dark H$_2$ envelope is always associated with CO-bright H$_2$ core. Hence, models combining components predict higher M(H$_{2}$)$_{\rm total}$/M(H$_{2}$)$_{\rm CO}$ at fixed \cii/CO (i.e., larger content of CO-dark gas), and a larger spread around the average linear relation.

\section{Results with statistical distributions of components}
\label{section_results_plaw}

As discussed in Section \ref{subsect_matching_co}, the architecture using power-law and broken power-law distributions for the $U$, $n$, and cut parameters performs better at reproducing the CO emission in our sample. We now focus only on the latter architecture to infer the integrated molecular gas masses and $\alphaCO$ values.

\subsection{Constraints on the power-laws and broken power-laws}

As described in Section \ref{broken_plaw_models}, we combine components following power-laws (defined by a slope over a given range of values between a lower and an upper bound) for the density and ionization parameter and a broken power-law (defined by two slopes over two ranges of values separated by a pivot point) for the cut parameter. 
We consider relatively weak priors for the slope and boundaries, defined as normal distributions centered on a value $\mu$ with a relatively large $\sigma$. We now examine the means of the posterior PDFs for the density, ionization parameter, and cut in our sample.

\begin{figure*}[h!]
\centering
\includegraphics[width=0.99\textwidth]{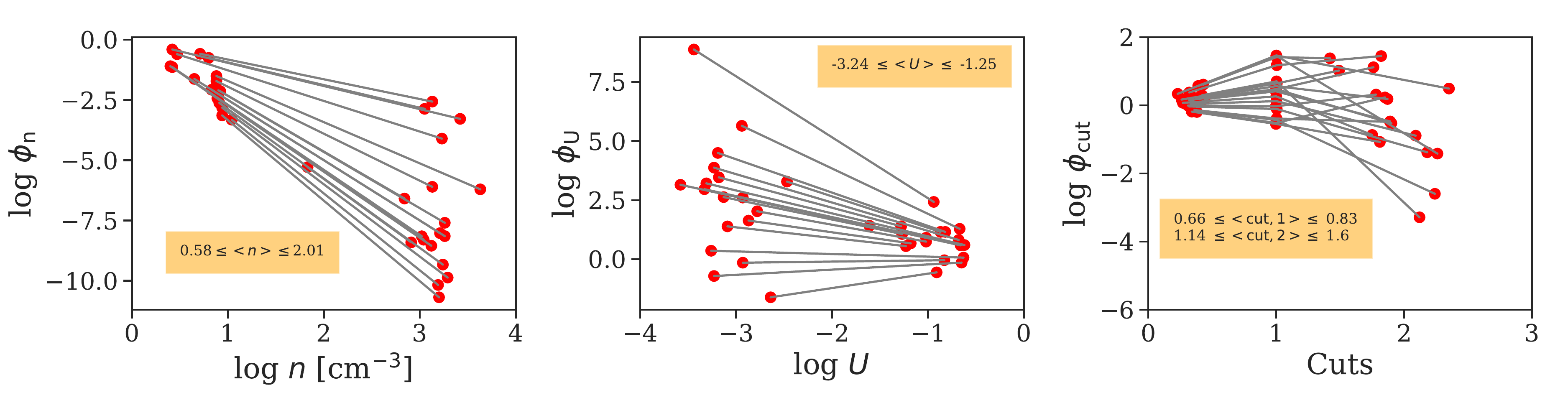}
\caption[Values of the power-law parameters for individual galaxies controlling the density, ionization parameter, and cut]{Values of the power-law parameters for individual galaxies controlling the density (left), ionization parameter (middle), and cut (right) for each galaxy. The red dots correspond to the means of the lower and upper boundaries of power-laws (and pivot point for the broken power-law) and the gray lines represent power-laws corresponding to the averages slope values, for each galaxy. }
\label{slopes_n_u}
\end{figure*}

In Figure \ref{slopes_n_u}, we show the means of the posterior PDFs obtained for the slopes and boundaries of the distributions for density, ionization parameter, and cut in each of the individual galaxies. We note that the power-laws controlling the density and ionization parameter are parameterized using the values of $U$ and $n$ at the illuminated face of the clouds, which may differ from the volume-averaged $U$ and $n$.  In Table \ref{table_mean_values}, we report the luminosity weighted-averaged of the corresponding parameters: the density and the ionization parameter at the illuminated edge of \hii\ regions, as well as the cuts in the ionized gas (\#1) and neutral gas (\#2).

In the left-hand side panel of Figure \ref{slopes_n_u}, the mean slopes obtained for the distribution of density are all negative, meaning that components with relatively low density dominate the emission with respect to denser regions. As reported in Table \ref{table_mean_values}, the averaged densities in the \hii\ regions are $\sim$10-100\,cm$^{-3}$, which is close to the typical values often considered for \hii\ regions in thermodynamic equilibrium. We note that three galaxies, NGC\,1140, NGC\,5253, and UM\,448, have mean densities lower than 10\,cm$^{-3}$. We speculate that the relatively high contribution of low density gas in those galaxies may be due to a disturbed morphology of the gas (e.g., ionization cones in NGC\,5253; \citealt{zastrow_ionization_2011}, interaction-induced inflow in a merger system for UM\,448; \citealt{James_2013}, complex ionized gas structure associated with superbubbles and shocked shells in NGC\,1140; \citealt{Westmoquette_ngc1140_2010}). On the other-hand, SBS\,0335-052 is associated with a larger mean density value ($\sim 100$\,cm$^{-3}$) than the other galaxies and a relatively high lower bound for the density ($\sim 67$\,cm$^{-3}$), which suggests that diffuse regions do not contribute much to the observed emission in this galaxy.
In the middle panel of Figure \ref{slopes_n_u}, we observe mostly negative medians for the slopes of the distribution of ionization parameters, meaning that regions with relatively low ionization parameters may dominate the total luminosity in most galaxies. The range of slopes covered by the whole sample is large, with several galaxies showing relatively flat slopes and three galaxies with a positive slopes (i.e, dominated by components of high ionization parameters). The means reported in Table \ref{table_mean_values} are between $\sim -3$ and $-1.5$, with three galaxies associated with relatively higher ionization parameters (II\,Zw\,40, Mrk\,209, and SBS\,0335-052) between -1.5 and -1.25.

On the right-hand side panel of Figure \ref{slopes_n_u}, we plot the broken power-law parameters for the cuts, which show a change of slopes at the ionization front (cut=1). 
This drastic change of slope is indicative of the presence of different populations of gas components, depending on their optical depth. Specifically, we find that the distribution of “naked” \hii\ regions, which are not associated with any neutral gas (i.e., cut $\leq$ 1), differs from the distribution of embedded \hii\ regions associated with a complete or partial PDR, and potentially with molecular gas (i.e., cut > 1). We observe a large spread in the slopes obtained for each individual galaxy, with both negative and positive slopes found in the ionized gas and in the neutral gas. 

Interestingly, if we focus only on galaxies associated with the largest upper bounds for cuts (e.g., cut $\geq$ 2, meaning that the models include components which are deep enough to exceed the dissociation front), we find that they are all associated with clearly negative slopes for the cuts within the neutral gas. Qualitatively, such negative slopes indicate that the component with relatively smaller optical depths are more numerous and dominate the total luminosity, while fewer components reaching larger optical depths contribute to the emission. These features could be associated with a “clumpy” distribution of dense clouds and will be further discussed in Section \ref{discussion_clumpiness}. In Table \ref{table_mean_values}, we identify in bold 6/18 galaxies associated with upper bounds larger than 2, which will be further examined in Section \ref{subsect_alpha_CO}. The averaged values for the cut in the ionized gas are always between 1 (ionization front) and 2 (dissociation front), which is indicative of galaxies dominated (in mass) by atomic neutral gas rather than molecular gas, as discussed in Section \ref{subsect_masses}. The mean cut values in the ionized gas are between $\sim 0.6$ and $\sim 0.8$, while one may expect mean cut values closer to unity, if most of the \hii\ regions were radiation-bounded. Such values may reflect an important contribution of density-bounded regions to the total luminosity. The presence of the latter regions, potentially associated with leakage of ionizing photons, was studied in more details in \cite{Ramambason2022}.

\subsection{Total H$_2$ masses}

\subsubsection{Tracers of M(H$_2$)}
\label{tracer_of_MH2}

\begin{figure*}[h!]
\centering
\includegraphics[height=0.32\textwidth]{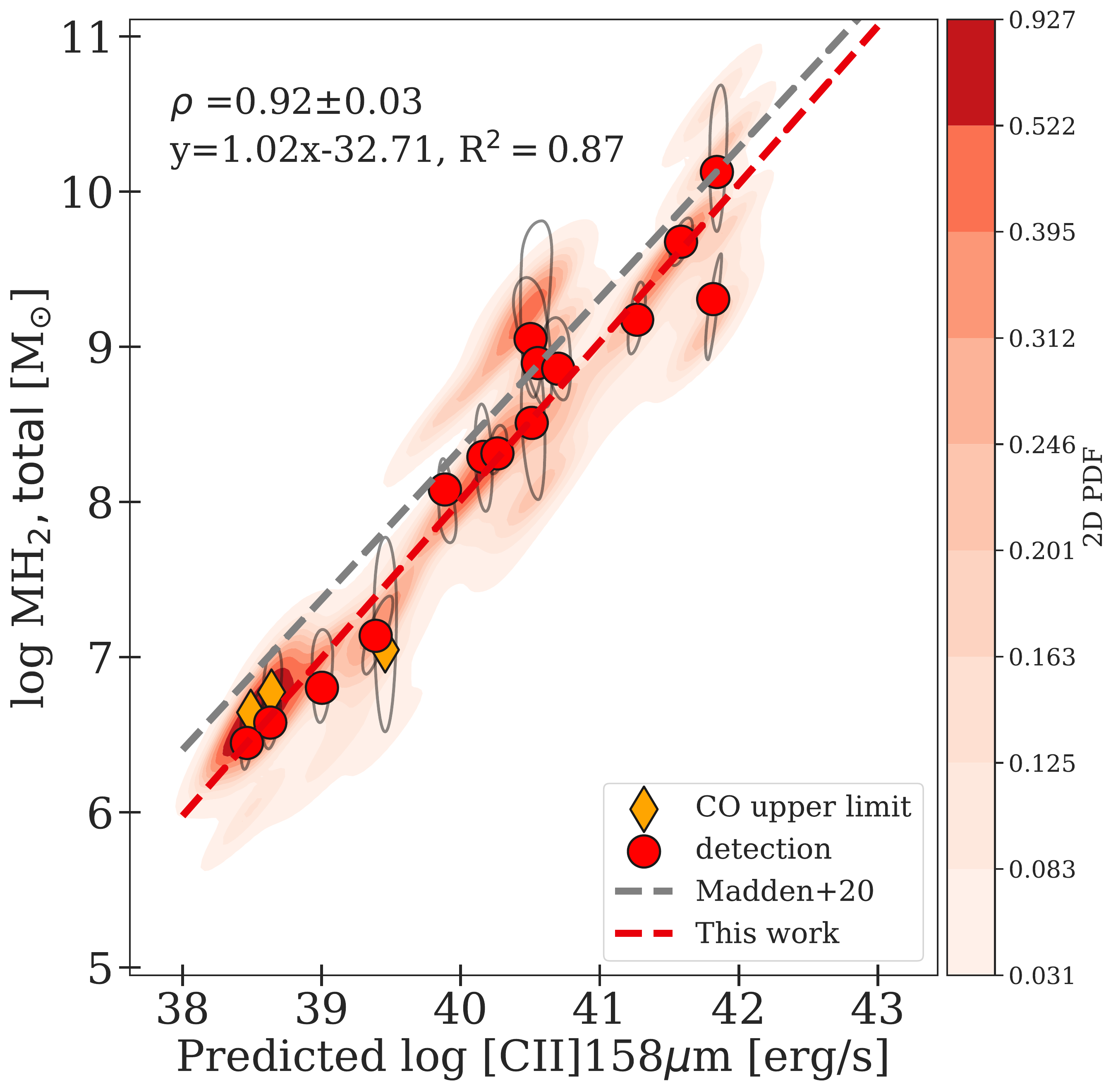}
\includegraphics[height=0.32\textwidth]{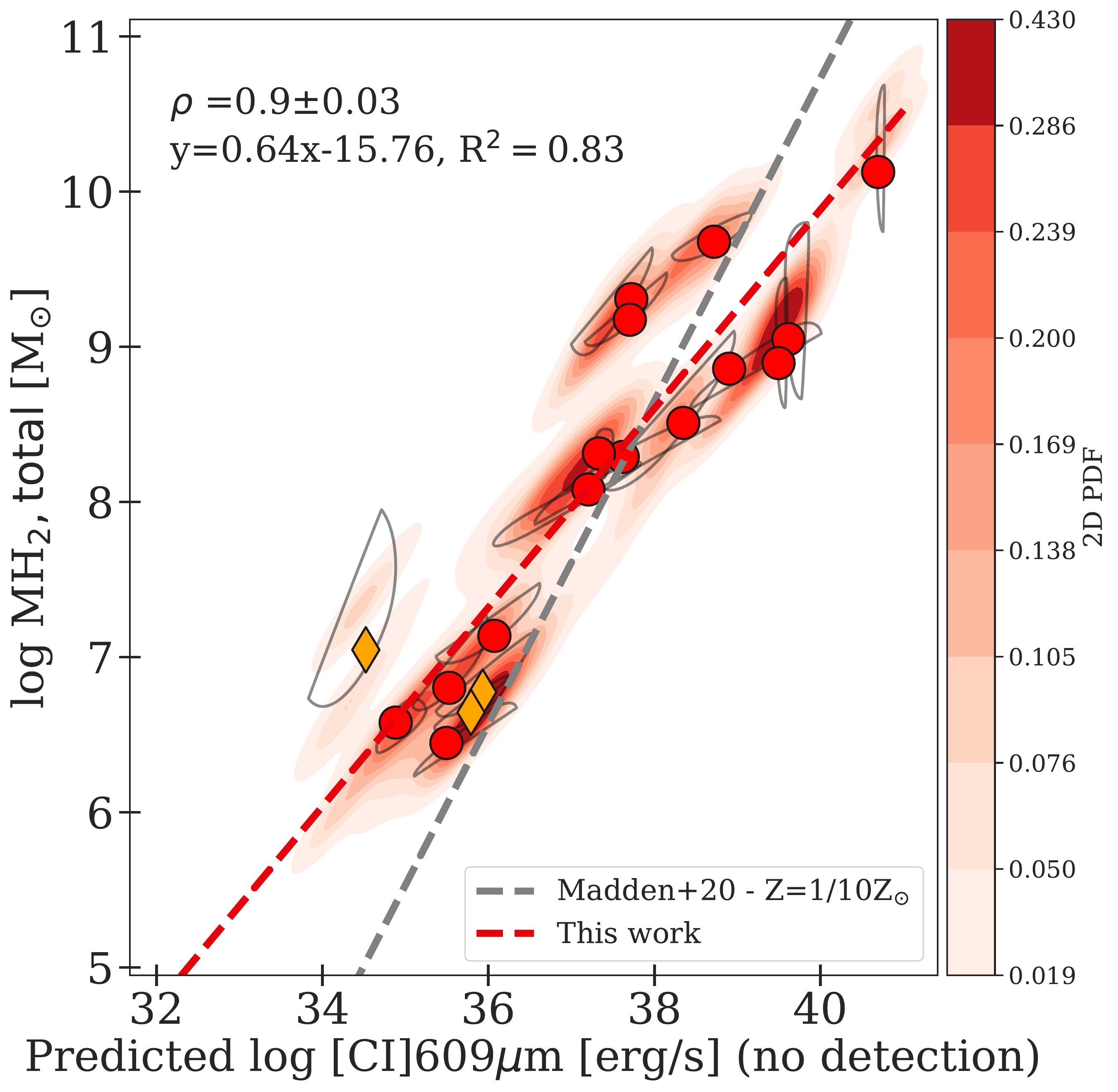}
\includegraphics[height=0.32\textwidth]{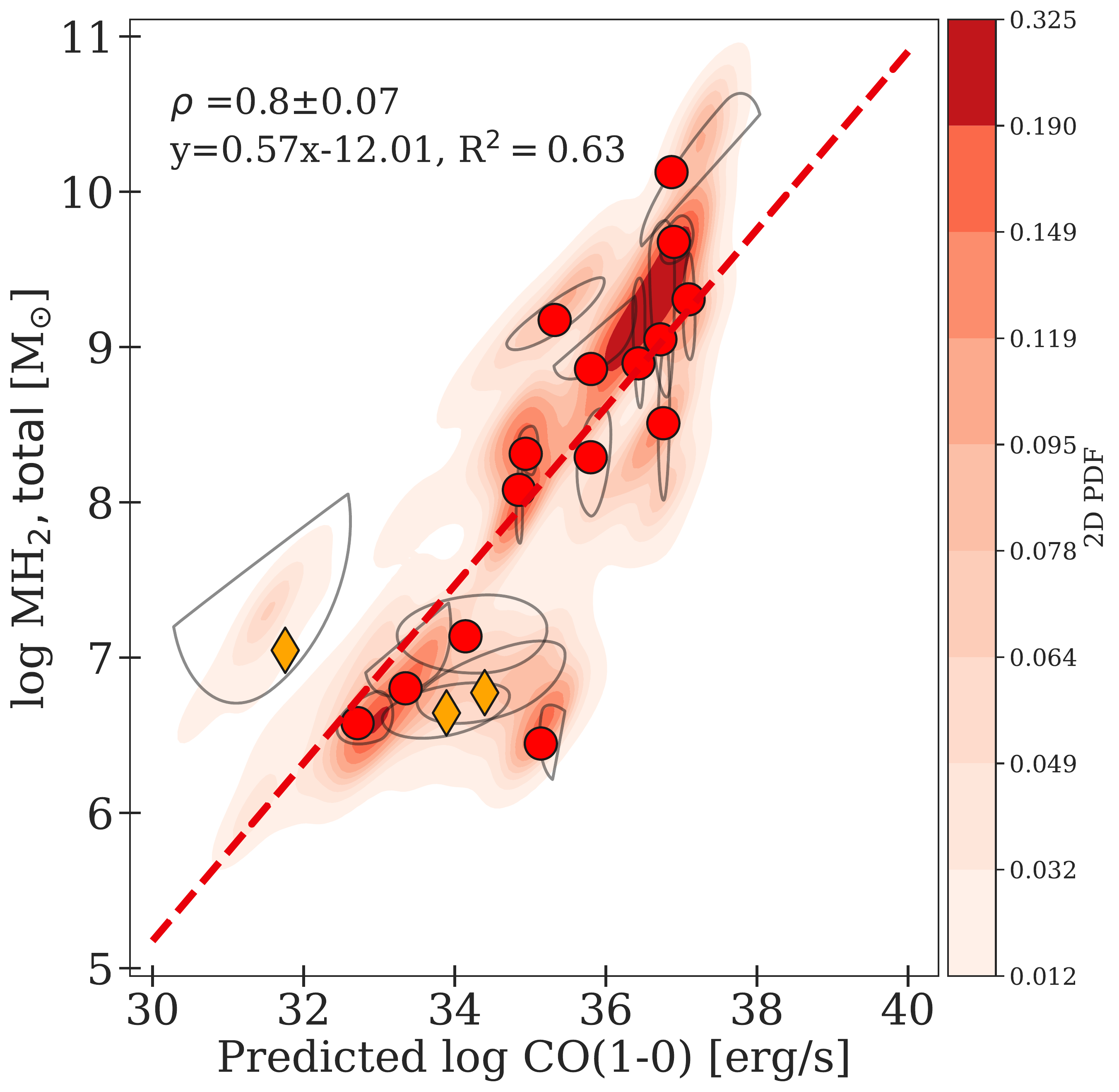}
\caption[Predicted line fluxes for \cii\, \ci, CO vs. MH$_{2, \rm total}$]{Predicted line fluxes for \cii\, \ci, and CO emission lines vs. MH$_{2, \rm total}$ for multicomponent models with power-law distributions. The predictions for \cii\ and CO match well the observed measurements, as shown in Figures \ref{pdr_lines_obs_pred} and \ref{CO_obs_pred}. The predictions for \ci\ are purely model-based, since no measurement is available. The dashed gray lines show the relations from \citetalias{madden_tracing_2020} while the dashed red lines show the linear regressions fitted to the combined posterior distribution for the entire sample. The shades, symbols, and legends are described in Figure \ref{masses_relation}. We indicate the equations of linear regressions fitted to the 2D-PDF of the whole sample, and the associated coefficients of determination R$^2$.}
\label{tracers_MH2}
\end{figure*}

In Figure \ref{tracers_MH2}, we plot the evolution of the total H$_2$ masses predicted by the power-law models with respect to different tracers: \cii, \ci, and CO(1-0). All the galaxies are detected in \cii\ and our line flux predictions are in excellent agreement with the observed values as shown in Figure \ref{pdr_lines_obs_pred}. For \ci, our plots show purely predicted luminosities since no detection is available in our sample. Nevertheless, as discussed in Section \ref{subsect_matching_co}, the power-law models are in good agreement with the few tracers arising from the PDRs, as shown by the P(3$\sigma$) in Table \ref{3sigma_values} and in Figure \ref{pdr_lines_obs_pred}. For the CO(1-0), our predictions are in good agreement within 0.5\,dex with all measurements, as shown in Figure \ref{CO_obs_pred}.

We find that the total H$_2$ mass correlates best with \cii 158$\mu$m, with a Spearman coefficient of $0.92 \pm 0.03$. As shown in the left-hand side panel of  Figure \ref{tracers_MH2}, the relation we derive is slightly below that of \citetalias{madden_tracing_2020}. This systematic offset comes from the nature of the power-law models (see Section \ref{subsect_matching_co}), which allows us to match the CO emission with numerous components rather than assuming a single component. As a result, the total H$_2$ masses derived for a fixed \cii\ value are systematically smaller than those derived in \citetalias{madden_tracing_2020}, on average by a factor 180. 
We note that the single component models (presented in Section \ref{subsect_single_comp}), predict masses in perfect agreement with the \citetalias{madden_tracing_2020} relation, indicating that this offset is merely due to the different distribution of gas (i.e., more independent components)  considered in the multicomponent models. 

In the middle panel of Figure \ref{tracers_MH2}, we show the relation between H$_2$ mass and \ci\ 609$\mu$m. We find that, based on our predictions, \ci\ 609$\mu$m also provides an excellent tracer of the total H$_2$ (with a Spearman correlation coefficient of $0.9 \pm 0.03$), although with a slightly larger dispersion than for the \cii\ line. We predict a shallower relation than that derived in \citetalias{madden_tracing_2020} based on single-sector models with a fixed metallicity of 1/10\,Z$_{\odot}$. 
On the right-hand side panel of Figure \ref{tracers_MH2}, we show the relation between M(H$_2$) and CO(1-0). Although the two quantities are correlated, we find a lower Spearman correlation coefficient ($\rho=0.8 \pm 0.07$) and a lower regression coefficient than for \cii\  and \ci. Our predictions suggest that both \cii\ and \ci\ are better tracers of the total H$_2$ mass than CO(1-0) at low-metallicity and should be preferred, provided they can be detected.

\subsubsection{Conversion factors}
\label{subsect_conversion_factors}

\begin{figure*}[h!]
\centering
\includegraphics[height=0.32\textwidth]{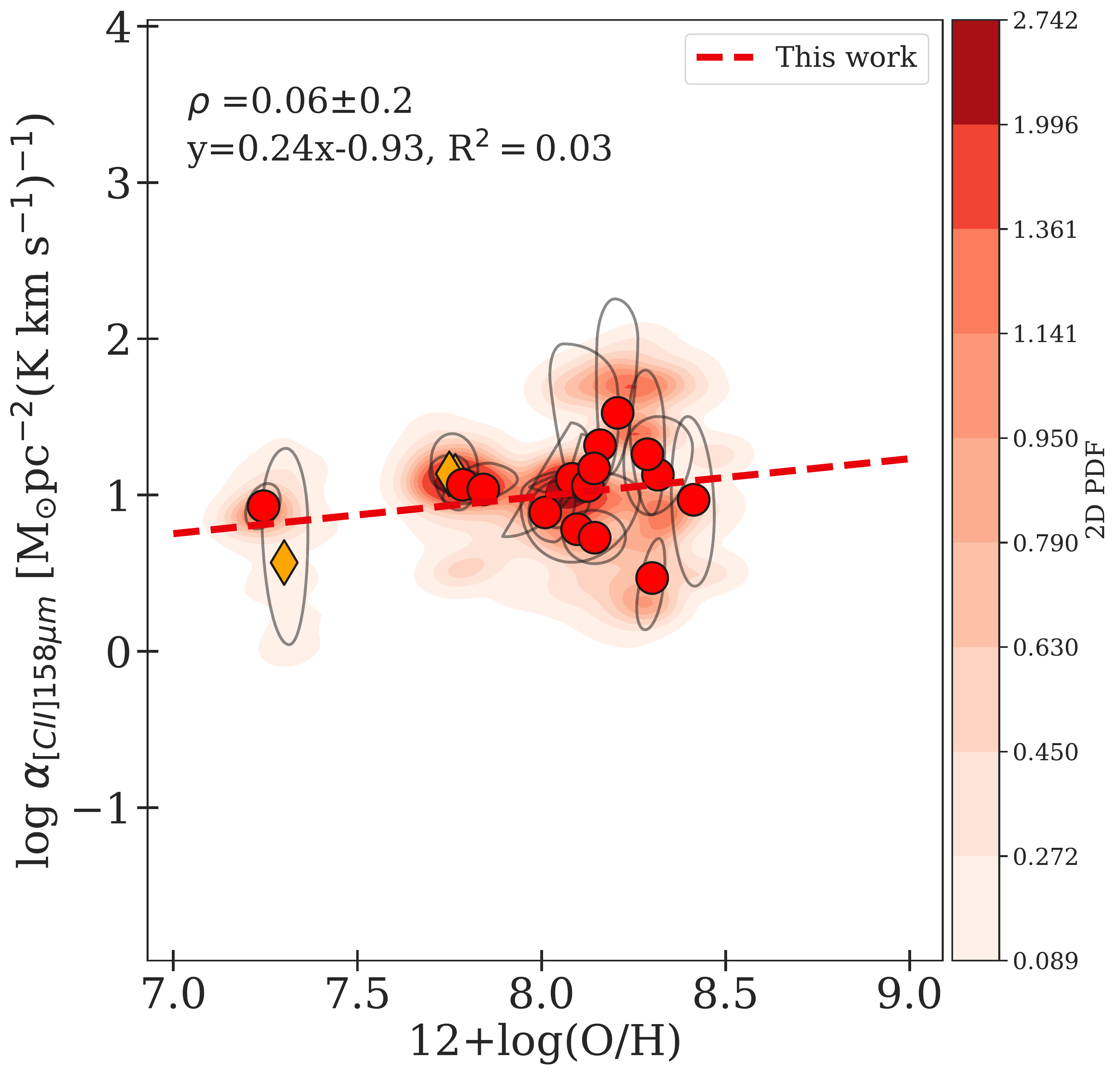}
\includegraphics[height=0.32\textwidth]{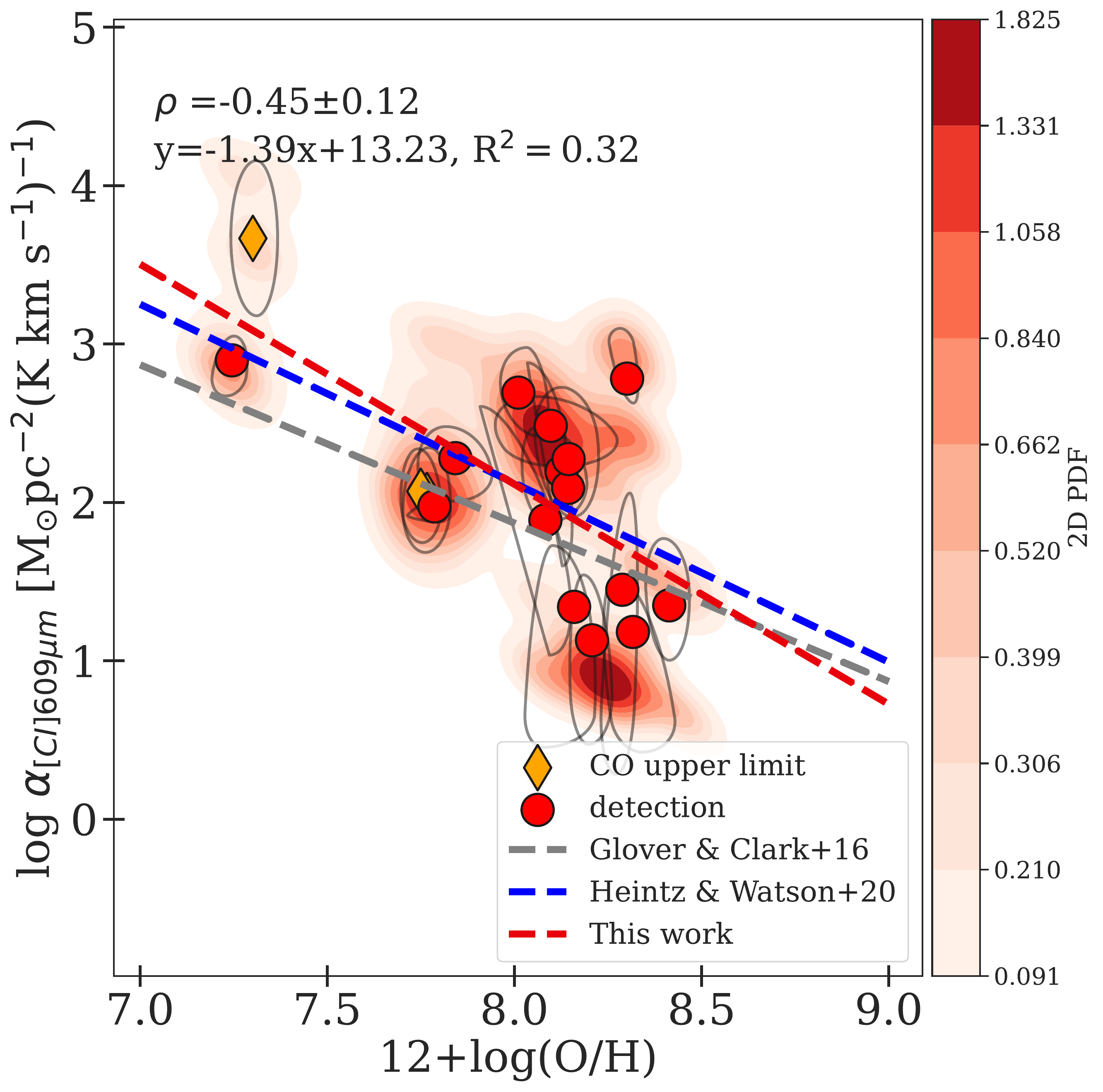}
\includegraphics[height=0.32\textwidth]{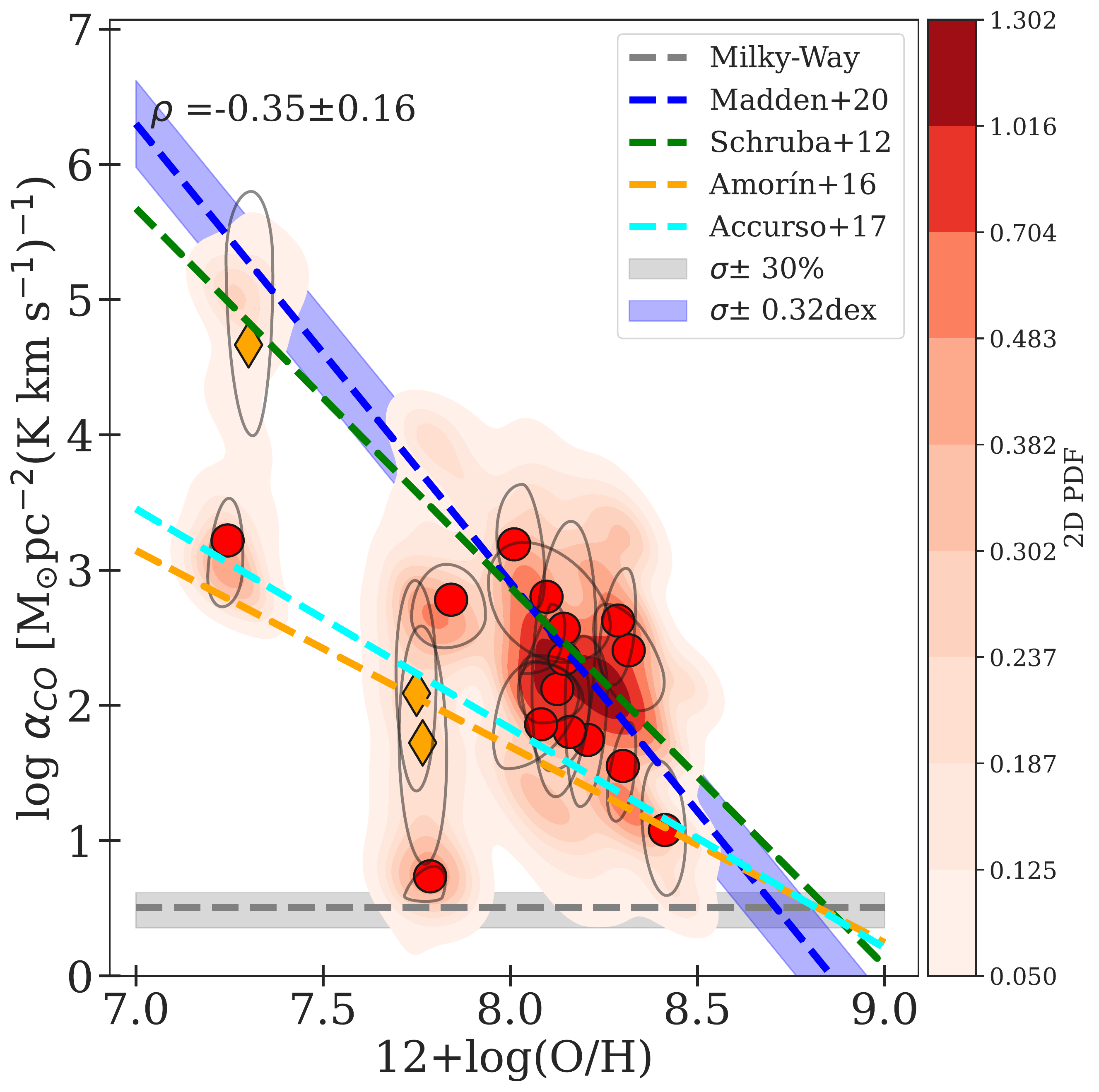}
\caption[$\alpha_{\rm CII}$, $\alpha_{\rm CI}$, $\alpha_{\rm CO}$ vs. metallicity]{$\alpha_{\rm CII}$ (left), $\alpha_{\rm CI}$ (middle), $\alpha_{\rm CO}$ (right) vs. metallicity for power-law models. The shades, symbols, and legends are described in Figure \ref{masses_relation}. We indicate the equations of linear regressions fitted to the 2D-PDF of the whole sample, and the associated coefficients of determination R$^2$. We also show the Galactic $\alpha_{\rm CO}$ value from \citetalias{Bolatto_2013} and the range of values compatible with it within 30\% (gray-shaded area), the $\alpha_{\rm CO}$ vs. metallicity relations from \citetalias{madden_tracing_2020}, \cite{Schruba_2012}, \cite{Amorin_2016}, and \cite{Accurso_2017}, and the $\alpha_{\rm [CI]}$ vs. metallicity relations from \cite{Glover_ci_2016} and \cite{Heintz_watson_2020}. We use a Galactic conversion factor of 3.2\,M$_{\odot}$ pc$^{-2}$(K\,km\,s$^{-1}$)$^{-1}$ to correct for the helium mass (factor of 1.36) when comparing to the model prediction of M(H$_2$). }
\label{alpha}
\end{figure*}

In Figure \ref{alpha}, we translate the relations from Section \ref{tracer_of_MH2} into conversion factors, assuming that the area of emission is the same for all tracers. \footnote{In practice, we use a conversion factor of erg\,s$^{-1}$ to K\,km\,s$^{-1}$pc$^{2}$ calculated for each frequency using the formula from on \cite{Solomon_1997}}. 
We then examine the variations on those conversion factors with metallicity. 
As shown in Figure \ref{alpha} (left and middle panel), we find a nearly constant $\alpha_{\rm [CII]}$ value at all metallicities while the $\alpha_{[CI]}$ values anticorrelate with metallicity, with a slope close to $-1$. We note, however, that there is a significant scatter ($\sim$2\,dex) around both relations, which is linked to the variety of gas geometry considered when combining many components. We find similar trends of  $\alpha_{\rm [CII]}$ and  $\alpha_{\rm [CI]}$ values with metallicity when considering lower numbers of components (from 1 to 4 components, see Section \ref{subsect_single_comp} and \ref{subsec_multi_comp}), with an increasing scatter as a function of the number of components.

Despite the large scatter, $\alpha_{\rm [CI]}$ (middle panel) shows a clear metallicity trend, with a Spearman coefficient of $-0.45 \pm 0.12$. We derive a negative slope of $-1.39$, which is in relatively good agreement, although slightly steeper, with the slopes derived in both observational and theoretical studies. In particular, our predictions are compatible with the study from \cite{Heintz_watson_2020}, which derived a slope of $-1.13$ based on samples of high-redshift $\gamma$-ray burst and quasar molecular gas absorbers. The slope we derive is  also close to the value of about $-1$ derived by \cite{Glover_ci_2016} based on simulations of star forming clouds, just before the onset of star formation. The latter study also points out that the exact dependence of $\alpha_{\rm [CI]}$ with metallicity is sensitive to dynamical effects and is likely to vary over time. Those effects are ignored in our stationary 1D models and will be further discussed in Section \ref{discussion_limits}. 

Finally, we examine the $\alpha_{\rm CO}$ versus metallicity dependence, shown in the right-hand side panel of Figure \ref{alpha}. While we observe a clear dependence of $\alpha_{\rm CO}$ with metallicity, we find a relatively low absolute value of the Spearman correlation (0.19), indicating that the monotonic trend is weak. This is especially striking when compared to the results from \citetalias{madden_tracing_2020} for which the $\alphaCO$ of the DGS galaxies followed a steep relation with metallicity with a narrow dispersion (0.32\,dex). 
We find instead a large scatter at fixed metallicity for the power-law models, which match best the observed emission lines (see Section \ref{subsect_matching_co}). While most of the galaxies in our sample are found in good agreement with the steep relation from \citetalias{madden_tracing_2020} and \cite{Schruba_2012}, several $\alpha_{\rm CO}$ values are in better agreement with flatter relations \citep[e.g.,][]{Amorin_2016,Accurso_2017}, and three galaxies have a predicted $\alphaCO$ values close to the Galactic value (within a factor 3 based on their 1$\sigma$ uncertainties), despite their subsolar metallicity. This finding indicates that the $\alphaCO$ may be driven by another physical parameter. In the next section, we further explore the impact of the geometry of our models on the derived $\alphaCO$ values, through the clumpiness of the gas.

\subsubsection{Impact of clumpiness on $\alpha_{\rm CO}$}

\label{subsect_alpha_CO}

\begin{figure}[h!]
\centering
\includegraphics[height=0.4\textwidth]{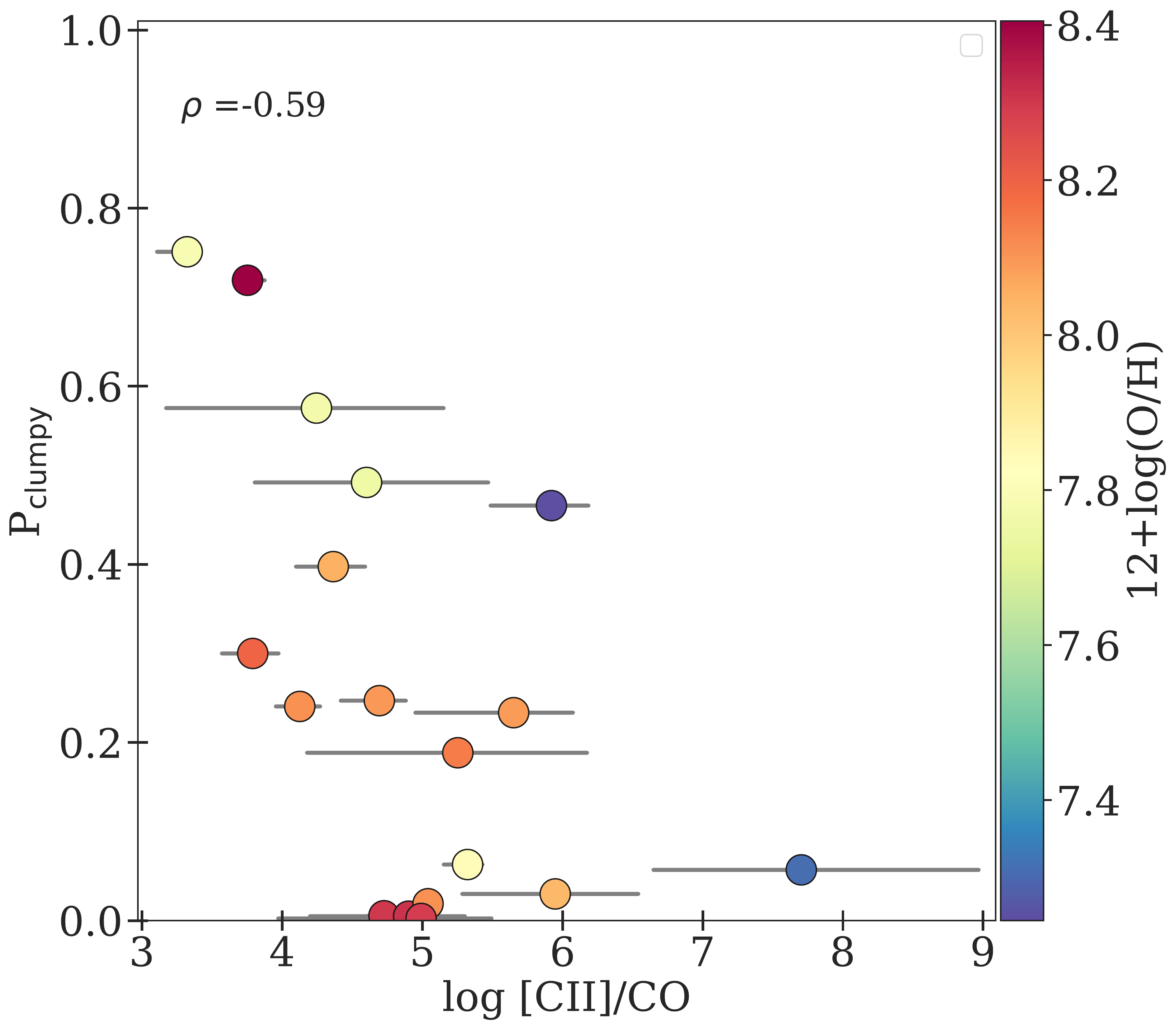}
\caption[Clumpiness parameter vs. \cii/CO]{Clumpiness parameter (defined by Equation \ref{eq_clumpy}) vs. predicted \cii/CO ratio, colorcoded as a function of metallicity. The clumpiness parameter anti-correlates with the \cii/CO ratio, and has a secondary dependence with metallicity.}
\label{clumpy_vs_cii_CO}
\end{figure}

\begin{figure*}[h!]
\centering
\includegraphics[height=0.4\textwidth]{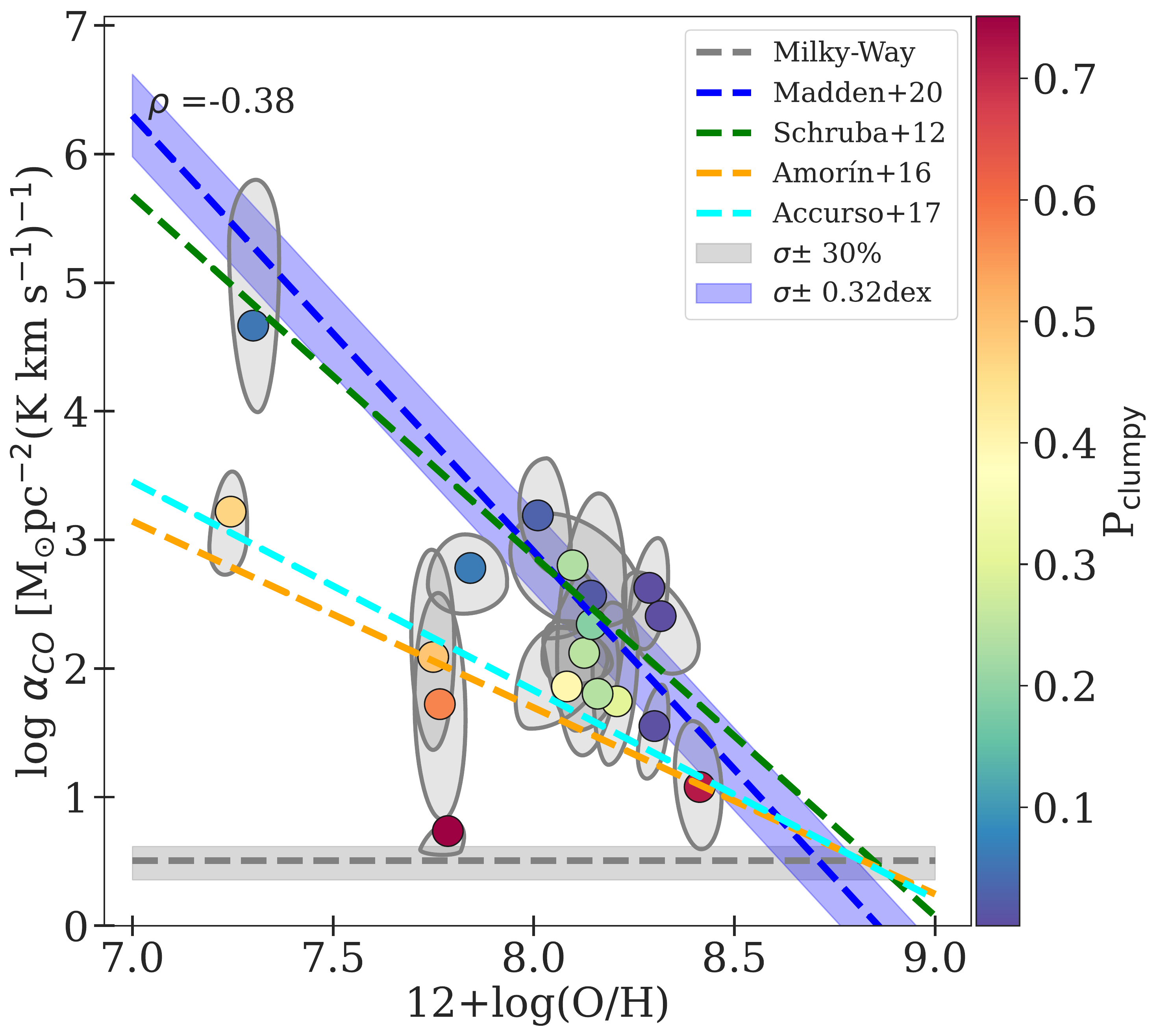}
\includegraphics[height=0.4\textwidth]{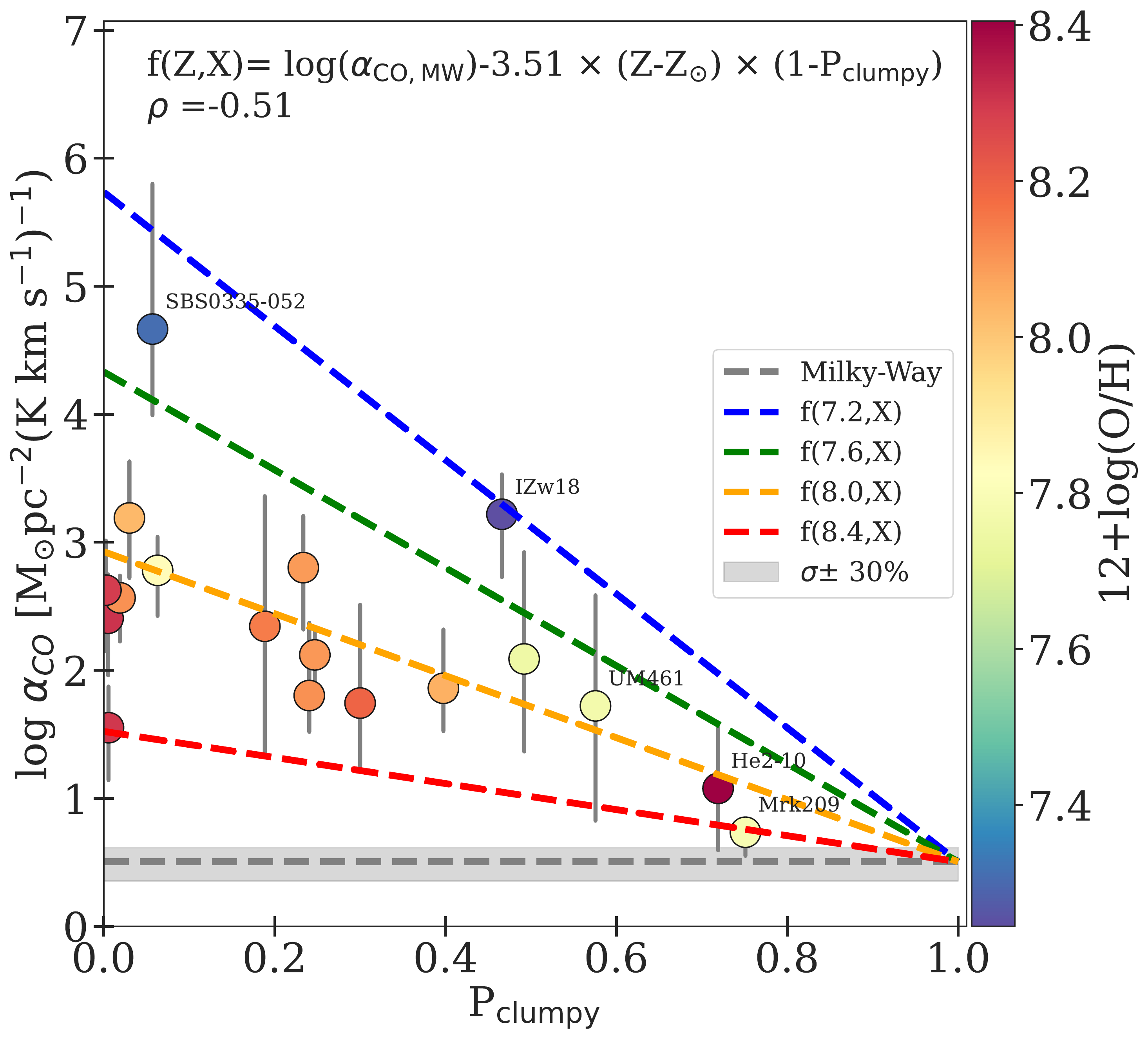}
\caption[$\alpha_{\rm CO}$ vs. Z colorcoded]{Relation between the $\alphaCO$ vs. the gas-phase metallicity (left panel) and the clumpiness parameter (right panel), defined by Equation \ref{eq_clumpy}. The Spearman correlation coefficients are calculated for the means of individual galaxies, shown by colored dots. We show the Galactic $\alphaCO$ value from \citetalias{Bolatto_2013} and the range of values compatible with it within 30\% (gray-shaded area).}
\label{dispersion_CO_Z}
\end{figure*}

To further examine the different morphologies of galaxies inferred by our code, we define a “clumpiness” probability parameter as follows:
\begin{equation}
    \label{eq_clumpy}
    P_{\rm clumpy} = < C >;\\
    {\rm with}\; C = \begin{cases}
        1\; \rm if\; \alpha_{\rm cut,2} < 0\; \&\; \rm cut_{max} > 2,\\
        0\; \rm else,
    \end{cases}
\end{equation}
where < $C$ > is the averaged value of the posterior PDF of the binary variable $C$, $\alpha_{\rm cut,2}$ and cut$_{\rm max}$ are respectively the slope and upper-bound of the distribution followed by the cut parameters (see Section \ref{broken_plaw_models}). 

The introduction of the clumpiness parameter allows us to qualitatively define two morphologies of CO-bright molecular gas:
\begin{itemize}
    \item \textit{Diffuse CO distribution}, corresponding to low P$_{\rm clumpy}$ parameters, in which the CO emission is dominated by numerous low-A$_V$ clouds\footnote{A cloud defines here a component drawn from our grid of models whose depth exceeds the ionization front (cut > 1)}. The inferred distributions favor low upper-bounds (cut$_{\rm max} < 2$) meaning that most clouds do not reach their dissociation front, and positive slopes, corresponding to relatively larger contribution of such clouds, with respect to deeper clouds, to the total gas mass.
    \item \textit{Clumpy CO distribution}, corresponding to large P$_{\rm clumpy}$ parameters, in which the CO emission is dominated by a few high-A$_V$ clouds. In such galaxies, the distribution of parameters inferred by our models favors large upper-bounds (cut > 2, meaning that some regions drawn during the sampling reach large A$_V$, after the dissociation front) and negative slopes ($\alpha_{\rm cut,2} < 0$), meaning that such deep clouds with large A$_V$ contribute relatively less to the total gas mass than the lower-A$_V$ components.
\end{itemize}
As the P$_{\rm clumpy}$ parameter increases, galaxies gradually evolve from a diffuse CO distribution to a clumpy distribution.

In Figure \ref{clumpy_vs_cii_CO}, we show the variation of the clumpiness parameter, P$_{\rm clumpy}$, as a function of the predicted luminosity ratio \cii/CO and find a relatively strong anti-correlation with a Spearman correlation coefficient of -0.59. This relation appears to shift with metallicity, with lower metallicities being associated with larger clumpiness parameter at fixed \cii/CO ratio. In Figure \ref{dispersion_CO_Z}, we then examine how the clumpiness parameter relates to $\alphaCO$ and the metallicity. In the left-hand side panel, we see that the dispersion in the $\alpha_{\rm CO}$ vs. metallicity relation is mainly linked to the clumpiness parameters. Galaxies corresponding to the diffuse CO distribution, with P$_{\rm clumpy} \lesssim 0.4$ are close to the steep relations from \cite{Schruba_2012} and \citetalias{madden_tracing_2020}. On the contrary, the clumpiest galaxies with P$_{\rm clumpy} \gtrsim 0.4$ deviate significantly from the latter relations and are associated with lower $\alpha_{\rm CO}$. The latter galaxies are flagged in bold in Table \ref{table_mean_values}. 

In the right-hand side panel of Figure \ref{dispersion_CO_Z}, we show $\alpha_{\rm CO}$ vs. P$_{\rm clumpy}$ which exhibits a clearer anti-correlation than the $\alpha_{\rm CO}$ vs. metallicity, with a Spearman coefficient of -0.51. In particular, the three clumpiest galaxies (P$_{\rm clumpy} > 0.5$) in our sample are found close to the Galactic $\alpha_{\rm CO}$ (within a factor of 3), despite their subsolar metallicities. While one of them, He\,2-10, has the highest metallicity in our sample (\oh=$8.43\pm0.01$, \citealt{Madden_2013}), the two other clumpiest galaxies, Mrk\,209 and UM\,461, both have a metallicity close to $\sim$1/10\,Z$_{\odot}$ (respectively \oh$=7.74\pm0.01$ and \oh$=7.73\pm0.01$; \citealt{Madden_2013}). 

Nevertheless, we still find that metallicity plays an important role, with the two lowest metallicity galaxies, I\,Zw\,18 and SBS\,0335-052, being clearly offset toward larger $\alpha_{\rm CO}$. The $\alpha_{\rm CO}$ variations can be relatively well described by fitting a simple linear relation, which depends on both the metallicity and clumpiness parameter:

\begin{multline}
         \log \alpha_{\rm CO}(\rm Z, P_{clumpy}) =\\ \rm \log(\alpha_{\rm CO, MW}) - \beta \times(\rm Z-Z_{\odot}) \times (1-\rm P_{clumpy});
         \label{eq_alphaCO_LR}
\end{multline}

where Z is the metallicity given as \oh, P$_{\rm clumpy}$ the clumpiness parameters (see Equation \ref{eq_clumpy}), and with an optimal slope of $\beta=-3.51 \pm 5 \times 10^{-5}$. In the latter equation, the clumpiness parameter modulates the slope of the $\alphaCO$ vs. metallicity anticorrelation which flattens with increasing clumpiness. For P$_{\rm clumpy}$=0, we find a slope of $-3.51$ which is close to that of \citetalias{madden_tracing_2020} ($-3.39$). We argue that the \citetalias{madden_tracing_2020} relation, derived assuming a diffuse CO emission provides an upper limit on \alphaCO\ while the actual \alphaCO\, accounting for the clumpiness of the medium, decreases with increasing P$_{\rm clumpy}$. We further discuss the possible physical mechanisms leading to either a clumpy or diffuse distribution of molecular gas in Section \ref{discussion_clumpiness}.

\section{Discussion}
\label{section_discussion}

\subsection{Clumpy vs. diffuse molecular gas at low-metallicity}
\label{discussion_clumpiness}

\subsubsection{Is the ISM clumpy at low-metallicity?}
In the current study, we aim at recovering information about the mass distribution of CO-emitting clouds in spatially unresolved galaxies. We introduce multicomponent models, described in Section \ref{broken_plaw_models}, that provide luminosity-weighted predictions for emission lines and masses, accounting for the relative contribution of diffuse low-A$_V$ clouds vs. denser larger-A$_V$ clouds (see Section \ref{subsect_alpha_CO}). Our sample of dwarf galaxies (see Section \ref{section_sample}) is best fitted by different luminosity-weighted topologies, going from a diffuse CO distribution (dominated by many low-A$_V$ clouds) to a clumpy CO distribution (dominated by a few high-A$_V$ clouds). Most of the DGS galaxies (12 out of 18 in this study) are associated with a “diffuse” scenario which matches the picture drawn from previous works of substantial CO-dark molecular gas reservoirs and large $\alphaCO$ values in low-metallicity environments \citep[e.g.,][]{lebouteiller_neutral_2017, Lebouteiller_2019, Chevance_30Dor_2020, madden_tracing_2020}. Nevertheless, our results predict that a clumpy molecular gas geometry (P$_{\rm clumpy} \geq 0.4$) is expected for 6 out of 18 galaxies in our sample.

Such results are consistent with recent observations, performed at high spatial resolution with ALMA, that have revealed CO clumps with sizes of $10-100$\,pc in several of the galaxies included in the current study (e.g., NGC\,625 at $\sim$20\,pc resolution; \citealt{Imara_ngc625_2020}, II\,Zw\,40 at $\sim$24\,pc resolution; \citealt{Kepley_2016}, He\,2-10 $\sim$26\,pc resolution; \citealt{Imara_faesi_2019}). 
Among the three clumpiest galaxies identified in our study, He\,2-10 is associated with a high P$_{\rm clumpy}$ of 0.75, which is consistent with the observations from \cite{Imara_faesi_2019} that reported the presence of 119 resolved giant molecular clouds, with average sizes of $\sim$26\,pc in which 45\% to 70\% of the total molecular mass is concentrated. The authors report an CO-to-H$_2$ conversion factor of $0.5$ to $13$ times the Milky-Way value, which is also in good agreement with our predictions. The two other clumpiest galaxies identified by our models (UM\,461 and Mrk\,209) have not been mapped in CO at high resolution. It is also possible that CO resides in even smaller clumps ($\lesssim$ 10\,pc), below the resolution of the existing ALMA observations. Indeed, recent observations performed at very high-angular resolution in other low metallicity dwarf galaxies have detected CO-clumps with sizes as small as a few parsecs \citep[e.g.,][]{Oey_2017, Shi_2020, Schruba_2017, Rubio_2015, Archer_2022}. In the Magellanic Clouds, recent studies have shown that CO emission has a complex, highly filamentary structure both in the LMC \citep{Wong_2019, Wong_2022} and in the SMC \citep{Ohno_2023}, and have identified CO sub-structures with equivalent radius down to $\sim$0.1\,pc.

Nevertheless, the latter observations remain scarce and were performed at different spatial resolutions. Observing a representative sample of dwarf galaxies at high spatial resolution is much needed to perform a meaningful comparison with the clumpiness predictions from the current study, and to better understand the physical and chemical mechanisms driving the clumpiness of molecular gas in the ISM. Meanwhile, magnetohydrodynamical simulations may help to gain insight into the structures that may form in cold neutral gas. In a recent work simulating galaxies with metallicities ranging from 0.2 to 1\,Z$_{\odot}$, \cite{Kobayashi_2023} report the presence of small cold neutral medium structures with sizes $\sim$0.1-1\,pc, which form naturally out of converging gas flows, although over longer formation timescales at lower metallicity. Regardless of their exact sizes, our results suggest that clumps may dominate the integrated CO luminosity, with important implications on the integrated gas masses and on the star formation laws derived at galactic-scales.

\subsubsection{Impact of clumpiness on integrated gas masses and $\alphaCO$ conversion factors}

As discussed in Section \ref{subsect_masses}, changes in terms of gas topologies (i.e., number and distribution of components) have little effect on the predicted values for the total gas masses associated with the ionized, neutral, and molecular gas reservoirs (see Figure \ref{mass_comparison_all}). However, the topology of our models strongly affects how H$_2$ is distributed between the warm CO-dark phase (associated with C$^{+}$ and C$^0$) and the cold CO-bright phase (associated with CO, see Figures \ref{mass_H2_all} and \ref{fdark}). While, the DGS galaxies are found to be largely dominated by CO-dark H$_2$, regardless of the architecture, we show that single-component models and models with only a few components tend to overestimate both the fractions of CO-dark H$_2$ gas (see Figure \ref{fdark}) and the total integrated H$_2$ mass (see Figure \ref{tracers_MH2}).

Using statistical distributions of components to mimic a clumpy medium, we find that all the galaxies in our sample are \hi-dominated, with M(H$_2$)/M(\hi) ranging from $\sim$5\% to 66\% (see Section \ref{overview_masses}). This result differs from those of \cite{Cormier_2014}, which predict several H$_2$-dominated galaxies using the same sample, based on an $\alphaCO$ prescription that scales with metallicity. Our predictions are in better agreement with the global mass distributions observed in larger surveys, such as the xGASS survey. In the latter survey, the atomic gas fractions are found to increase with decreasing stellar masses, leading to an average M(H$_2$)/M(\hi) ratio close to 10\% for stellar masses of $\sim$10$^9$M$_{\odot}$ \citep{Catinella_2018}.

We find a large dispersion around the \alphaCO\ vs. metallicity relation (see Figure \ref{alpha}), which we interpret in terms of clumpiness of the medium (see Figure \ref{dispersion_CO_Z}). The clumpiness parameter, defined by Equation \ref{eq_clumpy}, anti-correlates with the \cii/CO emission line ratio, as shown in Figure \ref{clumpy_vs_cii_CO}. In other words, our results recover the variations of \alphaCO\ as a function of \cii/CO, observed in large samples of galaxies up to z$\sim$2.5 \citep{Accurso_2017}, and provide a new physical interpretation in terms of clumpiness of the ISM. Our predicted \alphaCO\ values anti-correlate with both metallicity and clumpiness, and are well described by a linear equation involving both parameters (see Equation \ref{eq_alphaCO_LR}). As a result, low-metallicity galaxies may be associated with low \alphaCO\ values, close to the Galactic value, provided that they have a clumpy molecular gas distribution. 

This finding is in line with results from \cite{Gratier_2017} in the small spiral galaxy M33 that report an average $\alphaCO$ of twice the Galactic value, despite its half-solar metallicity. Similarly, $\alphaCO$ values of less than a factor ten larger than the Milky-Way value were reported in the Large and Small Magellanic clouds despite their subsolar metallicities \citep{Pineda_2017, Jameson_2018, Saldano_2023}. In particular, \cite{Pineda_2017} find that accounting for the small filling factor of CO emission tends to reduce the derived $\alphaCO$ values.

A complementary picture is provided in \cite{Hu_XCO_2022} that couple hydrodynamical simulations with time-dependent H$_2$ chemistry to study the effect of metallicity variations on the $\alphaCO$ derived from post-processed CO maps. The authors report significant spatial variations of $\alphaCO$ on parsec scales, with diffuse clouds being associated with high $\alphaCO$ values, while denser clouds, from where CO mainly arise, are associated with low $\alphaCO$ values. To account for those spatial variations, \cite{Hu_XCO_2022} introduce a multivariate $\alphaCO$ calibration that is not only a function of metallicity, but also depends on the beam size and line intensity.

\subsubsection{Impact on star-formation laws}

\begin{figure}[h!]
	\centering
	\includegraphics[height=0.4\textwidth]{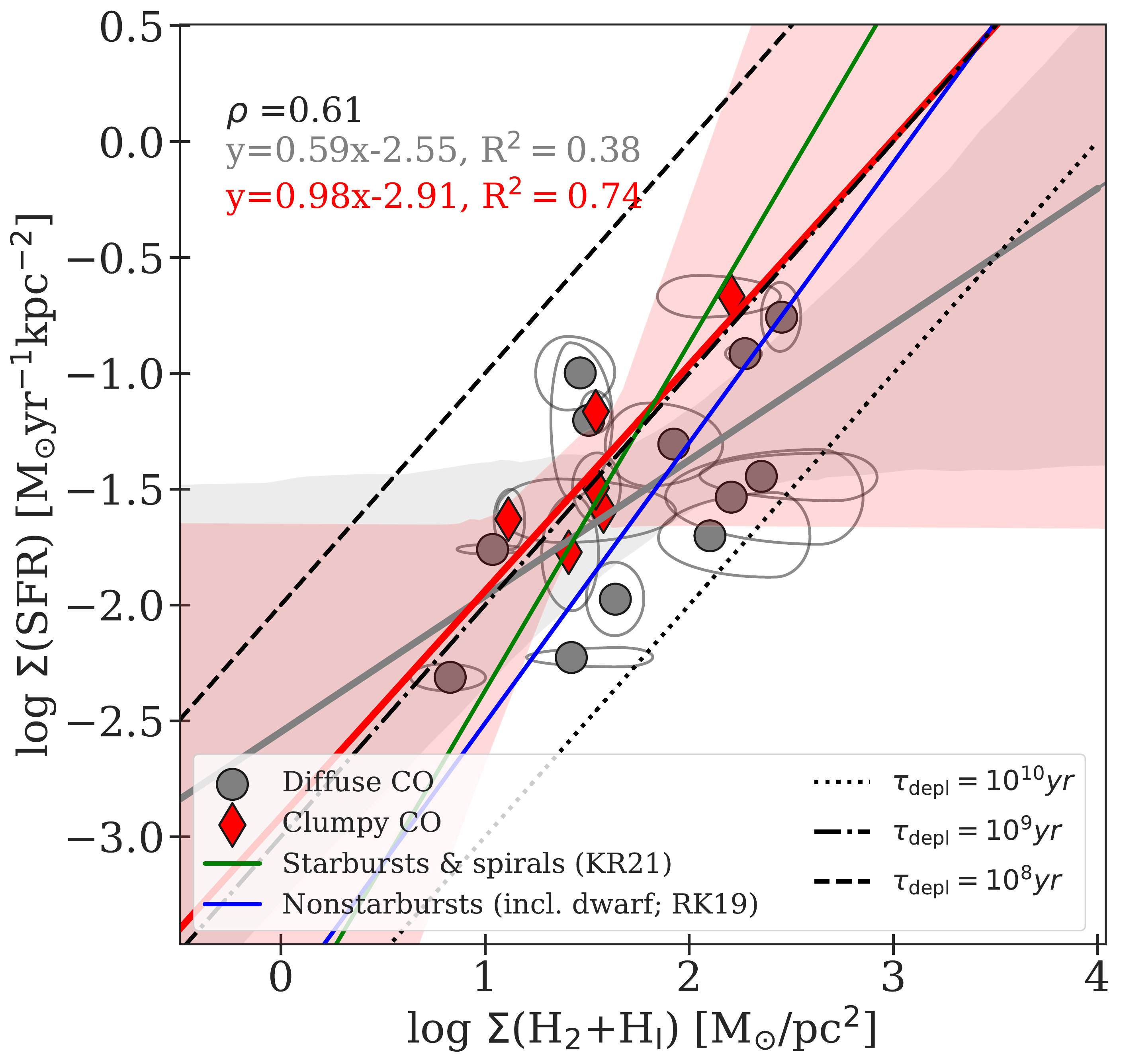}
	\caption[SF relations]{$\Sigma_{\rm SFR}$ vs. $\Sigma$(H$_2$+\hi). The galaxies associated with a “clumpy CO” distribution are reported in bold in Table \ref{table_mean_values} and correspond to P$_{\rm clumpy} \geq 0.4$ (see Eq. \ref{eq_clumpy}), while the others correspond to a “diffuse CO” distribution. Plain lines correspond to the best fits obtained for those two categories, with their 95\% confidence intervals shown as shaded-areas, as well as the relations derived for starbursting galaxies and nonstarbursting galaxies from \cite{Kennicutt_Reyes_2021} and \cite{Reyes_Kennicutt_2019}, respectively.}
    \label{sf_relations}
\end{figure}

While a few studies have examined the atomic gas-to-SFR relation in \hi-dominated dwarf galaxies \citep[see e.g.,][]{Bigiel_2008}, estimates of their total molecular gas surface density, including H$_2$, have remained challenging. An unprecedentedly large sample of dwarf galaxies is included in the recent analysis of \cite{Reyes_Kennicutt_2019} that revisit the integrated KS law in a wide range of environments, but the authors specifically excluded the blue compact dwarf galaxies (BCD), similar to the galaxies in our sample, to focus on nonstarbursting systems. In a subsequent paper, \cite{Kennicutt_Reyes_2021} provide an updated version of the KS law for spiral and starburst galaxies including BCDs, but most of their sample consists of massive galaxies. In addition, the necessity to adopt of a CO-to-H$_2$ conversion factor to estimate the total molecular gas surface density yields large uncertainties, especially for low-metallicity galaxies. 
Hence, it remains unclear whether star-forming dwarf galaxies are expected to follow the classical KS law, and predictions based on simulations \citep[such as e.g.,][]{Whitworth_KS_2022} or models such as e.g.,\citealt{Cormier_2014} or \citetalias{madden_tracing_2020} are much needed. 

We translate the predicted \hi\ and H$_2$ masses inferred by the power-law models into gas surface densities to examine the star-formation laws in the DGS. As previously done in \cite{Cormier_2014} and \citetalias{madden_tracing_2020}, we use the dust photometric apertures reported in \cite{remy-ruyer_linking_2015} to estimate the angular size of the emission. We note that the predicted \hi\ masses are globally in good agreement with the available measurements based on 21\,cm emission (as shown in Figure \ref{masses_relation}). The measured \hi\ masses were corrected so that they correspond to the exact same photometric apertures as the integrated line fluxes used in our modeling. This correction was performed by estimating the total \hi-emitting area and assuming a Gaussian distribution profile of the \hi\ 21\,cm emission. We stress that the uncertainties on the latter corrections remain quite large, and the Gaussian profile adopted to model the \hi\ radial extent may overestimate the amount of \hi\ located within the photometric aperture, meaning that \hi\ masses could be overestimated both in models and in observations. We assume that both H$_2$ masses and the corrected \hi\ masses broadly correspond to the same spatial area, and derive average gas surface densities over the dust apertures. In practice, the latter aperture may lead to an overestimation of $\Sigma$(\hi) (although this is mitigated by the correction of \hi\ masses), and to an underestimation of $\Sigma$(H$_2$), especially since H$_2$ clumps may have sizes much smaller than the photometric apertures ($\sim$1-500\,kpc$^2$).

Despite these uncertainties, we examine the position of our sample of star-forming low-metallicity dwarf galaxies with respect to classical star-formation laws in starbursting and nonstarbursting galaxies.
Figure \ref{sf_relations} shows that the DGS galaxies are globally in good agreement with the star formation law derived in \cite{Kennicutt_Reyes_2021} for an extended sample of starbursting galaxies and spiral. As previously reported in \citetalias{madden_tracing_2020}, accounting for the CO-dark H$_2$ gas solves the apparent offset of dwarf galaxies, which appeared shifted toward lower gas surface densities when accounting only for the CO-bright H$_2$ gas \citep[e.g.,][]{Cormier_2014}. For the “clumpy” subsample of galaxies we derive a nearly linear relation corresponding to a constant depletion time of $\sim 1$\,Gyr. Despite low statistics leading to large uncertainties on the fit, we find that the relation we derived for clumpy dwarf galaxies is compatible with the classical KS law derived for more massive for spiral and starburst galaxies in \cite{Kennicutt_Reyes_2021}.

On the other hand, we find that several galaxies among the ones predicted to have a diffuse CO distribution, are offset toward large gas surface densities and correspond to larger depletion times, between 1 to $10$\,Gyrs.
While similarly long depletion times have been reported for \hi-dominated irregular dwarf galaxies in \cite{Bigiel_2008} and for nonstarbursting dwarf galaxies (see e.g., the low surface density galaxies and faint irregular dwarf galaxies from \cite{Wyder_2009} and \cite{Roychowdhury_2017} shown in \cite{Reyes_Kennicutt_2019}), the latter timescales appear surprisingly large for starbursting galaxies. Galaxies associated with large depletion times (5 out of 18 galaxies with depletion time $> 1$\,Gyr) are responsible for a flattening of the slope that we derive, leading to a sublinear relation (slope of $0.59\pm0.28$) for dwarf galaxies with diffuse CO distribution. The latter relation is incompatible with the classical KS law for more massive starbursts, even within the large fit uncertainties. We speculate that this offset may be linked to the presence of diffuse molecular gas reservoirs that are not associated with any star forming regions, an explanation previously invoked in \cite{SKB_2013} and \cite{Shetty_2014} to account for the flattening of the KS. In particular, \cite{Shetty_2014} claim that the fraction of diffuse, nonstarforming CO-bright component could be at least of 30\% in extragalactic observations. The presence of a diffuse gas reservoir, shining in CO, is qualitatively consistent with our results, and the fact that most galaxies in our sample (12 out of 18) are associated with relatively low clumpiness parameters.

\subsection{Known caveats and potential improvements}
\label{discussion_limits}

\subsubsection{Limitations of the SFGX grid}

While the statistical framework provided by MULTIGRIS allows us to consider complex, more realistic geometries of the gas, its accuracy ultimately depends on the underlying grid of 1D models. In the current study, we used the SFGX grid (see Section \ref{section_models}) composed of spherical Cloudy models, computed at thermal equilibrium. A first caveat, already mentioned in Section \ref{sect_tracer_ism} is that the radial sampling of our grid does not enable to properly capture the sharp increase of H$_2$ cumulative emission (as shown in Figure \ref{H2_problem}). Mid-IR H$_2$ upper limits are overpredicted by our models, especially for the two lowest levels (H$_2$\,S(0) and H$_2$\,S(1)). This problem is somehow mitigated for the detections, which are found in relatively good agreement with predictions, but also tend to be overpredicted for H$_2$\,S(0) and H$_2$\,S(1). While this issue is restricted only to H$_2$ lines, associated with a sharp radial increase of the emission, it prevents us from using H$_2$ lines as constraints. The latter H$_2$ lines would provide additional information on the neutral atomic and molecular gas phase, for which few tracers are available (see Table \ref{tracers}). While resampling the grid using a finer step in cut or interpolating between different cut values may improve the solutions, properly recovering the fluxes of tracers with sharp radial variations remains challenging with our method. Carefully selecting the list of lines used in the analysis and excluding the problematic tracers remains, for now, the best option. 

A second caveat is that Cloudy models are static, while time-dependent effects have been shown to strongly affect both H$_2$ predictions and the associated CO-to-H$_2$ conversion factors \citep[e.g.,][]{Glover_clark_sf_2012}. Specifically, \cite{Hu_metdep_2021} show that steady-state models tend to overestimate H$_2$ and the CO-dark gas fraction with respect to time-dependent models. As a result, \cite{Hu_XCO_2022} find that the CO-to-H$_2$ conversion factors derived using time-dependent models are systematically lower than those derived using steady-state models. Based on a highly-resolved ($\sim$0.2\,pc) ISM simulation of an isolated low-metallicity (1/10\,$Z_{\odot}$) dwarf galaxy post-processed using a time-dependent chemistry network and a dust-evolution model, \cite{Hu_lowZ_2023} recently derive a CO-to-H$_2$ conversion factor close to the Milky-Way value. Accounting for out-of-equilibrium chemistry may result in even lower \alphaCO\ values than those predicted in the current study.

Additionally, several physical quantities are fixed in the SFGX grid, some of which may be important in order to correctly predict the emission in the neutral and molecular gas. First, the adopted cosmic ray ionization rate (CRIR) is an important parameter to accurately predict CO. The effect of an enhanced CRIR on the CO emission is not trivial as it depends on the density: at low densities (n $\sim 10^2$\,cm$^{-3}$), it causes a decrease in CO emissivity, while at higher densities, this effect is compensated by the rise in temperature \citep[e.g.,][]{Bisbas_Papadopoulos_Viti_2015, Vallini_2019}. Based on magnetohydrodynamical simulations, \cite{Gong_2020} found that a decrease of $\alphaCO$ is expected at higher CRIR. In the current study, we follow the prescription of \cite{cormier_herschel_2019} and adopt a CRIR value $\sim$3 times higher than the standard Galactic value from \cite{2007_indriolo_cosmics} in order to account for the recent star-formation history in the DGS. This value remains somewhat arbitrary, and is not varied in the SFGX grid. Nevertheless, the SFGX grid includes a variable contribution from X-ray sources with luminosities ranging from 0 to 10\% of the stellar luminosity, which may match the emission of galaxies with higher CRIR than the one considered in our models. We note that cosmic rays and X-rays have similar effects in terms of ionization and heating on the PDR tracers \citep[e.g.,][]{lebouteiller_neutral_2017}; hence, their effects cannot be disentangled with our set of constraints, which include only one line arising from the molecular gas. Disentangling the complex effects of enhanced CRIR and X-rays would require constraining the entire CO-sled \citep[e.g.,][]{Vallini_2019, Esposito_2022}, which is not possible with only CO(1-0).

Finally, as mentioned in Section \ref{subsection_radial_profile}, the radial density profile in our Cloudy models follows a physically-motivated prescription introduced in \cite{cormier_herschel_2019}, with density increasing linearly in the dense gas, above a given density threshold. This law controls the density profile of the PDR and molecular clouds associated with H$_2$ regions, and directly impacts the luminosity and mass estimates that we derive. The effect of choosing a different density law was examined in \cite{cormier_herschel_2019} which tried to fit constant pressure and constant density models to the DGS galaxies. They found that some variations are expected, in particular for \oi\ lines, for which the emission is boosted in constant pressure models and decreased in constant density models. Nevertheless, the authors report that such changes did not significantly affect the quality of their fit. The effect of changing density laws is not examined in the current study as it would require running numerous additional Cloudy models, and because it is not clear that this could be constrained within our multicomponent framework. Nevertheless, we point out that changing this prescription would likely change the distribution of parameters and the gas geometry inferred by our multicomponent models.

\subsubsection{Improving statistical distributions}
\label{discussion_powerlaw}
In Section \ref{broken_plaw_models}, we introduced combinations of models based on the use of power-laws and broken power-laws to describe three main parameters in the study: the density $n$, the ionization parameter $U$, and the radial depth controlled by the cut. While the use of power-laws to describe the distribution of such parameter is physically motivated, other PDFs could be considered based on the information provided by both theoretical studies and resolved observations.

In particular, the expected shape of the density PDF has been examined in several theoretical works. In a gravoturbulent ISM, the density of the gas is expected to follow a log-normal distribution at low densities (associated with a turbulent-dominated regime), while the high-density region is well-described by a power-law (associated with a self-gravity dominated regime) \citep[e.g.,][]{Offner_2014, Burkhart_2018, Burkhart_2019,Jaupart_Chabrier_2020, Appel_2023}. Hence, the power-law distribution adopted in this study is well-suited to describe the dense star-forming gas but may break down at lower density, where turbulence plays an important role. A possible improvement of the models would be to include a log-normal distribution to account for the diffuse ISM component. Nevertheless, we note that the exact transition between the turbulent-dominated and gravity-dominated regimes is not well-known (e.g., at the location of shock fronts; \citealt{Appel_2023}, at the phase transition between ionized and neutral gas, or between cold and warm neutral medium; \citealt{Kobayashi_2022}) and the transition point should be considered as an additional free parameter. 

The power-law distribution adopted for the ionization parameter is even less easily constrained. In the current study, we use the ionization parameter at the illuminated front as a proxy to control the intensity of the radiation fields. While the ionization parameter provides a convenient proxy to parameterize models, it can be defined differently (either at fixed radius or as a volumic average), and has known dependencies on several other parameters such as metallicity, stellar mass, and the spatial scales \citep[e.g.,][]{Ji_2022}. Deriving meaningful priors for the ionization parameter, based either on observations or simulations, may not be straightforward. 

A similar issue arises from the use of distribution of cuts rather than a parameterization based on the visual extinction A$_V$ or on column densities. While the cut parameter allows us to consider different distributions in the ionized gas and in the neutral gas, it prevents direct comparison of the inferred geometries with observation or simulation. The clumpiness parameter introduced in Section \ref{subsect_alpha_CO} can not be used outside the framework of this study. A more physically-motivated approach would consist in describing models based on either on column density of molecular clouds or on A$_V$-PDFs, as done for example in \cite{Bisbas_2019, Bisbas_2023}. Even with those quantities, the analytical functions remain debated (e.g., log-normal as in \cite{Bisbas_2019, Bisbas_2023} or power-laws as in \cite{Brunt_2015, Lombardi_2015}). Observational biases such as resolution, noise, boundaries, and superposition effects may also distort the true underlying PDFs of physical parameters, as discussed for example in \cite{Lombardi_2015}. 

Finally, while we have focused on refining the combination of gas parameters, the exploration of the parameter space associated with the stellar population and X-ray sources are beyond the scope of the current study. Nevertheless, distributions of parameters accounting for the spread in age and luminosity of the population of stellar clusters, as well as the effect of X-ray sources illuminating parts of the gas reservoirs, should also be explored. At any rate, statistics on the population of resolved \hii\ regions or on the stellar cluster distributions based on high angular resolution observations (e.g., with MUSE or JWST) in combination with predictions from high-resolution simulations of galaxies are much needed to introduce physically-motivated priors on physical parameters.

\section{Conclusion}
\label{section_conclusion}
In the current study, we revisit the results from \citetalias{madden_tracing_2020} motivated by the extreme \cii /CO(1-0) values observed in low-metallicity galaxies. We provide new predictions based on updated models and using a Bayesian statistical framework to account for the potentially complex structure of the neutral gas, through the use of statistical distribution of gas components. While the \cii/CO variations were interpreted as changes in the visual extinction A$_V$ in \citetalias{madden_tracing_2020}, we find instead that the latter variations result from the combined effects of both the metallicity and the clumpiness of the ISM. Our main conclusions are summarized below:
\begin{itemize}
    \item Models combining numerous components based on statistical distributions of parameters are the only models able to match simultaneously the suite of ISM tracers used in our study, in all the galaxies of our sample. Models based on a combination of a few components (1 to 4) significantly underpredict CO(1-0) in at least 3 out of 18 galaxies.
    \item Models combining a few components systematically overestimate the total H$_2$ mass. When more components are considered, we find that all the galaxies in our sample are \hi-dominated, which was not the case in previous studies.
    \item Regardless of the modeling architecture, we find that H$_2$ mass is always dominated by the reservoirs associated with C$^{+}$, and C$^0$ to a lesser extent, while the H$_2$ masses associated with CO are negligible. This leads to large predicted fractions of CO-dark H$_2$ gas in all galaxies, above 80\% for models based on statistical distributions. 
    \item We confirm that \cii 158$\mu$m is a good tracer of the total H$_2$ mass and that $\alpha_{\rm [CII]}$ does not depend on metallicity. We derive a \cii\ luminosity vs. M(H$_2$) relation, with a slope close to that of \citetalias{madden_tracing_2020} but with a systematic offset toward lower H$_2$ masses (a factor $\sim$180 lower). 
    \item Our models predict that \ci 609$\mu$m would also be a good tracer of the total H$_2$ mass, although $\alpha_{\rm [CI]}$ is metallicity-dependent. We find a large scatter around the $\alpha_{\rm [CII]}$, $\alpha_{\rm [CI]}$, and $\alphaCO$ vs. metallicity relations, associated with variations in the geometry of multicomponent models.
    \item We find that CO(1-0) is not a good tracer of the total H$_2$ mass at low-metallicity and that $\alphaCO$ strongly depends on both the metallicity and the clumpiness of the medium.
    \item We find that the $\alphaCO$ vs. metallicity relation derived in \citetalias{madden_tracing_2020}, assuming a simpler geometry of models, only provides an upper limit on the actual $\alphaCO$. While most of the galaxies in our sample (12 out of 18) are associated with a diffuse CO distribution and align on the \citetalias{madden_tracing_2020} relation, we predict a clumpy geometry of CO-emitting clouds associated with reduced $\alphaCO$ values in 6 out of 18 galaxies. Two out of these six clumpy galaxies are predicted with $\alphaCO$ values $\leq$3 times the Galactic value, indicating that $\alphaCO$ may reach values close to the Galactic $\alphaCO$ regardless of the metallicity of the galaxy.
\end{itemize}

\begin{acknowledgements}
The authors thank the anonymous referee for constructive feedback and useful comments. LR and MC gratefully acknowledge funding from the DFG through an Emmy Noether Research Group (grant number CH2137/1-1). VL, CR and LR thank the Transatlantic Partnership Program (grant number TJF21\_053). COOL Research DAO is a Decentralized Autonomous Organization supporting research in astrophysics aimed at uncovering our cosmic origins. CR acknowledges the support of the Elon University Japheth E. Rawls Professorship.
\end{acknowledgements}
\bibliographystyle{aa}
\bibliography{MyLibrary.bib}

\begin{thebibliography}{152}
\expandafter\ifx\csname natexlab\endcsname\relax\def\natexlab#1{#1}\fi

\bibitem[{{Accurso} {et~al.}(2017){Accurso}, {Saintonge}, {Catinella},
  {Cortese}, {Dav{\'e}}, {Dunsheath}, {Genzel}, {Gracia-Carpio}, {Heckman},
  {Jimmy}, {Kramer}, {Li}, {Lutz}, {Schiminovich}, {Schuster}, {Sternberg},
  {Sturm}, {Tacconi}, {Tran}, \& {Wang}}]{Accurso_2017}
{Accurso}, G., {Saintonge}, A., {Catinella}, B., {et~al.} 2017, \mnras, 470,
  4750

\bibitem[{{Ackermann} {et~al.}(2012){Ackermann}, {Ajello}, {Allafort},
  {Baldini}, {Ballet}, {Barbiellini}, {Bastieri}, {Bechtol}, {Bellazzini},
  {Berenji}, {Blandford}, {Bloom}, {Bonamente}, {Borgland}, {Bottacini},
  {Brandt}, {Bregeon}, {Brigida}, {Bruel}, {Buehler}, {Busetto}, {Buson},
  {Caliandro}, {Cameron}, {Caraveo}, {Casandjian}, {Cecchi}, {Charles},
  {Chekhtman}, {Chiang}, {Ciprini}, {Claus}, {Cohen-Tanugi}, {Conrad},
  {D'Ammando}, {de Angelis}, {de Palma}, {Dermer}, {Digel}, {Silva}, {Drell},
  {Drlica-Wagner}, {Falletti}, {Favuzzi}, {Fegan}, {Ferrara}, {Focke},
  {Fukazawa}, {Fukui}, {Funk}, {Fusco}, {Gargano}, {Gasparrini}, {Germani},
  {Giglietto}, {Giordano}, {Giroletti}, {Glanzman}, {Godfrey}, {Grenier},
  {Grondin}, {Grove}, {Guiriec}, {Hadasch}, {Hanabata}, {Harding}, {Hayashi},
  {Horan}, {Hou}, {Hughes}, {Itoh}, {Jackson}, {J{\'o}hannesson}, {Johnson},
  {Kamae}, {Katagiri}, {Kataoka}, {Kn{\"o}dlseder}, {Kuss}, {Lande}, {Larsson},
  {Lee}, {Lemoine-Goumard}, {Longo}, {Loparco}, {Lovellette}, {Lubrano},
  {Martin}, {Mazziotta}, {McEnery}, {Mehault}, {Michelson}, {Mitthumsiri},
  {Mizuno}, {Moiseev}, {Monte}, {Monzani}, {Morselli}, {Moskalenko}, {Murgia},
  {Naumann-Godo}, {Nemmen}, {Nishino}, {Norris}, {Nuss}, {Ohno}, {Ohsugi},
  {Okumura}, {Omodei}, {Orlando}, {Ormes}, {Ozaki}, {Paneque}, {Panetta},
  {Parent}, {Pesce-Rollins}, {Pierbattista}, {Piron}, {Pivato}, {Porter},
  {Rain{\`o}}, {Rando}, {Razzano}, {Reimer}, {Reimer}, {Romoli}, {Roth},
  {Sada}, {Sadrozinski}, {Sanchez}, {Sbarra}, {Sgr{\`o}}, {Siskind}, {Spandre},
  {Spinelli}, {Strong}, {Suson}, {Takahashi}, {Takahashi}, {Tanaka}, {Thayer},
  {Thayer}, {Thompson}, {Tibaldo}, {Tibolla}, {Tinivella}, {Torres}, {Tosti},
  {Tramacere}, {Troja}, {Uchiyama}, {Uehara}, {Usher}, {Vandenbroucke},
  {Vasileiou}, {Vianello}, {Vitale}, {Waite}, {Wang}, {Winer}, {Wood},
  {Yamamoto}, {Yang}, \& {Zimmer}}]{Ackermann_2012}
{Ackermann}, M., {Ajello}, M., {Allafort}, A., {et~al.} 2012, \apj, 755, 22

\bibitem[{{Allen} {et~al.}(2015){Allen}, {Hogg}, \& {Engelke}}]{Allen_2015}
{Allen}, R.~J., {Hogg}, D.~E., \& {Engelke}, P.~D. 2015, \aj, 149, 123

\bibitem[{{Amor{\'\i}n} {et~al.}(2016){Amor{\'\i}n}, {Mu{\~n}oz-Tu{\~n}{\'o}n},
  {Aguerri}, \& {Planesas}}]{Amorin_2016}
{Amor{\'\i}n}, R., {Mu{\~n}oz-Tu{\~n}{\'o}n}, C., {Aguerri}, J.~A.~L., \&
  {Planesas}, P. 2016, \aap, 588, A23

\bibitem[{{Appel} {et~al.}(2023){Appel}, {Burkhart}, {Semenov}, {Federrath},
  {Rosen}, \& {Tan}}]{Appel_2023}
{Appel}, S.~M., {Burkhart}, B., {Semenov}, V.~A., {et~al.} 2023, \apj, 954, 93

\bibitem[{{Archer} {et~al.}(2022){Archer}, {Hunter}, {Elmegreen}, {Cigan},
  {Jansen}, {Windhorst}, {Hunt}, \& {Rubio}}]{Archer_2022}
{Archer}, H.~N., {Hunter}, D.~A., {Elmegreen}, B.~G., {et~al.} 2022, \aj, 163,
  141

\bibitem[{{Asplund} {et~al.}(2009){Asplund}, {Grevesse}, {Sauval}, \&
  {Scott}}]{Asplund_2009}
{Asplund}, M., {Grevesse}, N., {Sauval}, A.~J., \& {Scott}, P. 2009, \araa, 47,
  481

\bibitem[{{Baldwin} {et~al.}(1995){Baldwin}, {Ferland}, {Korista}, \&
  {Verner}}]{Baldwin_1995}
{Baldwin}, J., {Ferland}, G., {Korista}, K., \& {Verner}, D. 1995, \apjl, 455,
  L119

\bibitem[{{B{\'e}thermin} {et~al.}(2020){B{\'e}thermin}, {Dessauges-Zavadsky},
  {Faisst}, {Ginolfi}, {Gruppioni}, {Jones}, {Khusanova}, {Lemaux}, {Capak},
  {Cassata}, {Le F{\`e}vre}, {Schaerer}, {Silverman}, {Yan}, \& {Alpine
  Collaboration}}]{Bethermin_2020}
{B{\'e}thermin}, M., {Dessauges-Zavadsky}, M., {Faisst}, A.~L., {et~al.} 2020,
  The Messenger, 180, 31

\bibitem[{{Bigiel} {et~al.}(2008){Bigiel}, {Leroy}, {Walter}, {Brinks}, {de
  Blok}, {Madore}, \& {Thornley}}]{Bigiel_2008}
{Bigiel}, F., {Leroy}, A., {Walter}, F., {et~al.} 2008, \aj, 136, 2846

\bibitem[{{Bisbas} {et~al.}(2015){Bisbas}, {Papadopoulos}, \&
  {Viti}}]{Bisbas_Papadopoulos_Viti_2015}
{Bisbas}, T.~G., {Papadopoulos}, P.~P., \& {Viti}, S. 2015, \apj, 803, 37

\bibitem[{{Bisbas} {et~al.}(2019){Bisbas}, {Schruba}, \& {van
  Dishoeck}}]{Bisbas_2019}
{Bisbas}, T.~G., {Schruba}, A., \& {van Dishoeck}, E.~F. 2019, \mnras, 485,
  3097

\bibitem[{{Bisbas} {et~al.}(2021){Bisbas}, {Tan}, \& {Tanaka}}]{Bisbas_2021}
{Bisbas}, T.~G., {Tan}, J.~C., \& {Tanaka}, K. E.~I. 2021, \mnras, 502, 2701

\bibitem[{{Bisbas} {et~al.}(2023){Bisbas}, {van Dishoeck}, {Hu}, \&
  {Schruba}}]{Bisbas_2023}
{Bisbas}, T.~G., {van Dishoeck}, E.~F., {Hu}, C.-Y., \& {Schruba}, A. 2023,
  \mnras, 519, 729

\bibitem[{{Bolatto} {et~al.}(2013){Bolatto}, {Wolfire}, \&
  {Leroy}}]{Bolatto_2013}
{Bolatto}, A.~D., {Wolfire}, M., \& {Leroy}, A.~K. 2013, \araa, 51, 207

\bibitem[{{Boogaard} {et~al.}(2023){Boogaard}, {Decarli}, {Walter}, {Wei{\ss}},
  {Popping}, {Neri}, {Aravena}, {Riechers}, {Ellis}, {Carilli}, {Cox}, \&
  {Pety}}]{Boogaard_2023}
{Boogaard}, L.~A., {Decarli}, R., {Walter}, F., {et~al.} 2023, \apj, 945, 111

\bibitem[{{Braine} {et~al.}(2010){Braine}, {Gratier}, {Kramer}, {Xilouris},
  {Rosolowsky}, {Buchbender}, {Boquien}, {Calzetti}, {Quintana-Lacaci},
  {Tabatabaei}, {Verley}, {Israel}, {van der Tak}, {Aalto}, {Combes},
  {Garcia-Burillo}, {Gonzalez}, {Henkel}, {Koribalski}, {Mookerjea}, {Roellig},
  {Schuster}, {Rela{\~n}o}, {Bertoldi}, {van der Werf}, \&
  {Wiedner}}]{Braine_2010}
{Braine}, J., {Gratier}, P., {Kramer}, C., {et~al.} 2010, \aap, 518, L69

\bibitem[{{Brunt}(2015)}]{Brunt_2015}
{Brunt}, C.~M. 2015, \mnras, 449, 4465

\bibitem[{{Burkhart}(2018)}]{Burkhart_2018}
{Burkhart}, B. 2018, \apj, 863, 118

\bibitem[{{Burkhart} \& {Mocz}(2019)}]{Burkhart_2019}
{Burkhart}, B. \& {Mocz}, P. 2019, \apj, 879, 129

\bibitem[{{Catinella} {et~al.}(2018){Catinella}, {Saintonge}, {Janowiecki},
  {Cortese}, {Dav{\'e}}, {Lemonias}, {Cooper}, {Schiminovich}, {Hummels},
  {Fabello}, {Ger{\'e}b}, {Kilborn}, \& {Wang}}]{Catinella_2018}
{Catinella}, B., {Saintonge}, A., {Janowiecki}, S., {et~al.} 2018, \mnras, 476,
  875

\bibitem[{{Chevance} {et~al.}(2023){Chevance}, {Krumholz}, {McLeod},
  {Ostriker}, {Rosolowsky}, \& {Sternberg}}]{Chevance_life_MC_2022}
{Chevance}, M., {Krumholz}, M.~R., {McLeod}, A.~F., {et~al.} 2023, 534, 1

\bibitem[{{Chevance} {et~al.}(2020){Chevance}, {Madden}, {Fischer}, {Vacca},
  {Lebouteiller}, {Fadda}, {Galliano}, {Indebetouw}, {Kruijssen}, {Lee},
  {Poglitsch}, {Polles}, {Cormier}, {Hony}, {Iserlohe}, {Krabbe}, {Meixner},
  {Sabbi}, \& {Zinnecker}}]{Chevance_30Dor_2020}
{Chevance}, M., {Madden}, S.~C., {Fischer}, C., {et~al.} 2020, \mnras, 494,
  5279

\bibitem[{Ching \& Chen(2007)}]{Ching_2007}
Ching, J. \& Chen, Y.-C. 2007, Journal of Engineering Mechanics, 133, 816

\bibitem[{Cormier {et~al.}(2019)Cormier, Abel, Hony, Lebouteiller, Madden,
  Polles, Galliano, De~Looze, Galametz, \&
  Lambert-Huyghe}]{cormier_herschel_2019}
Cormier, D., Abel, N.~P., Hony, S., {et~al.} 2019, A\&A, 626, A23

\bibitem[{{Cormier} {et~al.}(2017){Cormier}, {Bendo}, {Hony}, {Lebouteiller},
  {Madden}, {Galliano}, {Glover}, {Klessen}, {Abel}, {Bigiel}, \&
  {Clark}}]{Cormier_2017}
{Cormier}, D., {Bendo}, G.~J., {Hony}, S., {et~al.} 2017, \mnras, 468, L87

\bibitem[{{Cormier} {et~al.}(2012){Cormier}, {Lebouteiller}, {Madden}, {Abel},
  {Hony}, {Galliano}, {Baes}, {Barlow}, {Cooray}, {De Looze}, {Galametz},
  {Karczewski}, {Parkin}, {R{\'e}my}, {Sauvage}, {Spinoglio}, {Wilson}, \&
  {Wu}}]{2012_Cormier}
{Cormier}, D., {Lebouteiller}, V., {Madden}, S.~C., {et~al.} 2012, \aap, 548,
  A20

\bibitem[{{Cormier} {et~al.}(2015){Cormier}, {Madden}, {Lebouteiller}, {Abel},
  {Hony}, {Galliano}, {R{\'e}my-Ruyer}, {Bigiel}, {Baes}, {Boselli},
  {Chevance}, {Cooray}, {De Looze}, {Doublier}, {Galametz}, {Hughes},
  {Karczewski}, {Lee}, {Lu}, \& {Spinoglio}}]{Cormier_2015}
{Cormier}, D., {Madden}, S.~C., {Lebouteiller}, V., {et~al.} 2015, \aap, 578,
  A53

\bibitem[{{Cormier} {et~al.}(2014){Cormier}, {Madden}, {Lebouteiller}, {Hony},
  {Aalto}, {Costagliola}, {Hughes}, {R{\'e}my-Ruyer}, {Abel}, {Bayet},
  {Bigiel}, {Cannon}, {Cumming}, {Galametz}, {Galliano}, {Viti}, \&
  {Wu}}]{Cormier_2014}
{Cormier}, D., {Madden}, S.~C., {Lebouteiller}, V., {et~al.} 2014, \aap, 564,
  A121

\bibitem[{{Crocker} {et~al.}(2019){Crocker}, {Pellegrini}, {Smith}, {Draine},
  {Wilson}, {Wolfire}, {Armus}, {Brinks}, {Dale}, {Groves}, {Herrera-Camus},
  {Hunt}, {Kennicutt}, {Murphy}, {Sandstrom}, {Schinnerer}, {Rigopoulou},
  {Rosolowsky}, \& {van der Werf}}]{Crocker_2019}
{Crocker}, A.~F., {Pellegrini}, E., {Smith}, J. D.~T., {et~al.} 2019, \apj,
  887, 105

\bibitem[{{Davies} {et~al.}(2017){Davies}, {Baes}, {Bianchi}, {Jones},
  {Madden}, {Xilouris}, {Bocchio}, {Casasola}, {Cassara}, {Clark}, {De Looze},
  {Evans}, {Fritz}, {Galametz}, {Galliano}, {Lianou}, {Mosenkov}, {Smith},
  {Verstocken}, {Viaene}, {Vika}, {Wagle}, \& {Ysard}}]{Davies_dustpedia_2017}
{Davies}, J.~I., {Baes}, M., {Bianchi}, S., {et~al.} 2017, \pasp, 129, 044102

\bibitem[{{de los Reyes} \& {Kennicutt}(2019)}]{Reyes_Kennicutt_2019}
{de los Reyes}, M. A.~C. \& {Kennicutt}, Robert~C., J. 2019, \apj, 872, 16

\bibitem[{{den Brok} {et~al.}(2023){den Brok}, {Bigiel}, {Chastenet},
  {Sandstrom}, {Leroy}, {Usero}, {Schinnerer}, {Rosolowsky}, {Koch}, {Chiang},
  {Barnes}, {Puschnig}, {Saito}, {Be{\v{s}}li{\'c}}, {Chevance}, {Dale},
  {Eibensteiner}, {Glover}, {Jim{\'e}nez-Donaire}, {Teng}, \&
  {Williams}}]{den_Brok_2023}
{den Brok}, J.~S., {Bigiel}, F., {Chastenet}, J., {et~al.} 2023, \aap, 676, A93

\bibitem[{{Dessauges-Zavadsky} {et~al.}(2020){Dessauges-Zavadsky}, {Ginolfi},
  {Pozzi}, {B{\'e}thermin}, {Le F{\`e}vre}, {Fujimoto}, {Silverman}, {Jones},
  {Vallini}, {Schaerer}, {Faisst}, {Khusanova}, {Fudamoto}, {Cassata},
  {Loiacono}, {Capak}, {Yan}, {Amorin}, {Bardelli}, {Boquien}, {Cimatti},
  {Gruppioni}, {Hathi}, {Ibar}, {Koekemoer}, {Lemaux}, {Narayanan}, {Oesch},
  {Rodighiero}, {Romano}, {Talia}, {Toft}, {Vergani}, {Zamorani}, \&
  {Zucca}}]{Dessauges-Zavadsky_2020}
{Dessauges-Zavadsky}, M., {Ginolfi}, M., {Pozzi}, F., {et~al.} 2020, \aap, 643,
  A5

\bibitem[{{Dickman} {et~al.}(1986){Dickman}, {Snell}, \&
  {Schloerb}}]{Dickman_1986}
{Dickman}, R.~L., {Snell}, R.~L., \& {Schloerb}, F.~P. 1986, \apj, 309, 326

\bibitem[{{Dunne} {et~al.}(2022){Dunne}, {Maddox}, {Papadopoulos}, {Ivison}, \&
  {Gomez}}]{Dunne_2022}
{Dunne}, L., {Maddox}, S.~J., {Papadopoulos}, P.~P., {Ivison}, R.~J., \&
  {Gomez}, H.~L. 2022, \mnras, 517, 962

\bibitem[{{Dunne} {et~al.}(2021){Dunne}, {Maddox}, {Vlahakis}, \&
  {Gomez}}]{Dunne_2021}
{Dunne}, L., {Maddox}, S.~J., {Vlahakis}, C., \& {Gomez}, H.~L. 2021, \mnras,
  501, 2573

\bibitem[{{Eldridge} {et~al.}(2017){Eldridge}, {Stanway}, {Xiao}, {McClelland},
  {Taylor}, {Ng}, {Greis}, \& {Bray}}]{eldridge_binary_2017}
{Eldridge}, J.~J., {Stanway}, E.~R., {Xiao}, L., {et~al.} 2017, \pasa, 34, e058

\bibitem[{{Esposito} {et~al.}(2022){Esposito}, {Vallini}, {Pozzi}, {Casasola},
  {Mingozzi}, {Vignali}, {Gruppioni}, \& {Salvestrini}}]{Esposito_2022}
{Esposito}, F., {Vallini}, L., {Pozzi}, F., {et~al.} 2022, \mnras, 512, 686

\bibitem[{{Ferguson} {et~al.}(1997){Ferguson}, {Korista}, {Baldwin}, \&
  {Ferland}}]{Ferguson_1997}
{Ferguson}, J.~W., {Korista}, K.~T., {Baldwin}, J.~A., \& {Ferland}, G.~J.
  1997, \apj, 487, 122

\bibitem[{{Ferland} {et~al.}(2017){Ferland}, {Chatzikos}, {Guzm{\'a}n},
  {Lykins}, {van Hoof}, {Williams}, {Abel}, {Badnell}, {Keenan}, {Porter}, \&
  {Stancil}}]{2017_Cloudy_v17}
{Ferland}, G.~J., {Chatzikos}, M., {Guzm{\'a}n}, F., {et~al.} 2017, \rmxaa, 53,
  385

\bibitem[{{Freundlich} {et~al.}(2019){Freundlich}, {Combes}, {Tacconi},
  {Genzel}, {Garcia-Burillo}, {Neri}, {Contini}, {Bolatto}, {Lilly},
  {Salom{\'e}}, {Bicalho}, {Boissier}, {Boone}, {Bouch{\'e}}, {Bournaud},
  {Burkert}, {Carollo}, {Cooper}, {Cox}, {Feruglio}, {F{\"o}rster Schreiber},
  {Juneau}, {Lippa}, {Lutz}, {Naab}, {Renzini}, {Saintonge}, {Sternberg},
  {Walter}, {Weiner}, {Wei{\ss}}, \& {Wuyts}}]{Freundlich_2019}
{Freundlich}, J., {Combes}, F., {Tacconi}, L.~J., {et~al.} 2019, \aap, 622,
  A105

\bibitem[{{Galliano} {et~al.}(2021){Galliano}, {Nersesian}, {Bianchi}, {De
  Looze}, {Roychowdhury}, {Baes}, {Casasola}, {Cassar{\'a}}, {Dobbels},
  {Fritz}, {Galametz}, {Jones}, {Madden}, {Mosenkov}, {Xilouris}, \&
  {Ysard}}]{galliano_nearby_2021}
{Galliano}, F., {Nersesian}, A., {Bianchi}, S., {et~al.} 2021, \aap, 649, A18

\bibitem[{{Gao} {et~al.}(2022){Gao}, {Gu}, {Shi}, {Zhou}, {Bao}, {Yu}, {Zhang},
  {Wang}, {Madden}, {Hayes}, {Lu}, \& {Xu}}]{Gao_Haro11_2022}
{Gao}, Y., {Gu}, Q., {Shi}, Y., {et~al.} 2022, \aap, 661, A136

\bibitem[{{Genzel} {et~al.}(2015){Genzel}, {Tacconi}, {Lutz}, {Saintonge},
  {Berta}, {Magnelli}, {Combes}, {Garc{\'\i}a-Burillo}, {Neri}, {Bolatto},
  {Contini}, {Lilly}, {Boissier}, {Boone}, {Bouch{\'e}}, {Bournaud}, {Burkert},
  {Carollo}, {Colina}, {Cooper}, {Cox}, {Feruglio}, {F{\"o}rster Schreiber},
  {Freundlich}, {Gracia-Carpio}, {Juneau}, {Kovac}, {Lippa}, {Naab}, {Salome},
  {Renzini}, {Sternberg}, {Walter}, {Weiner}, {Weiss}, \&
  {Wuyts}}]{Genzel_2015}
{Genzel}, R., {Tacconi}, L.~J., {Lutz}, D., {et~al.} 2015, \apj, 800, 20

\bibitem[{{Ginolfi} {et~al.}(2017){Ginolfi}, {Maiolino}, {Nagao}, {Carniani},
  {Belfiore}, {Cresci}, {Hatsukade}, {Mannucci}, {Marconi}, {Pallottini},
  {Schneider}, \& {Santini}}]{Ginolfi_2017}
{Ginolfi}, M., {Maiolino}, R., {Nagao}, T., {et~al.} 2017, \mnras, 468, 3468

\bibitem[{{Glover} \& {Clark}(2012{\natexlab{a}})}]{Glover_Clark_necessay_2012}
{Glover}, S. C.~O. \& {Clark}, P.~C. 2012{\natexlab{a}}, \mnras, 421, 9

\bibitem[{{Glover} \& {Clark}(2012{\natexlab{b}})}]{Glover_clark_sf_2012}
{Glover}, S. C.~O. \& {Clark}, P.~C. 2012{\natexlab{b}}, \mnras, 426, 377

\bibitem[{{Glover} \& {Clark}(2016)}]{Glover_ci_2016}
{Glover}, S. C.~O. \& {Clark}, P.~C. 2016, \mnras, 456, 3596

\bibitem[{{Glover} \& {Mac Low}(2011)}]{Glover_MacLow_2011}
{Glover}, S.~C.~O. \& {Mac Low}, M.~M. 2011, \mnras, 412, 337

\bibitem[{{Gong} {et~al.}(2020){Gong}, {Ostriker}, {Kim}, \& {Kim}}]{Gong_2020}
{Gong}, M., {Ostriker}, E.~C., {Kim}, C.-G., \& {Kim}, J.-G. 2020, \apj, 903,
  142

\bibitem[{{Gratier} {et~al.}(2017){Gratier}, {Braine}, {Schuster},
  {Rosolowsky}, {Boquien}, {Calzetti}, {Combes}, {Kramer}, {Henkel}, {Herpin},
  {Israel}, {Koribalski}, {Mookerjea}, {Tabatabaei}, {R{\"o}llig}, {van der
  Tak}, {van der Werf}, \& {Wiedner}}]{Gratier_2017}
{Gratier}, P., {Braine}, J., {Schuster}, K., {et~al.} 2017, \aap, 600, A27

\bibitem[{{Grenier} {et~al.}(2005){Grenier}, {Casandjian}, \&
  {Terrier}}]{Grenier_2005}
{Grenier}, I.~A., {Casandjian}, J.-M., \& {Terrier}, R. 2005, Science, 307,
  1292

\bibitem[{{Greve} {et~al.}(1996){Greve}, {Becker}, {Johansson}, \&
  {McKeith}}]{Greve_1996}
{Greve}, A., {Becker}, R., {Johansson}, L.~E.~B., \& {McKeith}, C.~D. 1996,
  \aap, 312, 391

\bibitem[{{Hayashi} {et~al.}(2019){Hayashi}, {Okamoto}, {Yamamoto}, {Hayakawa},
  {Tachihara}, \& {Fukui}}]{Hayashi_2019}
{Hayashi}, K., {Okamoto}, R., {Yamamoto}, H., {et~al.} 2019, \apj, 878, 131

\bibitem[{{Heintz} \& {Watson}(2020)}]{Heintz_watson_2020}
{Heintz}, K.~E. \& {Watson}, D. 2020, \apjl, 889, L7

\bibitem[{{Hernandez} {et~al.}(2023){Hernandez}, {Jones}, {Smith}, {Togi},
  {Aloisi}, {Blair}, {Hirschauer}, {Hunt}, {James}, {Kumari}, {Lebouteiller},
  {Mingozzi}, \& {Ramambason}}]{Hernandez_2023}
{Hernandez}, S., {Jones}, L., {Smith}, L.~J., {et~al.} 2023, \apj, 948, 124

\bibitem[{{Hirashita}(2023)}]{Hirashita_2023}
{Hirashita}, H. 2023, \mnras, 522, 4612

\bibitem[{{Hosokawa} \& {Inutsuka}(2005)}]{Hosokawa_Inutsuka_2005}
{Hosokawa}, T. \& {Inutsuka}, S.-i. 2005, \apj, 623, 917

\bibitem[{{Hosokawa} \& {Inutsuka}(2006)}]{Hosokawa_Inutsuka_2006}
{Hosokawa}, T. \& {Inutsuka}, S.-i. 2006, \apj, 646, 240

\bibitem[{{Hu} {et~al.}(2022){Hu}, {Schruba}, {Sternberg}, \& {van
  Dishoeck}}]{Hu_XCO_2022}
{Hu}, C.-Y., {Schruba}, A., {Sternberg}, A., \& {van Dishoeck}, E.~F. 2022,
  \apj, 931, 28

\bibitem[{{Hu} {et~al.}(2021){Hu}, {Sternberg}, \& {van
  Dishoeck}}]{Hu_metdep_2021}
{Hu}, C.-Y., {Sternberg}, A., \& {van Dishoeck}, E.~F. 2021, \apj, 920, 44

\bibitem[{{Hu} {et~al.}(2023){Hu}, {Sternberg}, \& {van
  Dishoeck}}]{Hu_lowZ_2023}
{Hu}, C.-Y., {Sternberg}, A., \& {van Dishoeck}, E.~F. 2023, \apj, 952, 140

\bibitem[{{Hunt} {et~al.}(2023){Hunt}, {Belfiore}, {Lelli}, {Draine},
  {Marasco}, {Garc{\'\i}a-Burillo}, {Venturi}, {Combes}, {Wei{\ss}}, {Henkel},
  {Menten}, {Annibali}, {Casasola}, {Cignoni}, {McLeod}, {Tosi}, {Beltr{\'a}n},
  {Concas}, {Cresci}, {Ginolfi}, {Kumari}, \& {Mannucci}}]{Hunt_2023}
{Hunt}, L.~K., {Belfiore}, F., {Lelli}, F., {et~al.} 2023, \aap, 675, A64

\bibitem[{{Hunt} {et~al.}(2015){Hunt}, {Garc{\'\i}a-Burillo}, {Casasola},
  {Caselli}, {Combes}, {Henkel}, {Lundgren}, {Maiolino}, {Menten}, {Testi}, \&
  {Weiss}}]{Hunt_2015}
{Hunt}, L.~K., {Garc{\'\i}a-Burillo}, S., {Casasola}, V., {et~al.} 2015, \aap,
  583, A114

\bibitem[{{Hunt} {et~al.}(2014){Hunt}, {Testi}, {Casasola},
  {Garc{\'\i}a-Burillo}, {Combes}, {Nikutta}, {Caselli}, {Henkel}, {Maiolino},
  {Menten}, {Sauvage}, \& {Weiss}}]{Hunt_2014}
{Hunt}, L.~K., {Testi}, L., {Casasola}, V., {et~al.} 2014, \aap, 561, A49

\bibitem[{{Imara} {et~al.}(2020){Imara}, {De Looze}, {Faesi}, \&
  {Cormier}}]{Imara_ngc625_2020}
{Imara}, N., {De Looze}, I., {Faesi}, C.~M., \& {Cormier}, D. 2020, \apj, 895,
  21

\bibitem[{{Imara} \& {Faesi}(2019)}]{Imara_faesi_2019}
{Imara}, N. \& {Faesi}, C.~M. 2019, \apj, 876, 141

\bibitem[{{Indriolo} {et~al.}(2007){Indriolo}, {Geballe}, {Oka}, \&
  {McCall}}]{2007_indriolo_cosmics}
{Indriolo}, N., {Geballe}, T.~R., {Oka}, T., \& {McCall}, B.~J. 2007, \apj,
  671, 1736

\bibitem[{{James} {et~al.}(2013){James}, {Tsamis}, {Barlow}, {Walsh}, \&
  {Westmoquette}}]{James_2013}
{James}, B.~L., {Tsamis}, Y.~G., {Barlow}, M.~J., {Walsh}, J.~R., \&
  {Westmoquette}, M.~S. 2013, \mnras, 428, 86

\bibitem[{{Jameson} {et~al.}(2018){Jameson}, {Bolatto}, {Wolfire}, {Warren},
  {Herrera-Camus}, {Croxall}, {Pellegrini}, {Smith}, {Rubio}, {Indebetouw},
  {Israel}, {Meixner}, {Roman-Duval}, {van Loon}, {Muller}, {Verdugo},
  {Zinnecker}, \& {Okada}}]{Jameson_2018}
{Jameson}, K.~E., {Bolatto}, A.~D., {Wolfire}, M., {et~al.} 2018, \apj, 853,
  111

\bibitem[{{Jaupart} \& {Chabrier}(2020)}]{Jaupart_Chabrier_2020}
{Jaupart}, E. \& {Chabrier}, G. 2020, \apjl, 903, L2

\bibitem[{{Ji} \& {Yan}(2022)}]{Ji_2022}
{Ji}, X. \& {Yan}, R. 2022, \aap, 659, A112

\bibitem[{{Jiao} {et~al.}(2019){Jiao}, {Zhao}, {Lu}, {Gao}, {Salak}, {Zhu},
  {Zhang}, {Jiang}, \& {Tan}}]{Jiao_2019}
{Jiao}, Q., {Zhao}, Y., {Lu}, N., {et~al.} 2019, \apj, 880, 133

\bibitem[{{Kennicutt} \& {De Los Reyes}(2021)}]{Kennicutt_Reyes_2021}
{Kennicutt}, Robert~C., J. \& {De Los Reyes}, M. A.~C. 2021, \apj, 908, 61

\bibitem[{{Kepley} {et~al.}(2016){Kepley}, {Leroy}, {Johnson}, {Sandstrom}, \&
  {Chen}}]{Kepley_2016}
{Kepley}, A.~A., {Leroy}, A.~K., {Johnson}, K.~E., {Sandstrom}, K., \& {Chen},
  C. H.~R. 2016, \apj, 828, 50

\bibitem[{{Kobayashi} {et~al.}(2022){Kobayashi}, {Inoue}, {Tomida}, {Iwasaki},
  \& {Nakatsugawa}}]{Kobayashi_2022}
{Kobayashi}, M. I.~N., {Inoue}, T., {Tomida}, K., {Iwasaki}, K., \&
  {Nakatsugawa}, H. 2022, \apj, 930, 76

\bibitem[{{Kobayashi} {et~al.}(2023){Kobayashi}, {Iwasaki}, {Tomida}, {Inoue},
  {Omukai}, \& {Tokuda}}]{Kobayashi_2023}
{Kobayashi}, M. I.~N., {Iwasaki}, K., {Tomida}, K., {et~al.} 2023, \apj, 954,
  38

\bibitem[{{Kobulnicky} {et~al.}(1995){Kobulnicky}, {Dickey}, {Sargent}, {Hogg},
  \& {Conti}}]{Kobulnicky_1995}
{Kobulnicky}, H.~A., {Dickey}, J.~M., {Sargent}, A.~I., {Hogg}, D.~E., \&
  {Conti}, P.~S. 1995, \aj, 110, 116

\bibitem[{{Lagos} \& {Papaderos}(2013)}]{Lagos_Papaderos_2013}
{Lagos}, P. \& {Papaderos}, P. 2013, Advances in Astronomy, 2013, 631943

\bibitem[{{Lebouteiller} {et~al.}(2019){Lebouteiller}, {Cormier}, {Madden},
  {Galametz}, {Hony}, {Galliano}, {Chevance}, {Lee}, {Braine}, {Polles},
  {Reque{\~n}a-Torres}, {Indebetouw}, {Hughes}, \& {Abel}}]{Lebouteiller_2019}
{Lebouteiller}, V., {Cormier}, D., {Madden}, S.~C., {et~al.} 2019, \aap, 632,
  A106

\bibitem[{Lebouteiller {et~al.}(2017)Lebouteiller, Péquignot, Cormier, Madden,
  Pakull, Kunth, Galliano, Chevance, Heap, Lee, \&
  Polles}]{lebouteiller_neutral_2017}
Lebouteiller, V., Péquignot, D., Cormier, D., {et~al.} 2017, A\&A, 602, A45

\bibitem[{{Lebouteiller} \& {Ramambason}(2022{\natexlab{a}})}]{MULTIGRIS2022}
{Lebouteiller}, V. \& {Ramambason}, L. 2022{\natexlab{a}}, {MULTIGRIS:
  Multicomponent probabilistic grid search}, Astrophysics Source Code Library,
  record ascl:2207.001

\bibitem[{{Lebouteiller} \&
  {Ramambason}(2022{\natexlab{b}})}]{LebouteillerRamambason2022}
{Lebouteiller}, V. \& {Ramambason}, L. 2022{\natexlab{b}}, \aap, 667, A34

\bibitem[{{Leroy} {et~al.}(2005){Leroy}, {Bolatto}, {Simon}, \&
  {Blitz}}]{Leroy_2005}
{Leroy}, A., {Bolatto}, A.~D., {Simon}, J.~D., \& {Blitz}, L. 2005, \apj, 625,
  763

\bibitem[{{Leroy} {et~al.}(2007){Leroy}, {Cannon}, {Walter}, {Bolatto}, \&
  {Weiss}}]{Leroy_2007}
{Leroy}, A., {Cannon}, J., {Walter}, F., {Bolatto}, A., \& {Weiss}, A. 2007,
  \apj, 663, 990

\bibitem[{{Leroy} {et~al.}(2021){Leroy}, {Schinnerer}, {Hughes}, {Rosolowsky},
  {Pety}, {Schruba}, {Usero}, {Blanc}, {Chevance}, {Emsellem}, {Faesi},
  {Herrera}, {Liu}, {Meidt}, {Querejeta}, {Saito}, {Sandstrom}, {Sun},
  {Williams}, {Anand}, {Barnes}, {Behrens}, {Belfiore}, {Benincasa},
  {Be{\v{s}}li{\'c}}, {Bigiel}, {Bolatto}, {den Brok}, {Cao}, {Chandar},
  {Chastenet}, {Chiang}, {Congiu}, {Dale}, {Deger}, {Eibensteiner}, {Egorov},
  {Garc{\'\i}a-Rodr{\'\i}guez}, {Glover}, {Grasha}, {Henshaw}, {Ho}, {Kepley},
  {Kim}, {Klessen}, {Kreckel}, {Koch}, {Kruijssen}, {Larson}, {Lee}, {Lopez},
  {Machado}, {Mayker}, {McElroy}, {Murphy}, {Ostriker}, {Pan}, {Pessa},
  {Puschnig}, {Razza}, {S{\'a}nchez-Bl{\'a}zquez}, {Santoro}, {Sardone},
  {Scheuermann}, {Sliwa}, {Sormani}, {Stuber}, {Thilker}, {Turner}, {Utomo},
  {Watkins}, \& {Whitmore}}]{Leroy_CO_phangs_2021}
{Leroy}, A.~K., {Schinnerer}, E., {Hughes}, A., {et~al.} 2021, \apjs, 257, 43

\bibitem[{{Leroy} {et~al.}(2009){Leroy}, {Walter}, {Bigiel}, {Usero}, {Weiss},
  {Brinks}, {de Blok}, {Kennicutt}, {Schuster}, {Kramer}, {Wiesemeyer}, \&
  {Roussel}}]{Leroy_heracles_2009}
{Leroy}, A.~K., {Walter}, F., {Bigiel}, F., {et~al.} 2009, \aj, 137, 4670

\bibitem[{{Liszt} \& {Lucas}(1996)}]{Liszt_Lucas_oh_1996}
{Liszt}, H. \& {Lucas}, R. 1996, \aap, 314, 917

\bibitem[{{Lombardi} {et~al.}(2015){Lombardi}, {Alves}, \&
  {Lada}}]{Lombardi_2015}
{Lombardi}, M., {Alves}, J., \& {Lada}, C.~J. 2015, \aap, 576, L1

\bibitem[{{Lucas} \& {Liszt}(1996)}]{Lucas_Liszt_hco_1996}
{Lucas}, R. \& {Liszt}, H. 1996, \aap, 307, 237

\bibitem[{{Madden} {et~al.}(2020){Madden}, {Cormier}, {Hony}, {Lebouteiller},
  {Abel}, {Galametz}, {De Looze}, {Chevance}, {Polles}, {Lee}, {Galliano},
  {Lambert-Huyghe}, {Hu}, \& {Ramambason}}]{madden_tracing_2020}
{Madden}, S.~C., {Cormier}, D., {Hony}, S., {et~al.} 2020, \aap, 643, A141

\bibitem[{{Madden} {et~al.}(1997){Madden}, {Poglitsch}, {Geis}, {Stacey}, \&
  {Townes}}]{Madden_1997}
{Madden}, S.~C., {Poglitsch}, A., {Geis}, N., {Stacey}, G.~J., \& {Townes},
  C.~H. 1997, \apj, 483, 200

\bibitem[{{Madden} {et~al.}(2013){Madden}, {R{\'e}my-Ruyer}, {Galametz},
  {Cormier}, {Lebouteiller}, {Galliano}, {Hony}, {Bendo}, {Smith}, {Pohlen},
  {Roussel}, {Sauvage}, {Wu}, {Sturm}, {Poglitsch}, {Contursi}, {Doublier},
  {Baes}, {Barlow}, {Boselli}, {Boquien}, {Carlson}, {Ciesla}, {Cooray},
  {Cortese}, {de Looze}, {Irwin}, {Isaak}, {Kamenetzky}, {Karczewski}, {Lu},
  {MacHattie}, {O'Halloran}, {Parkin}, {Rangwala}, {Schirm}, {Schulz},
  {Spinoglio}, {Vaccari}, {Wilson}, \& {Wozniak}}]{Madden_2013}
{Madden}, S.~C., {R{\'e}my-Ruyer}, A., {Galametz}, M., {et~al.} 2013, \pasp,
  125, 600

\bibitem[{{Magnelli} {et~al.}(2012){Magnelli}, {Saintonge}, {Lutz}, {Tacconi},
  {Berta}, {Bournaud}, {Charmandaris}, {Dannerbauer}, {Elbaz},
  {F{\"o}rster-Schreiber}, {Graci{\'a}-Carpio}, {Ivison}, {Maiolino}, {Nordon},
  {Popesso}, {Rodighiero}, {Santini}, \& {Wuyts}}]{Magnelli_2012}
{Magnelli}, B., {Saintonge}, A., {Lutz}, D., {et~al.} 2012, \aap, 548, A22

\bibitem[{Meng(1994)}]{Meng_PPP_1994}
Meng, X.-L. 1994, The Annals of Statistics, 22, 1142

\bibitem[{Minson {et~al.}(2013)Minson, Simons, \& Beck}]{Minson_2013}
Minson, S.~E., Simons, M., \& Beck, J.~L. 2013, Geophysical Journal
  International, 194, 1701

\bibitem[{{Montoya Arroyave} {et~al.}(2023){Montoya Arroyave}, {Cicone},
  {Makroleivaditi}, {Weiss}, {Lundgren}, {Severgnini}, {De Breuck},
  {Baumschlager}, {Schimek}, {Shen}, \& {Aravena}}]{Montoya_Arroyave_2023}
{Montoya Arroyave}, I., {Cicone}, C., {Makroleivaditi}, E., {et~al.} 2023,
  \aap, 673, A13

\bibitem[{{Nguyen} {et~al.}(2018){Nguyen}, {Dawson}, {Miville-Desch{\^e}nes},
  {Tang}, {Li}, {Heiles}, {Murray}, {Stanimirovi{\'c}}, {Gibson},
  {McClure-Griffiths}, {Troland}, {Bronfman}, \& {Finger}}]{Nguyen_2018}
{Nguyen}, H., {Dawson}, J.~R., {Miville-Desch{\^e}nes}, M.~A., {et~al.} 2018,
  \apj, 862, 49

\bibitem[{{Oey} {et~al.}(2017){Oey}, {Herrera}, {Silich}, {Reiter}, {James},
  {Jaskot}, \& {Micheva}}]{Oey_2017}
{Oey}, M.~S., {Herrera}, C.~N., {Silich}, S., {et~al.} 2017, \apjl, 849, L1

\bibitem[{{Offner} {et~al.}(2014){Offner}, {Clark}, {Hennebelle}, {Bastian},
  {Bate}, {Hopkins}, {Moraux}, \& {Whitworth}}]{Offner_2014}
{Offner}, S.~S.~R., {Clark}, P.~C., {Hennebelle}, P., {et~al.} 2014, in
  Protostars and Planets VI, ed. H.~{Beuther}, R.~S. {Klessen}, C.~P.
  {Dullemond}, \& T.~{Henning}, 53

\bibitem[{{Ohno} {et~al.}(2023){Ohno}, {Tokuda}, {Konishi}, {Matsumoto},
  {Sewi{\l}o}, {Kondo}, {Sano}, {Tsuge}, {Zahorecz}, {Goto}, {Neelamkodan},
  {Wong}, {Fukushima}, {Takekoshi}, {Muraoka}, {Kawamura}, {Tachihara},
  {Fukui}, \& {Onishi}}]{Ohno_2023}
{Ohno}, T., {Tokuda}, K., {Konishi}, A., {et~al.} 2023, \apj, 949, 63

\bibitem[{{P{\'e}quignot}(2008)}]{pequignot_heating_nodate}
{P{\'e}quignot}, D. 2008, \aap, 478, 371

\bibitem[{{Pineda} {et~al.}(2017){Pineda}, {Langer}, {Goldsmith}, {Horiuchi},
  {Kuiper}, {Muller}, {Hughes}, {Ott}, {Requena-Torres}, {Velusamy}, \&
  {Wong}}]{Pineda_2017}
{Pineda}, J.~L., {Langer}, W.~D., {Goldsmith}, P.~F., {et~al.} 2017, \apj, 839,
  107

\bibitem[{{Poglitsch} {et~al.}(1995){Poglitsch}, {Krabbe}, {Madden}, {Nikola},
  {Geis}, {Johansson}, {Stacey}, \& {Sternberg}}]{Poglitsch_1995}
{Poglitsch}, A., {Krabbe}, A., {Madden}, S.~C., {et~al.} 1995, \apj, 454, 293

\bibitem[{Polles {et~al.}(2019)Polles, Madden, Lebouteiller, Cormier, Abel,
  Galliano, Hony, Karczewski, Lee, Chevance, Galametz, \&
  Lianou}]{polles_modeling_2019}
Polles, F.~L., Madden, S.~C., Lebouteiller, V., {et~al.} 2019, A\&A, 622, A119

\bibitem[{{Ramambason} {et~al.}(2022){Ramambason}, {Lebouteiller}, {Bik},
  {Richardson}, {Galliano}, {Schaerer}, {Morisset}, {Polles}, {Madden},
  {Chevance}, \& {De Looze}}]{Ramambason2022}
{Ramambason}, L., {Lebouteiller}, V., {Bik}, A., {et~al.} 2022, \aap, 667, A35

\bibitem[{{Remy} {et~al.}(2018){Remy}, {Grenier}, {Marshall}, \&
  {Casandjian}}]{Remy_2018}
{Remy}, Q., {Grenier}, I.~A., {Marshall}, D.~J., \& {Casandjian}, J.~M. 2018,
  \aap, 616, A71

\bibitem[{{R{\'e}my-Ruyer} {et~al.}(2013){R{\'e}my-Ruyer}, {Madden},
  {Galliano}, {Hony}, {Sauvage}, {Bendo}, {Roussel}, {Pohlen}, {Smith},
  {Galametz}, {Cormier}, {Lebouteiller}, {Wu}, {Baes}, {Barlow}, {Boquien},
  {Boselli}, {Ciesla}, {De Looze}, {Karczewski}, {Panuzzo}, {Spinoglio},
  {Vaccari}, \& {Wilson}}]{Remy-Ruyer_2013}
{R{\'e}my-Ruyer}, A., {Madden}, S.~C., {Galliano}, F., {et~al.} 2013, \aap,
  557, A95

\bibitem[{Richardson {et~al.}(2014)Richardson, Allen, Baldwin, Hewett, \&
  Ferland}]{richardson_interpreting_2014}
Richardson, C.~T., Allen, J.~T., Baldwin, J.~A., Hewett, P.~C., \& Ferland,
  G.~J. 2014, Monthly Notices of the Royal Astronomical Society, 437, 2376

\bibitem[{Richardson {et~al.}(2016)Richardson, Allen, Baldwin, Hewett, Ferland,
  Crider, \& Meskhidze}]{richardson_interpreting_2016}
Richardson, C.~T., Allen, J.~T., Baldwin, J.~A., {et~al.} 2016, Mon. Not. R.
  Astron. Soc., 458, 988

\bibitem[{{Richardson} {et~al.}(2019){Richardson}, {Polimera}, {Kannappan},
  {Moffett}, \& {Bittner}}]{richardson_addressing_2019}
{Richardson}, C.~T., {Polimera}, M.~S., {Kannappan}, S.~J., {Moffett}, A.~J.,
  \& {Bittner}, A.~S. 2019, \mnras, 486, 3541

\bibitem[{{Roueff} {et~al.}(2019){Roueff}, {Abgrall}, {Czachorowski},
  {Pachucki}, {Puchalski}, \& {Komasa}}]{Roueff_2019}
{Roueff}, E., {Abgrall}, H., {Czachorowski}, P., {et~al.} 2019, \aap, 630, A58

\bibitem[{{Roychowdhury} {et~al.}(2017){Roychowdhury}, {Chengalur}, \&
  {Shi}}]{Roychowdhury_2017}
{Roychowdhury}, S., {Chengalur}, J.~N., \& {Shi}, Y. 2017, \aap, 608, A24

\bibitem[{{Rubio} {et~al.}(2015){Rubio}, {Elmegreen}, {Hunter}, {Brinks},
  {Cort{\'e}s}, \& {Cigan}}]{Rubio_2015}
{Rubio}, M., {Elmegreen}, B.~G., {Hunter}, D.~A., {et~al.} 2015, \nat, 525, 218

\bibitem[{Rémy-Ruyer {et~al.}(2014)Rémy-Ruyer, Madden, Galliano, Galametz,
  Takeuchi, Asano, Zhukovska, Lebouteiller, Cormier, Jones, Bocchio, Baes,
  Bendo, Boquien, Boselli, DeLooze, Doublier-Pritchard, Hughes, Karczewski, \&
  Spinoglio}]{remy-ruyer_gas--dust_2014}
Rémy-Ruyer, A., Madden, S.~C., Galliano, F., {et~al.} 2014, A\&A, 563, A31

\bibitem[{Rémy-Ruyer {et~al.}(2015)Rémy-Ruyer, Madden, Galliano,
  Lebouteiller, Baes, Bendo, Boselli, Ciesla, Cormier, Cooray, Cortese,
  De~Looze, Doublier-Pritchard, Galametz, Jones, Karczewski, Lu, \&
  Spinoglio}]{remy-ruyer_linking_2015}
Rémy-Ruyer, A., Madden, S.~C., Galliano, F., {et~al.} 2015, A\&A, 582, A121

\bibitem[{{Sage} {et~al.}(1992){Sage}, {Salzer}, {Loose}, \&
  {Henkel}}]{Sage_1992}
{Sage}, L.~J., {Salzer}, J.~J., {Loose}, H.~H., \& {Henkel}, C. 1992, \aap,
  265, 19

\bibitem[{{Saintonge} {et~al.}(2017){Saintonge}, {Catinella}, {Tacconi},
  {Kauffmann}, {Genzel}, {Cortese}, {Dav{\'e}}, {Fletcher},
  {Graci{\'a}-Carpio}, {Kramer}, {Heckman}, {Janowiecki}, {Lutz}, {Rosario},
  {Schiminovich}, {Schuster}, {Wang}, {Wuyts}, {Borthakur}, {Lamperti}, \&
  {Roberts-Borsani}}]{Saintonge_xcoldgass_2017}
{Saintonge}, A., {Catinella}, B., {Tacconi}, L.~J., {et~al.} 2017, \apjs, 233,
  22

\bibitem[{{Salda{\~n}o} {et~al.}(2023){Salda{\~n}o}, {Rubio}, {Bolatto},
  {Verdugo}, {Jameson}, \& {Leroy}}]{Saldano_2023}
{Salda{\~n}o}, H.~P., {Rubio}, M., {Bolatto}, A.~D., {et~al.} 2023, \aap, 672,
  A153

\bibitem[{{Salvatier} {et~al.}(2016{\natexlab{a}}){Salvatier}, {Wiecki}, \&
  {Fonnesbeck}}]{pymc3_paper}
{Salvatier}, J., {Wiecki}, T.~V., \& {Fonnesbeck}, C. 2016{\natexlab{a}}, PeerJ
  Computer Science, 2

\bibitem[{{Salvatier} {et~al.}(2016{\natexlab{b}}){Salvatier}, {Wiecki}, \&
  {Fonnesbeck}}]{PyMC3}
{Salvatier}, J., {Wiecki}, T.~V., \& {Fonnesbeck}, C. 2016{\natexlab{b}},
  {PyMC3: Python probabilistic programming framework}, Astrophysics Source Code
  Library, record ascl:1610.016

\bibitem[{{Sandstrom} {et~al.}(2013){Sandstrom}, {Leroy}, {Walter}, {Bolatto},
  {Croxall}, {Draine}, {Wilson}, {Wolfire}, {Calzetti}, {Kennicutt}, {Aniano},
  {Donovan Meyer}, {Usero}, {Bigiel}, {Brinks}, {de Blok}, {Crocker}, {Dale},
  {Engelbracht}, {Galametz}, {Groves}, {Hunt}, {Koda}, {Kreckel}, {Linz},
  {Meidt}, {Pellegrini}, {Rix}, {Roussel}, {Schinnerer}, {Schruba}, {Schuster},
  {Skibba}, {van der Laan}, {Appleton}, {Armus}, {Brandl}, {Gordon}, {Hinz},
  {Krause}, {Montiel}, {Sauvage}, {Schmiedeke}, {Smith}, \&
  {Vigroux}}]{Sandstrom_2013}
{Sandstrom}, K.~M., {Leroy}, A.~K., {Walter}, F., {et~al.} 2013, \apj, 777, 5

\bibitem[{{Schruba} {et~al.}(2017){Schruba}, {Leroy}, {Kruijssen}, {Bigiel},
  {Bolatto}, {de Blok}, {Tacconi}, {van Dishoeck}, \& {Walter}}]{Schruba_2017}
{Schruba}, A., {Leroy}, A.~K., {Kruijssen}, J.~M.~D., {et~al.} 2017, \apj, 835,
  278

\bibitem[{{Schruba} {et~al.}(2012){Schruba}, {Leroy}, {Walter}, {Bigiel},
  {Brinks}, {de Blok}, {Kramer}, {Rosolowsky}, {Sandstrom}, {Schuster},
  {Usero}, {Weiss}, \& {Wiesemeyer}}]{Schruba_2012}
{Schruba}, A., {Leroy}, A.~K., {Walter}, F., {et~al.} 2012, \aj, 143, 138

\bibitem[{{Shetty} {et~al.}(2014){Shetty}, {Clark}, \& {Klessen}}]{Shetty_2014}
{Shetty}, R., {Clark}, P.~C., \& {Klessen}, R.~S. 2014, \mnras, 442, 2208

\bibitem[{{Shetty} {et~al.}(2013){Shetty}, {Kelly}, \& {Bigiel}}]{SKB_2013}
{Shetty}, R., {Kelly}, B.~C., \& {Bigiel}, F. 2013, \mnras, 430, 288

\bibitem[{{Shi} {et~al.}(2020){Shi}, {Wang}, {Zhang}, {Zhang}, {Gao}, {Zhou},
  {Gu}, {Qiu}, {Xia}, {Hao}, \& {Chen}}]{Shi_2020}
{Shi}, Y., {Wang}, J., {Zhang}, Z.-Y., {et~al.} 2020, \apj, 892, 147

\bibitem[{{Solomon} {et~al.}(1997){Solomon}, {Downes}, {Radford}, \&
  {Barrett}}]{Solomon_1997}
{Solomon}, P.~M., {Downes}, D., {Radford}, S.~J.~E., \& {Barrett}, J.~W. 1997,
  \apj, 478, 144

\bibitem[{{Sun} {et~al.}(2022){Sun}, {Leroy}, {Rosolowsky}, {Hughes},
  {Schinnerer}, {Schruba}, {Koch}, {Blanc}, {Chiang}, {Groves}, {Liu}, {Meidt},
  {Pan}, {Pety}, {Querejeta}, {Saito}, {Sandstrom}, {Sardone}, {Usero},
  {Utomo}, {Williams}, {Barnes}, {Benincasa}, {Bigiel}, {Bolatto}, {Boquien},
  {Chevance}, {Dale}, {Deger}, {Emsellem}, {Glover}, {Grasha}, {Henshaw},
  {Klessen}, {Kreckel}, {Kruijssen}, {Ostriker}, \&
  {Thilker}}]{Sun_prop_clouds_2022}
{Sun}, J., {Leroy}, A.~K., {Rosolowsky}, E., {et~al.} 2022, \aj, 164, 43

\bibitem[{{Sun} {et~al.}(2020){Sun}, {Leroy}, {Schinnerer}, {Hughes},
  {Rosolowsky}, {Querejeta}, {Schruba}, {Liu}, {Saito}, {Herrera}, {Faesi},
  {Usero}, {Pety}, {Kruijssen}, {Ostriker}, {Bigiel}, {Blanc}, {Bolatto},
  {Boquien}, {Chevance}, {Dale}, {Deger}, {Emsellem}, {Glover}, {Grasha},
  {Groves}, {Henshaw}, {Jimenez-Donaire}, {Kim}, {Klessen}, {Kreckel}, {Lee},
  {Meidt}, {Sandstrom}, {Sardone}, {Utomo}, \& {Williams}}]{Sun_2020}
{Sun}, J., {Leroy}, A.~K., {Schinnerer}, E., {et~al.} 2020, \apjl, 901, L8

\bibitem[{{Tacconi} {et~al.}(2018){Tacconi}, {Genzel}, {Saintonge}, {Combes},
  {Garc{\'\i}a-Burillo}, {Neri}, {Bolatto}, {Contini}, {F{\"o}rster Schreiber},
  {Lilly}, {Lutz}, {Wuyts}, {Accurso}, {Boissier}, {Boone}, {Bouch{\'e}},
  {Bournaud}, {Burkert}, {Carollo}, {Cooper}, {Cox}, {Feruglio}, {Freundlich},
  {Herrera-Camus}, {Juneau}, {Lippa}, {Naab}, {Renzini}, {Salome}, {Sternberg},
  {Tadaki}, {{\"U}bler}, {Walter}, {Weiner}, \& {Weiss}}]{Tacconi_2018}
{Tacconi}, L.~J., {Genzel}, R., {Saintonge}, A., {et~al.} 2018, \apj, 853, 179

\bibitem[{{Tacconi} \& {Young}(1987)}]{Tacconi_1987}
{Tacconi}, L.~J. \& {Young}, J.~S. 1987, \apj, 322, 681

\bibitem[{{Taylor} {et~al.}(1998){Taylor}, {Kobulnicky}, \&
  {Skillman}}]{Taylor_1998}
{Taylor}, C.~L., {Kobulnicky}, H.~A., \& {Skillman}, E.~D. 1998, \aj, 116, 2746

\bibitem[{{Teng} {et~al.}(2022){Teng}, {Sandstrom}, {Sun}, {Leroy}, {Johnson},
  {Bolatto}, {Kruijssen}, {Schruba}, {Usero}, {Barnes}, {Bigiel}, {Blanc},
  {Groves}, {Israel}, {Liu}, {Rosolowsky}, {Schinnerer}, {Smith}, \&
  {Walter}}]{Teng_2022}
{Teng}, Y.-H., {Sandstrom}, K.~M., {Sun}, J., {et~al.} 2022, \apj, 925, 72

\bibitem[{{Thronson} \& {Bally}(1987)}]{Thronson_Bally_1987}
{Thronson}, Harley~A., J. \& {Bally}, J. 1987, in NASA Conference Publication,
  Vol. 2466, NASA Conference Publication, ed. C.~J. {Lonsdale Persson},
  267--270

\bibitem[{{Togi} \& {Smith}(2016)}]{Togi_2016}
{Togi}, A. \& {Smith}, J.~D.~T. 2016, \apj, 830, 18

\bibitem[{{Tokuda} {et~al.}(2021){Tokuda}, {Kondo}, {Ohno}, {Konishi}, {Sano},
  {Tsuge}, {Zahorecz}, {Goto}, {Neelamkodan}, {Wong}, {Sewi{\l}o}, {Fukushima},
  {Takekoshi}, {Muraoka}, {Kawamura}, {Tachihara}, {Fukui}, \&
  {Onishi}}]{Tokuda_2021}
{Tokuda}, K., {Kondo}, H., {Ohno}, T., {et~al.} 2021, \apj, 922, 171

\bibitem[{{Turner} {et~al.}(2015){Turner}, {Beck}, {Benford}, {Consiglio},
  {Ho}, {Kov{\'a}cs}, {Meier}, \& {Zhao}}]{Turner_2015}
{Turner}, J.~L., {Beck}, S.~C., {Benford}, D.~J., {et~al.} 2015, \nat, 519, 331

\bibitem[{{Vallini} {et~al.}(2019){Vallini}, {Tielens}, {Pallottini},
  {Gallerani}, {Gruppioni}, {Carniani}, {Pozzi}, \& {Talia}}]{Vallini_2019}
{Vallini}, L., {Tielens}, A.~G.~G.~M., {Pallottini}, A., {et~al.} 2019, \mnras,
  490, 4502

\bibitem[{{Vizgan} {et~al.}(2022){Vizgan}, {Greve}, {Olsen}, {Zanella},
  {Narayanan}, {Dav{\`e}}, {Magdis}, {Popping}, {Valentino}, \&
  {Heintz}}]{Vizgan_2022}
{Vizgan}, D., {Greve}, T.~R., {Olsen}, K.~P., {et~al.} 2022, \apj, 929, 92

\bibitem[{{Westmoquette} {et~al.}(2010){Westmoquette}, {Gallagher}, \& {de
  Poitiers}}]{Westmoquette_ngc1140_2010}
{Westmoquette}, M.~S., {Gallagher}, J.~S., \& {de Poitiers}, L. 2010, \mnras,
  403, 1719

\bibitem[{{Whitworth} {et~al.}(2022){Whitworth}, {Smith}, {Tress}, {Kay},
  {Glover}, {Sormani}, \& {Klessen}}]{Whitworth_KS_2022}
{Whitworth}, D.~J., {Smith}, R.~J., {Tress}, R., {et~al.} 2022, \mnras, 510,
  4146

\bibitem[{{Wolfire} {et~al.}(2010){Wolfire}, {Hollenbach}, \&
  {McKee}}]{Wolfire_2010}
{Wolfire}, M.~G., {Hollenbach}, D., \& {McKee}, C.~F. 2010, \apj, 716, 1191

\bibitem[{{Wong} {et~al.}(2019){Wong}, {Hughes}, {Tokuda}, {Indebetouw},
  {Onishi}, {Bandurski}, {Chen}, {Fukui}, {Glover}, {Klessen}, {Pineda},
  {Roman-Duval}, {Sewi{\l}o}, {Wojciechowski}, \& {Zahorecz}}]{Wong_2019}
{Wong}, T., {Hughes}, A., {Tokuda}, K., {et~al.} 2019, \apj, 885, 50

\bibitem[{{Wong} {et~al.}(2022){Wong}, {Oudshoorn}, {Sofovich}, {Green},
  {Shah}, {Indebetouw}, {Meixner}, {Hacar}, {Nayak}, {Tokuda}, {Bolatto},
  {Chevance}, {De Marchi}, {Fukui}, {Hirschauer}, {Jameson}, {Kalari},
  {Lebouteiller}, {Looney}, {Madden}, {Onishi}, {Roman-Duval}, {Rubio}, \&
  {Tielens}}]{Wong_2022}
{Wong}, T., {Oudshoorn}, L., {Sofovich}, E., {et~al.} 2022, \apj, 932, 47

\bibitem[{{Wyder} {et~al.}(2009){Wyder}, {Martin}, {Barlow}, {Foster},
  {Friedman}, {Morrissey}, {Neff}, {Neill}, {Schiminovich}, {Seibert},
  {Bianchi}, {Donas}, {Heckman}, {Lee}, {Madore}, {Milliard}, {Rich}, {Szalay},
  \& {Yi}}]{Wyder_2009}
{Wyder}, T.~K., {Martin}, D.~C., {Barlow}, T.~A., {et~al.} 2009, \apj, 696,
  1834

\bibitem[{{Young} {et~al.}(1995){Young}, {Xie}, {Tacconi}, {Knezek}, {Viscuso},
  {Tacconi-Garman}, {Scoville}, {Schneider}, {Schloerb}, {Lord}, {Lesser},
  {Kenney}, {Huang}, {Devereux}, {Claussen}, {Case}, {Carpenter}, {Berry}, \&
  {Allen}}]{Young_1995}
{Young}, J.~S., {Xie}, S., {Tacconi}, L., {et~al.} 1995, \apjs, 98, 219

\bibitem[{{Zanella} {et~al.}(2018){Zanella}, {Daddi}, {Magdis}, {Diaz Santos},
  {Cormier}, {Liu}, {Cibinel}, {Gobat}, {Dickinson}, {Sargent}, {Popping},
  {Madden}, {Bethermin}, {Hughes}, {Valentino}, {Rujopakarn}, {Pannella},
  {Bournaud}, {Walter}, {Wang}, {Elbaz}, \& {Coogan}}]{Zanella_2018}
{Zanella}, A., {Daddi}, E., {Magdis}, G., {et~al.} 2018, \mnras, 481, 1976

\bibitem[{Zastrow {et~al.}(2011)Zastrow, Oey, Veilleux, McDonald, \&
  Martin}]{zastrow_ionization_2011}
Zastrow, J., Oey, M.~S., Veilleux, S., McDonald, M., \& Martin, C.~L. 2011,
  ApJ, 741, L17

\bibitem[{{Zavala} {et~al.}(2022){Zavala}, {Casey}, {Spilker}, {Tadaki},
  {Tsujita}, {Champagne}, {Iono}, {Kohno}, {Manning}, \&
  {Monta{\~n}a}}]{Zavala_2022}
{Zavala}, J.~A., {Casey}, C.~M., {Spilker}, J., {et~al.} 2022, \apj, 933, 242

\bibitem[{{Zhou} {et~al.}(2021){Zhou}, {Shi}, {Zhang}, \&
  {Wang}}]{Zhou_co_2021}
{Zhou}, L., {Shi}, Y., {Zhang}, Z.-Y., \& {Wang}, J. 2021, \aap, 653, L10

\end{thebibliography}

\begin{appendix}
\section{Additional figures}
\label{appendix}

\begin{figure}[h!]
\centering
\includegraphics[width=0.45\textwidth]{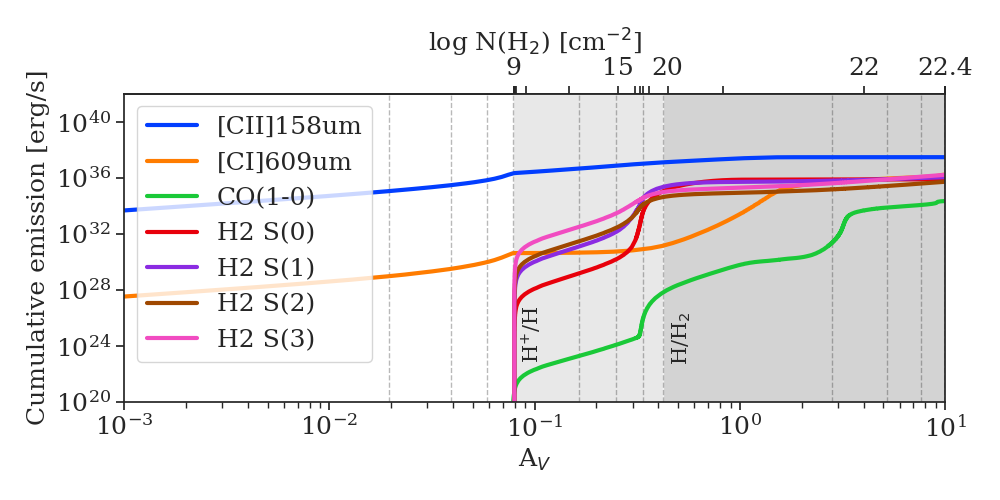}
\includegraphics[width=0.45\textwidth]{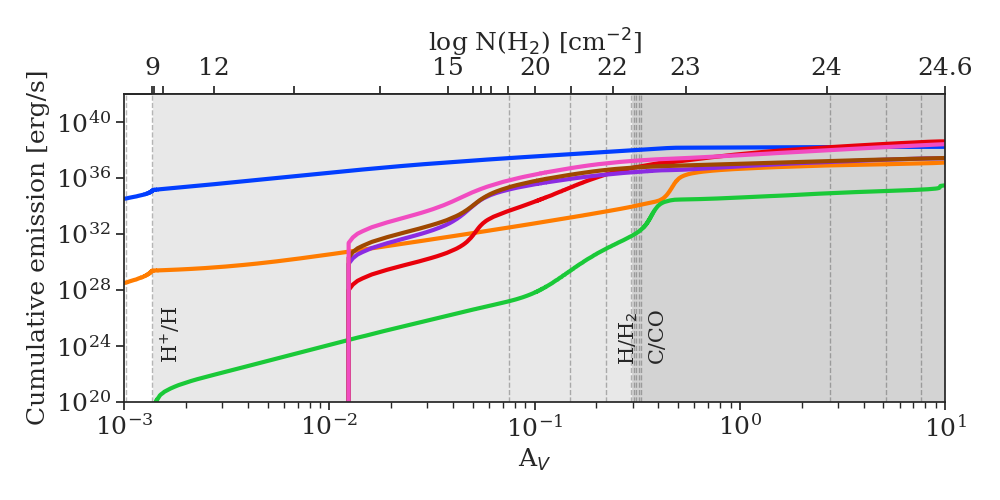}
\caption[Cumulative emission lines profiles of some chosen tracers.]{Cumulative emission lines profiles of some chosen tracers, including the H$_2$ lines for the two models described in Figure \ref{density_profile} and \ref{sampling_cut}.}
\label{H2_problem}
\end{figure}

\begin{figure}[h!]
\centering
\includegraphics[height=0.23\textwidth]{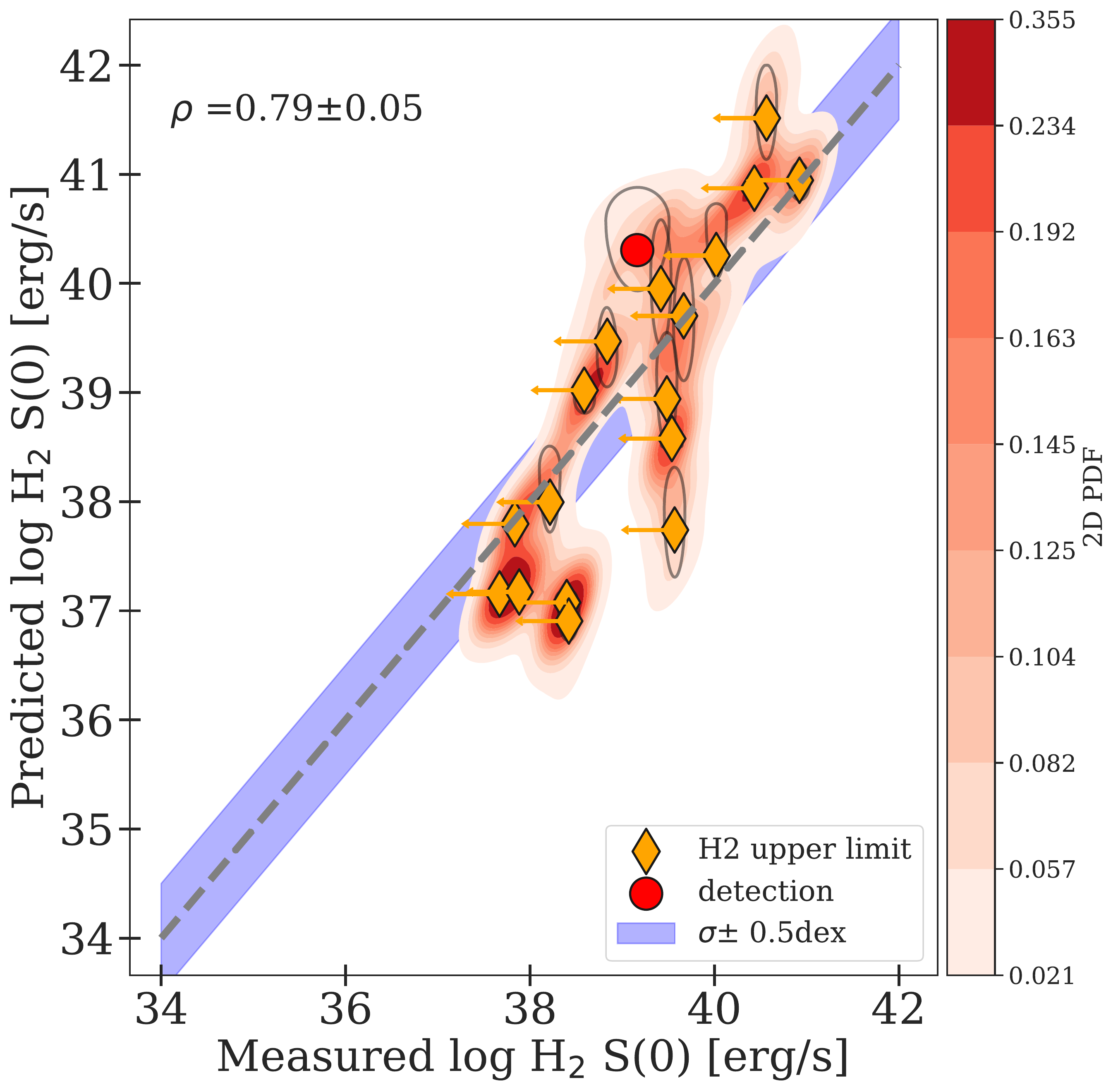}
\includegraphics[height=0.23\textwidth]{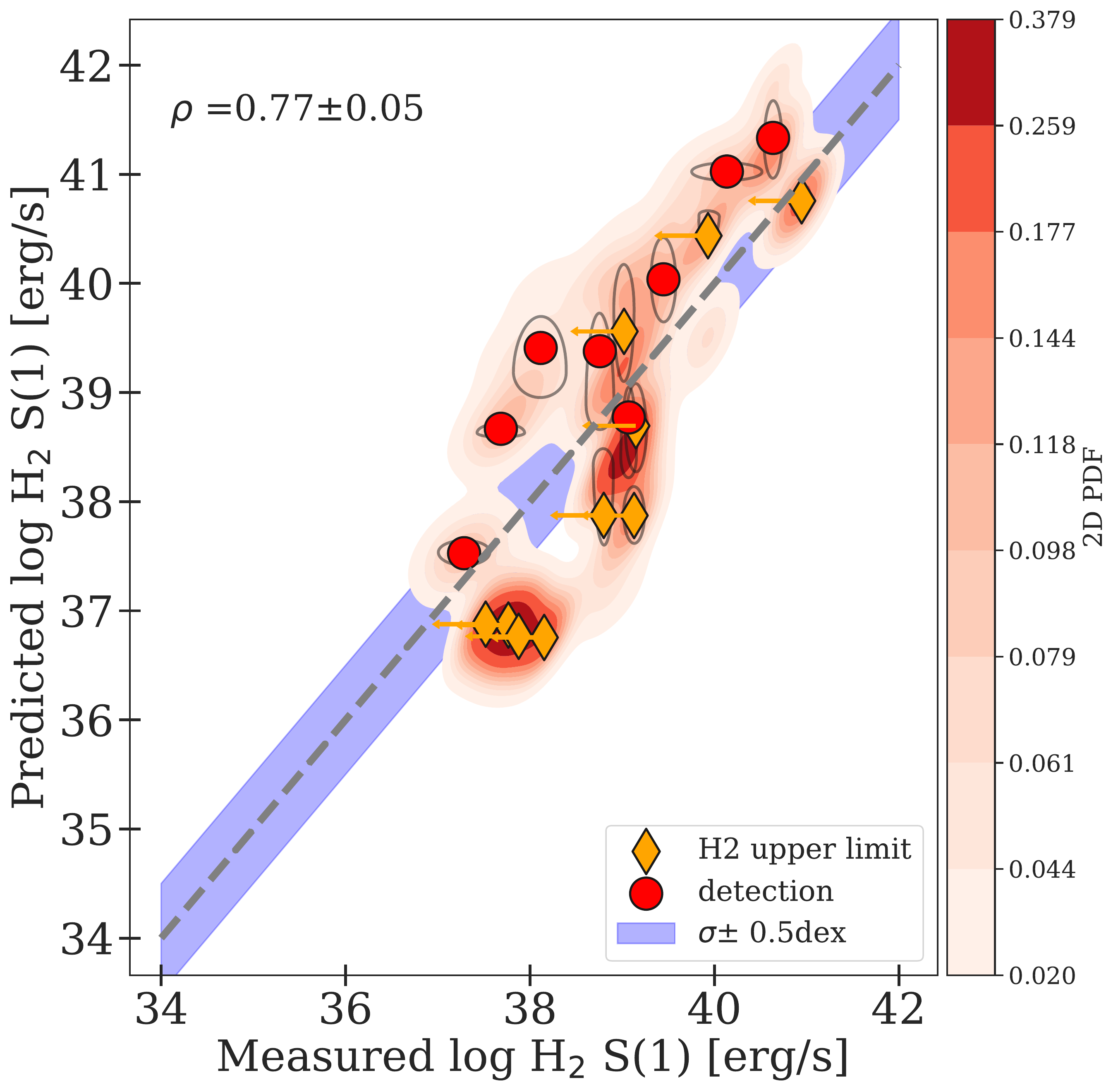}\\
\includegraphics[height=0.23\textwidth]{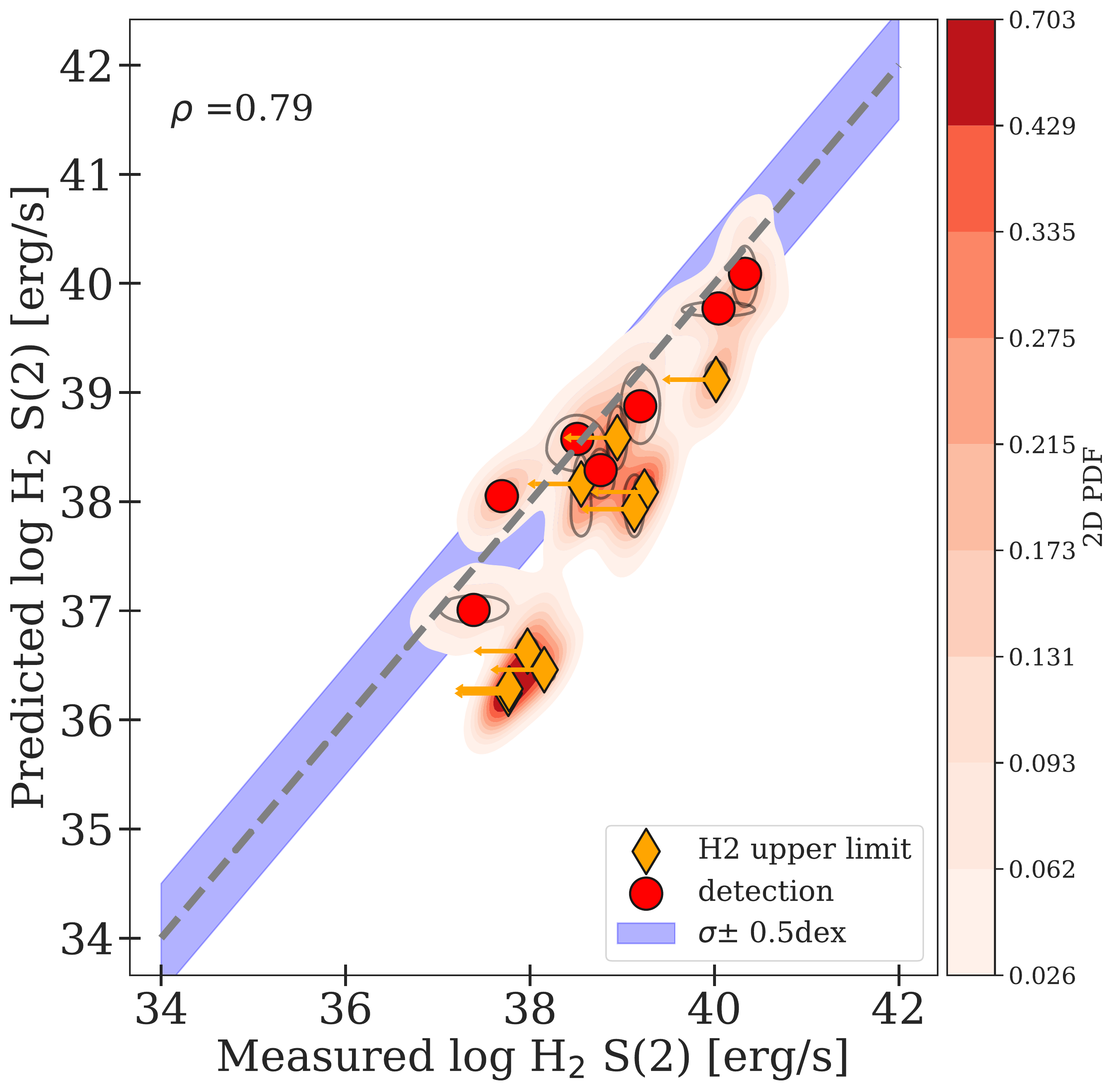}
\includegraphics[height=0.23\textwidth]{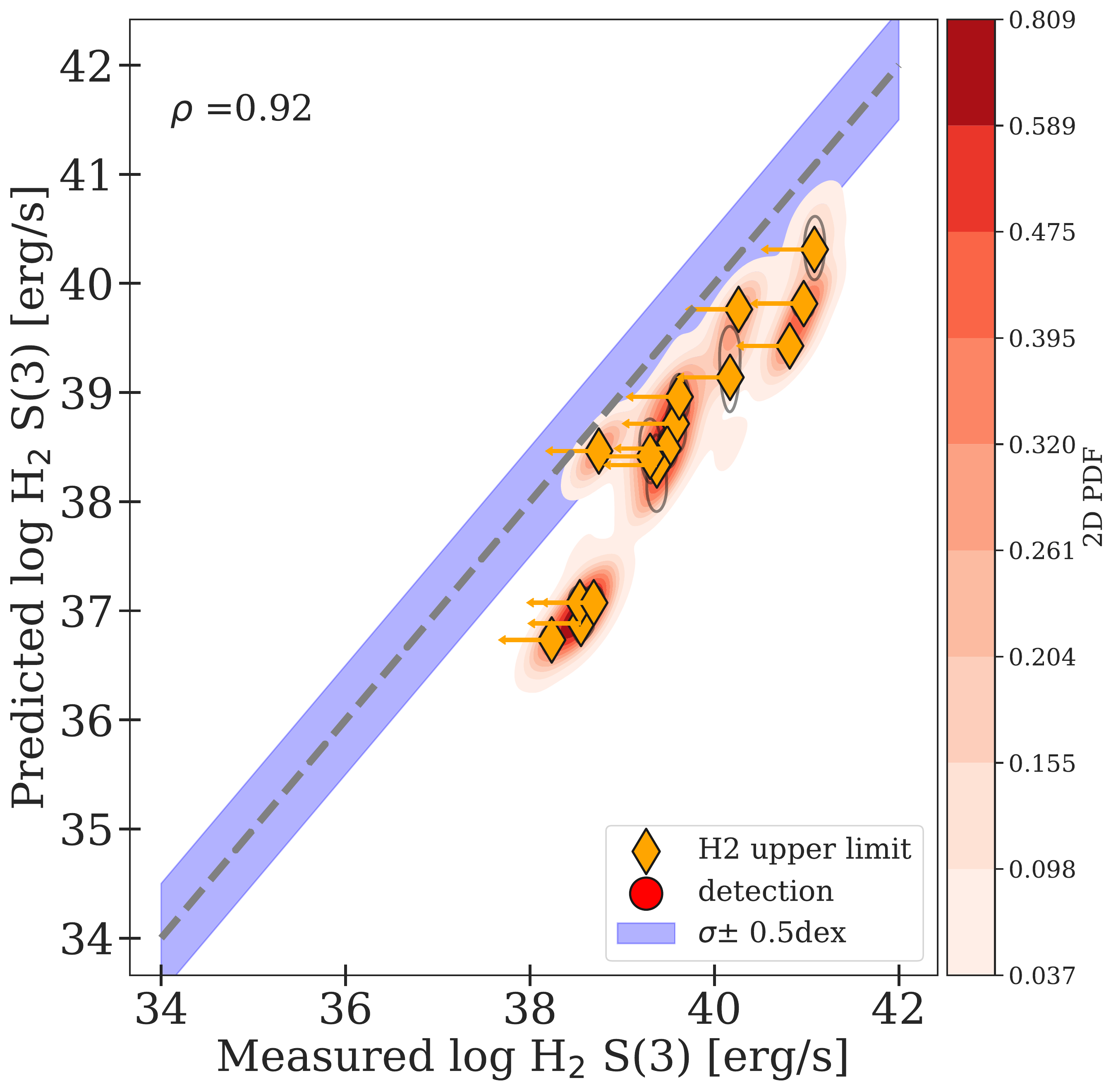}

\caption[Predicted vs. observed H$_2$ lines for power-law models]{Predicted vs. observed H$_2$ lines for power-law models.}
\label{H2_obs_pred}
\end{figure}

\begin{figure}[h!]
\centering
\includegraphics[height=0.32\textwidth]{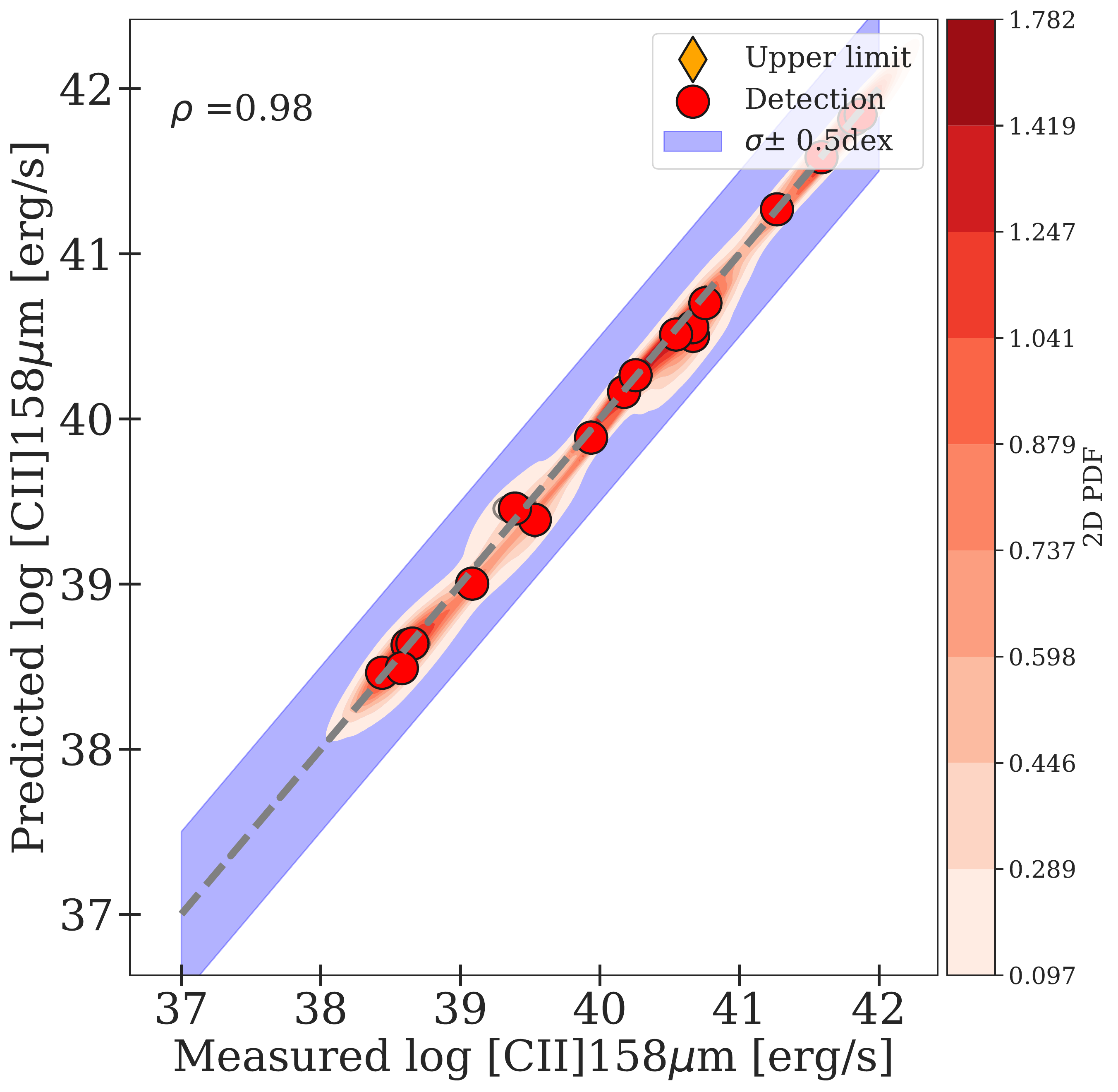}
\includegraphics[height=0.32\textwidth]{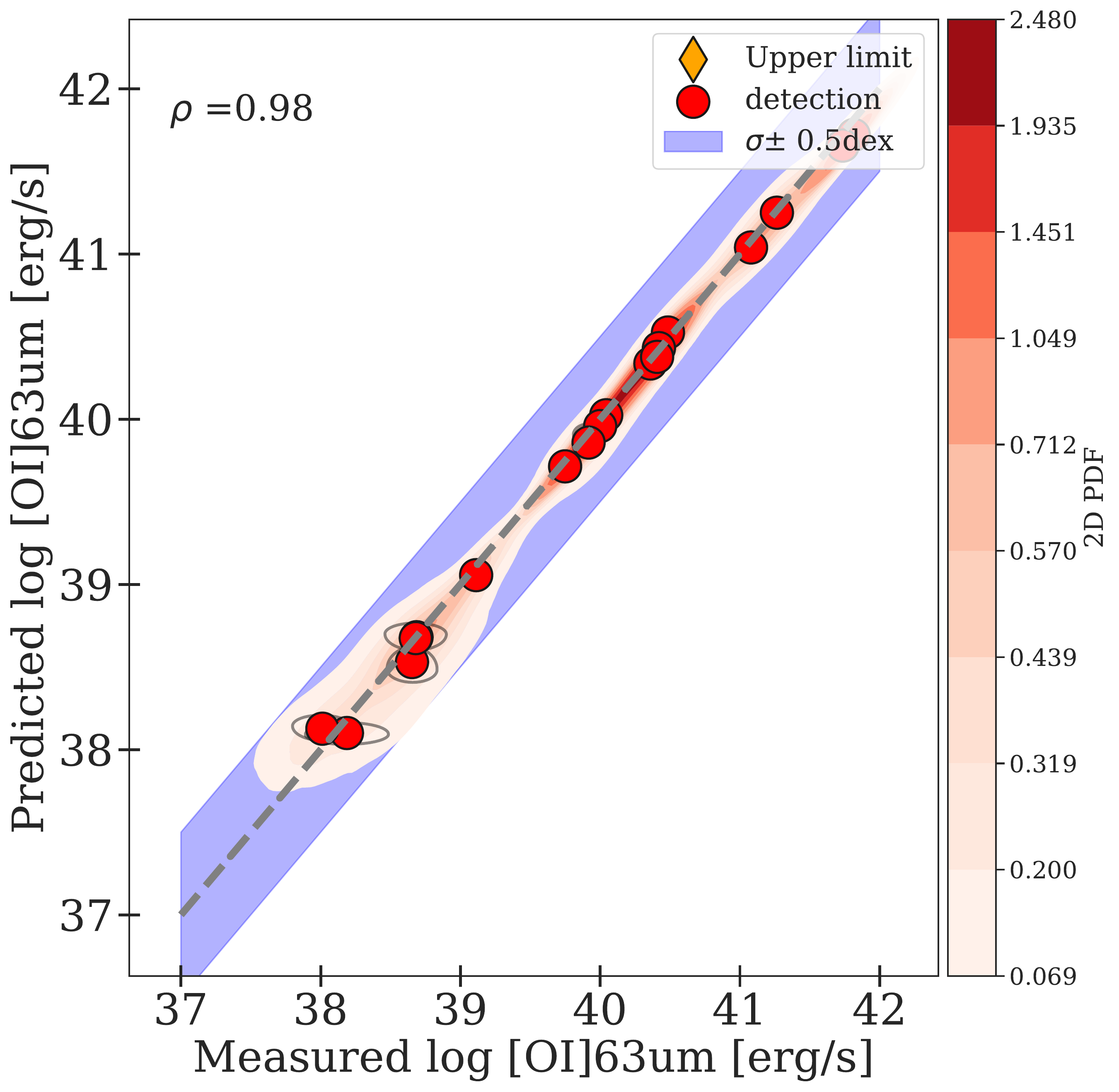}
\includegraphics[height=0.32\textwidth]{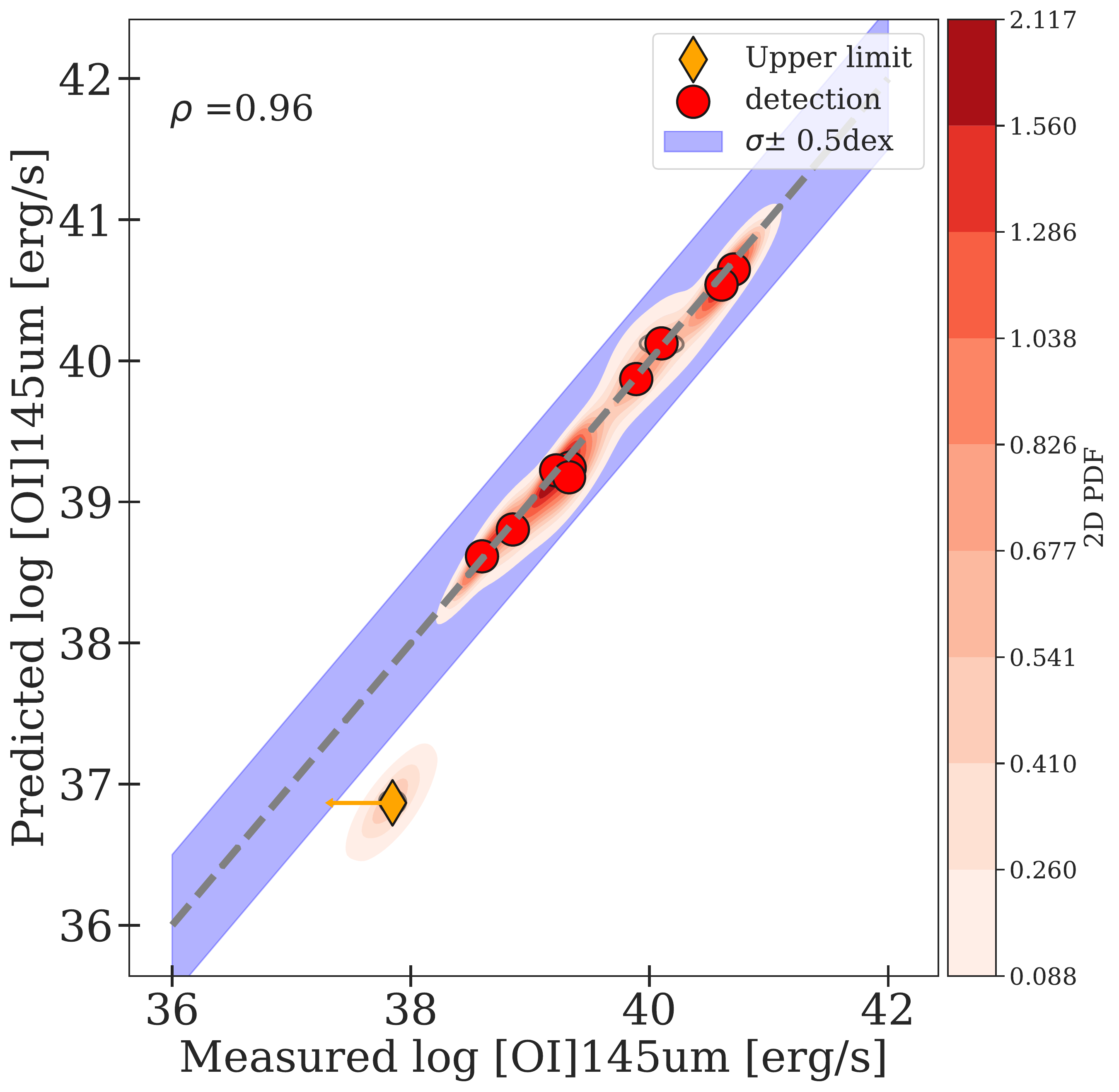}

\caption[Predicted vs. observed PDR lines for power-law models]{Predicted vs. observed PDR lines for power-law models.}
\label{pdr_lines_obs_pred}
\end{figure}

\begin{table*}[h!]
\centering
\label{table_lm}
\begin{threeparttable}
\caption[Best configuration selected for multicomponent modeling with 1 to 4 components.]{Best configuration selected for multicomponent modeling with 1 to 4 components.} 
\label{table_lm}
\begin{tabular}{|p{0.15\textwidth}|p{0.07\textwidth}|p{0.07\textwidth}|p{0.07\textwidth}|p{0.07\textwidth}|p{0.07\textwidth}|p{0.07\textwidth}|}
\hline
Galaxy & N$_{\rm sectors, best}$ & $\mathcal{L}_{\rm M,best}$$^{1}$ & $\Delta \mathcal{L}_{M,1}^{2}$ & $\Delta \mathcal{L}_{M,2}^{2}$& $\Delta \mathcal{L}_{M,3}^{2}$& $\Delta \mathcal{L}_{M,4}^{2}$  \\ \hline 
Haro2&4.0&-52.18&-7.58&-2.41&-0.18&0.0\\
Haro3&3.0&-81.34&-16.74&-3.97&0.0&-3.6\\
Haro11&3.0&-67.03&-4.9&-1.17&0.0&-1.1\\
He2-10&4.0&-86.47&-13.16&-4.37&-2.71&0.0\\
IIZw40&3.0&-52.11&-3.55&-2.93&0.0&-0.8\\
IZw18&2.0&-41.78&-7.98&0.0&-0.37&-1.79\\
Mrk209&2.0&-41.34&-0.36&0.0&-0.19&-2.66\\
Mrk930&1.0&-32.0&0.0&-3.19&-1.59&-4.85\\
Mrk1089&2.0&-49.19&-6.15&0.0&-0.27&-2.26\\
NGC1140&3.0&-64.04&-1.17&-0.83&0.0&-1.09\\
NGC1569&1.0&-70.78&0.0&-2.13&-2.38&-3.58\\
NGC1705&1.0&-48.73&0.0&-5.45&-7.5&-9.09\\
NGC5253&4.0&-61.3&-7.17&-3.44&-1.68&0.0\\
NGC625&2.0&-62.23&-5.93&0.0&-0.04&-1.38\\
SBS0335-052&2.0&-42.4&-1.72&0.0&-1.86&-1.98\\
UM448&3.0&-60.33&-8.61&-1.84&0.0&-4.98\\
UM461&1.0&-36.54&0.0&-0.63&-0.17&-1.01\\
VIIZw403&3.0&-31.06&-6.45&-2.49&0.0&-0.36\\
\hline
\end{tabular}
\begin{tablenotes}
   \footnotesize
   \item We report the absolute value of the logarithm of the marginal likelihood for the best configuration corresponding to N$_{\rm sectors, best}$ and the relative marginal likelihoods $\Delta \mathcal{L}_{M,\rm i}$ = ln($\mathcal{L}_{M, i}$/$\mathcal{L}_{M,\rm best}$) for all architectures.
\end{tablenotes}
\end{threeparttable}
\end{table*}

\begin{table*}[h!]
\begin{threeparttable}
\centering
\caption[Luminosity weighted-averaged and slopes of the $U$, $n$, and cut parameters of the power-law models.]{Luminosity weighted-averaged and slopes of the $U$, $n$, and cut$_1$ (ionized gas), and cut$_2$ (neutral gas) parameters of the power-law models.}
\label{table_mean_values}

\begin{tabular}{|p{0.08\textwidth}|p{0.08\textwidth}p{0.09\textwidth}p{0.08\textwidth}p{0.08\textwidth}|p{0.08\textwidth}p{0.08\textwidth}p{0.08\textwidth}p{0.08\textwidth}|p{0.05\textwidth}|}

\hline
Galaxy & $\log$ <n>$^{(1)}$&$\log$ <U>$^{(2)}$&$\alpha_n^{(3)}$&$\alpha_U^{(3)}$& <cut$_1$>$^{(4)}$ & <cut$_2$>$^{(5)}$ & $\alpha_{\rm cut_1}$$^{(3)}$&$\alpha_{\rm cut_2}^{(3)}$ & P$_{\rm clumpy}$\\ \hline 

\hline 
Haro2&1.29$_{0.42}^{3.42}$&-2.71$_{-3.19}^{-0.82}$&-0.96$_{-1.16}^{-0.79}$&-1.41$_{-1.56}^{-1.24}$&0.74$_{0.32}^{1}$&1.36$_{1}^{1.9}$&0.47$_{-1.0}^{1.63}$&-1.11$_{-2.4}^{0.56}$&0.30 \\
Haro3&1.05$_{0.47}^{3.23}$&-2.41$_{-3.13}^{-1.27}$&-1.27$_{-2.97}^{-0.55}$&-0.84$_{-1.41}^{-0.04}$&0.71$_{0.27}^{1}$&1.27$_{1}^{1.75}$&0.26$_{-1.73}^{1.79}$&-1.51$_{-4.53}^{1.11}$& 0.24\\
Haro11&1.24$_{0.88}^{3.63}$&-2.51$_{-3.31}^{-0.62}$&-1.71$_{-1.78}^{-1.65}$&-0.97$_{-1.12}^{-0.76}$&0.8$_{0.32}^{1}$&1.51$_{1}^{1.82}$&1.17$_{-0.15}^{2.51}$&0.33$_{-1.1}^{1.32}$&0.005\\
\textbf{He2-10}&1.61$_{0.71}^{3.13}$&-2.59$_{-3.18}^{-1.28}$&-0.82$_{-0.92}^{-0.78}$&-1.09$_{-1.48}^{-0.88}$&0.81$_{0.23}^{1}$&1.6$_{1}^{\textbf{2.35}}$&1.46$_{0.59}^{2.22}$&\textbf{-0.72}$_{-1.09}^{-0.06}$&\textbf{0.71}\\
\textbf{IIZw40}&1.55$_{0.8}^{3.05}$&-1.28$_{-2.64}^{-0.91}$&-0.94$_{-1.56}^{0.38}$&0.61$_{-0.52}^{1.8}$&0.7$_{0.32}^{1}$&1.45$_{1}^{\textbf{2.18}}$&-0.11$_{-1.54}^{1.86}$&\textbf{-1.08}$_{-4.39}^{0.39}$&\textbf{0.40}\\
\textbf{IZw18}&1.19$_{1.04}^{3.19}$&-2.63$_{-2.94}^{-0.67}$&-3.19$_{-3.8}^{-2.38}$&-1.92$_{-3.07}^{-0.6}$&0.79$_{0.42}^{1}$&1.14$_{1}^{\textbf{2.12}}$&0.7$_{-0.82}^{2.03}$&\textbf{-3.57}$_{-5.17}^{-1.89}$&\textbf{0.47}\\
\textbf{Mrk209}&1.08$_{0.89}^{3.04}$&-1.25$_{-3.23}^{-0.65}$&-2.73$_{-3.74}^{-1.35}$&0.22$_{-0.6}^{0.97}$&0.82$_{0.39}^{1}$&1.24$_{1}^{\textbf{2.26}}$&1.45$_{-0.59}^{2.55}$&\textbf{-2.28}$_{-2.91}^{-0.15}$&\textbf{0.75}\\
Mrk930&1.08$_{0.91}^{3.24}$&-2.03$_{-2.78}^{-1.02}$&-2.88$_{-3.97}^{-1.7}$&-0.73$_{-1.87}^{0.38}$&0.78$_{0.44}^{1}$&1.52$_{1}^{1.76}$&0.47$_{-0.7}^{1.89}$&0.85$_{-0.18}^{2.14}$&0.03\\
Mrk1089&1.15$_{0.92}^{2.84}$&-2.85$_{-3.58}^{-1.61}$&-2.32$_{-2.9}^{-1.72}$&-0.88$_{-1.08}^{-0.61}$&0.76$_{0.38}^{1}$&1.44$_{1}^{1.87}$&0.56$_{-2.98}^{3.0}$&-0.43$_{-2.37}^{2.26}$&0.25\\
NGC1140&0.61$_{0.42}^{3.02}$&-2.06$_{-2.87}^{-1.18}$&-2.7$_{-3.84}^{-1.52}$&-0.57$_{-2.45}^{0.69}$&0.75$_{0.26}^{1}$&1.31$_{1}^{1.49}$&0.64$_{-0.91}^{2.25}$&0.77$_{-0.85}^{4.63}$&0.01\\
NGC1569&1.08$_{0.94}^{3.2}$&-2.13$_{-3.09}^{-1.23}$&-3.34$_{-4.38}^{-2.58}$&-0.45$_{-2.4}^{0.29}$&0.66$_{0.34}^{1}$&1.39$_{1}^{1.81}$&-0.55$_{-2.17}^{1.02}$&-0.65$_{-2.54}^{1.15}$&0.06\\
NGC1705&1.1$_{0.94}^{3.29}$&-3.24$_{-3.44}^{-0.94}$&-3.0$_{-3.82}^{-2.15}$&-2.58$_{-3.2}^{-2.24}$&0.77$_{0.44}^{1}$&1.37$_{1}^{1.89}$
&0.41$_{-1.73}^{2.42}$&-0.99$_{-3.34}^{2.77}$&0.23\\
NGC5253&0.86$_{0.65}^{3.21}$&-2.44$_{-3.33}^{-0.66}$&-2.5$_{-2.91}^{-1.57}$&-0.89$_{-1.73}^{-0.17}$&0.7$_{0.31}^{1}$&1.49$_{1}^{1.78}$
&-0.03$_{-0.52}^{0.57}$&0.44$_{-0.93}^{1.59}$&0.02\\
NGC625&1.14$_{0.9}^{3.26}$&-2.02$_{-2.47}^{-0.87}$&-2.33$_{-3.14}^{-1.48}$&-1.33$_{-2.41}^{-0.48}$&0.69$_{0.36}^{1}$&1.51$_{1}^{1.89}$&-0.39$_{-1.37}^{0.71}$&-0.11$_{-1.28}^{1.48}$&0.19\\
SBS0335&2.0$_{1.83}^{2.91}$&-1.48$_{-2.93}^{-0.83}$&-2.89$_{-4.45}^{-1.31}$&0.05$_{-1.03}^{1.78}$&0.69$_{0.38}^{1}$&1.59$_{1}^{1.85}$&-0.52$_{-2.7}^{1.76}$&0.88$_{-0.33}^{2.36}$&0.05\\
UM448&0.59$_{0.4}^{3.12}$&-2.22$_{-2.93}^{-1.02}$&-2.74$_{-3.91}^{-1.8}$&-0.89$_{-2.64}^{0.22}$&0.83$_{0.43}^{1}$&1.22$_{1}^{1.42}$&1.41$_{-0.36}^{3.05}$&-0.08$_{-1.84}^{1.43}$&0.003\\
\textbf{UM461}&1.04$_{0.83}^{3.26}$&-1.51$_{-3.26}^{-0.63}$&-2.5$_{-3.59}^{-1.64}$&-0.11$_{-0.66}^{0.63}$&0.69$_{0.37}^{1}$&1.31$_{1}^{\textbf{2.24}}$&-0.43$_{-1.98}^{1.21}$&\textbf{-1.75}$_{-2.46}^{-1.0}$&\textbf{0.58}\\
\textbf{VIIZw403}&1.17$_{0.88}^{3.13}$&-2.62$_{-3.23}^{-0.68}$&-1.95$_{-3.17}^{-1.16}$&-1.2$_{-1.46}^{-0.9}$&0.72$_{0.33}^{1}$&1.45$_{1}^{\textbf{2.09}}$&0.12$_{-1.36}^{1.41}$&\textbf{-0.93}$_{-1.95}^{-0.02}$&\textbf{0.49}\\
\hline
\end{tabular}
\begin{tablenotes}
   \footnotesize
   \item The reported errorbars for weighted-averages correspond to the upper bound and lower boundaries of the power-law, and to the High Density Probability Interval at 94\% for the slopes. Galaxies flagged in bold with maximum cuts larger than 2 and negative $\alpha_{\rm cut_2}$ slopes correspond to clumpy galaxies, as defined by Equation \ref{eq_clumpy}.
\end{tablenotes}
\end{threeparttable}
\end{table*}

\begin{table*}[h!]
\label{table_pnsigma}
\begin{threeparttable}
\caption[P(3$\sigma$) for all lines, PDR lines, and CO(1-0) line]{P(3$\sigma$) probabilities and averaged p-values for the power-law models, multicomponent and single component.}
\label{3sigma_values}

\begin{tabular}{|p{0.15\textwidth}|p{0.07\textwidth}p{0.07\textwidth}p{0.07\textwidth}|p{0.07\textwidth}p{0.07\textwidth}p{0.07\textwidth}|p{0.07\textwidth}p{0.07\textwidth}p{0.07\textwidth}|}
\hline
Galaxy& All$_{\rm p-law}$&All$_{\rm multi}$&All$_{\rm single}$&PDR$_{\rm p-law}$&PDR$_{\rm multi}$&PDR$_{\rm single}$&CO$_{\rm p-law}$&CO$_{\rm multi}$&CO$_{\rm single}$\\
\hline
Haro2&91.50 (0.749)&88.00 (0.750)&78.30 (0.750)&55.54 (0.754)&57.83 (0.576)&53.59 (0.896)&92.00 (0.816)&\textbf{\textcolor{red}{4.60 (1.000)}}&\textbf{\textcolor{red}{20.20 (1.000)}}\\
Haro3&74.20 (0.750)&81.6 (0.750)&71.70 (0.747)&68.62 (0.583)&75.62 (0.414)&79.07 (0.334)&86.20 (0.010)&\textbf{35.05 (0.571)}&\textbf{\textcolor{red}{3.90 (0.994)}}\\
Haro11&87.60 (0.750)&85.00 (0.750)&72.80 (0.7500)&\textbf{48.90 (0.478)}&\textbf{45.45 (0.515)}&53.00 (0.720)&57.23 (0.752)&67.77 (0.390)&70.62 (0.811)\\
He2-10&77.30 (0.749)&79.20 (0.750)&65.60 (0.750)&63.90 (0.502)&61.97 (0.389)&55.18 (0.729)&51.55 (0.230)&\textbf{45.65 (0.237)}&97.20 (0.754)\\
IIZw40&95.60 (0.750)&92.80 (0.750)&83.90 (0.750)&80.75 (0.415)&80.39 (0.383)&81.18 (0.565)&85.60 (0.080)&52.20 (0.162)&\textbf{\textcolor{red}{0.00 (1.000)}}\\
IZw18&93.00 (0.700)&89.00 (0.702)&80.70 (0.750)&79.64 (0.629)&68.56 (0.724)&79.68 (0.615)&87.75 (0.150)&90.95 (0.756)&\textbf{\textcolor{red}{0.00 (1.000)}}\\
Mrk209&86.80 (0.744)&80.60 (0.750)&78.70 (0.750)&79.18 (0.559)&79.53 (0.454)&81.26 (0.460)&71.50 (0.318)&\textbf{\textcolor{red}{0.00 (1.000)}}&\textbf{\textcolor{red}{0.00 (1.000)}}\\
Mrk930&98.70 (0.750)&99.30 (0.750)&99.3 (0.750)&90.53 (0.371)&87.08 (0.576)&87.08 (0.576)&99.95 (0.950)&100.0$^*$ (1.000)&100.0$^*$ (1.000)\\
Mrk1089&88.00 (0.750)&84.30 (0.746)&79.30 (0.658)&64.35 (0.358)&67.25 (0.384)&63.31 (0.614)&76.30 (0.559)&55.15 (0.842)&57.70 (0.421)\\
NGC1140&77.4 (0.742)&81.70 (0.704)&75.20 (0.748)&61.21 (0.567)&65.66 (0.486)&66.12 (0.615)&53.35 (0.165)&58.25 (0.129)&\textbf{\textcolor{red}{0.55 (0.997)}}\\
NGC1569&76.40 (0.750)&74.60 (0.750)&74.60 (0.750)&\textbf{49.02 (0.403)}&51.26 (0.678)&51.26 (0.678)&82.70 (0.396)&98.80 (0.940)&98.80 (0.940)\\
NGC1705&87.60 (0.740)&84.70 (0.750)&84.70 (0.750)&99.95 (0.981)&100.0 (1.000)&100.0 (1.000)&\textbf{45.94 (0.433)}&\textbf{45.37 (0.430)}&\textbf{45.37 (0.430)}\\
NGC5253&80.10 (0.750)&82.30 (0.750)&79.40 (0.750)&97.40 (0.204)&72.65 (0.463)&67.95 (0.128)&65.23 (0.624)&62.55 (0.429)&58.12 (0.577)\\
NGC625&70.90 (0.749)&70.90 (0.749)&64.10 (0.750)&\textbf{41.88 (0.777)}&\textbf{44.4 (0.705)}&\textbf{44.96 (0.785)}&52.20 (0.828)&\textbf{\textcolor{red}{10.25 (0.999)}}&\textbf{\textcolor{red}{0.00 (1.000)}}\\
SBS0335-052&70.60 (0.749)&77.20 (0.750)&74.40 (0.725)&79.13 (0.543)&82.56 (0.397)&82.90 (0.590)&100.0$^*$ (0.500)&100.0$^*$ (0.500)&100.0$^*$ (0.500)\\
UM448&78.80 (0.750)&88.30 (0.750)&80.30 (0.698)&71.33 (0.754)&79.83 (0.345)&61.11 (0.603)&69.25 (0.200)&91.5 (0.448)&64.15 (0.896)\\
UM461&90.70 (0.734)&88.8 (0.750)&88.8 (0.750)&82.83 (0.385)&83.22 (0.615)&83.22 (0.615)&100.0$^*$ (0.500)&100.0$^*$ (0.500)&100.0$^*$ (0.500)\\
VIIZw403&92.5 (0.748)&95.00 (0.742)&72.90 (0.731)&86.67 (0.546)&88.46 (0.569)&83.47 (0.610)&100.0$^*$ (0.499)&100.0$^*$ (0.500)&100.0$^*$ (0.500)\\
\hline
\end{tabular}
\begin{tablenotes}
   \footnotesize
   \item P(3$\sigma$) represents the probability that the inferred value from MCMC sampling fall within 3$\sigma$ from the observed value. We report P(3$\sigma$) as averaged percentages for all lines, PDR lines, and the CO(1-0) line alone. We provide averaged values for the following PDR lines: \cii\ 158$\mu$m, \silii\ 34$\mu$m, \oi\ 63$\mu$m, \oi\ 145$\mu$m, \feii 17$\mu$m, and \feii\ 25$\mu$m. P(3$\sigma$) below 50\% are shown in bold and below 25\% in red. The averaged p-values are reported between parenthesis; they should ideally be close to 0.5. Very low p-value (near 0) or very high (near 1) can be used to identify overfitting/underfitting.
   \item * : P(3$\sigma$)= 100\% correspond to observed upper limits which are well-matched by models. 
\end{tablenotes}
\end{threeparttable}
\end{table*}

\end{appendix}
\end{document}